\documentclass[twocolumn]{aastex62}

\usepackage{rotating}
\usepackage{appendix}
\usepackage{graphicx}
\usepackage{epstopdf}
\usepackage{natbib}
\usepackage{xspace}
\usepackage{color}
\usepackage{amsmath}

\newcommand{\kms}     {~km~s$^{-1}$\xspace}

\newcommand{\mjy}     {~mJy~beam$^{-1}$\xspace}
\newcommand{\mujy}    {~$\mu$Jy~beam$^{-1}$\xspace}

\newcommand{\msun}    {~$M_{\sun}$\xspace}
\newcommand{\msunyr}{~$M_{\sun}$~yr$^{-1}$\xspace}

\newcommand{\cmd}     {~cm$^{-2}$\xspace}

\newcommand{\dechms}[4]{$#1^{\rm h}#2^{\rm m}#3\mbox{$^{\rm s}\mskip-7.6mu.\,$}#4$} 
 
\newcommand{\decdms}[4]{$-#1^{\circ}#2'#3\mbox{$''\mskip-7.6mu.\,$}#4$}

\newcommand{\R}[1]{\ensuremath{R_\mathrm{#1}}}

\newcommand{\I}[1]{\ensuremath{I_\mathrm{#1}}}
\renewcommand{\H}[1]{\ensuremath{H_\mathrm{#1}}}
\newcommand{\beq} {\begin{equation}}
\newcommand{\eeq} {\end{equation}}
\newcommand{\beqa}{\begin{eqnarray}}
\newcommand{\eeqa}{\end{eqnarray}}
\newcommand{\ad}[1]{#1}

\shorttitle{Disk and Envelope Streamers of the GGD27-MM1 Massive Protostar}
\shortauthors{Fern\'andez L\'opez et al.}
\begin{document}

\title{Disk and Envelope Streamers of the GGD27-MM1 Massive Protostar}


\author[0000-0001-5811-0454]{M. Fern\'andez-L\'opez}
\affiliation{Instituto Argentino de Radioastronom\'\i a (CCT-La Plata, CONICET; CICPBA), C.C. No. 5, 1894, Villa Elisa, Buenos Aires, Argentina}
\email{manferna@gmail.com}

\author[0000-0002-3829-5591]{J. M. Girart}
\affiliation{Institut de Ciencies de l’Espai (ICE-CSIC), Campus UAB, Carrer de Can Magrans S/N, E-08193 Cerdanyola del Valles, Catalonia}
\affiliation{Institut d'Estudis Espacials de Catalunya (IEEC), c/ Gran Capità, 2-4, 08034 Barcelona, Spain}

\author[0000-0002-5845-8722]{J. A. L\'opez-V\'azquez}
\affiliation{Academia Sinica Institute of Astronomy and Astrophysics, No. 1, Sec. 4, Roosvelt Road, Taipei 10617, Taiwan}

\author[0000-0001-7341-8641]{R. Estalella}
\affiliation{Departament de Física Quàntica i Astrofísica (FQA), Universitat de Barcelona (UB), c/ Martí i Franquès 1, 08028 Barcelona, Spain}

\author[0000-0002-2189-6278]{G. Busquet}
\affiliation{Departament de Física Quàntica i Astrofísica (FQA), Universitat de Barcelona (UB), c/ Martí i Franquès 1, 08028 Barcelona, Spain}
\affiliation{Institut de Ciències del Cosmos (ICCUB), Universitat de Barcelona (UB), c/ Martí i Franquès 1, 08028 Barcelona, Spain}
\affiliation{Institut d'Estudis Espacials de Catalunya (IEEC), c/ Gran Capità, 2-4, 08034 Barcelona, Spain}

\author[0000-0003-4576-0436]{S. Curiel}
\affiliation{Instituto de Astronom\'{i}a, Universidad Nacional Aut\'onoma de M\'exico (UNAM), Apartado Postal 70-264, DF 04510 M\'exico}

\author[0000-0002-3255-2665]{N. A\~nez-L\'opez}
\affiliation{Université Paris-Saclay, Université Paris Cité, CEA, CNRS, AIM, 91191, Gif-sur-Yvette, France}



\begin{abstract}
We present new Atacama Large (sub)Millimeter Array 0.98\,mm observations of the continuum emission and several molecular lines toward the high-mass protostellar system GGD27-MM1, driving the HH\,80-81 radio-jet. The detailed analysis of the continuum and the CH$_3$CN molecular emission allows us to separate the contributions from the dust content of the disk (extending up to 190\,au), the molecular content of the disk (extending from 140 to 360\,au), and the content of the envelope, revealing the presence of several possible accretion streamers (also seen in other molecular tracers, such as CH$_3$OH). We analyze the physical properties of the system, producing temperature and column density maps, and radial profiles for the disk and the envelope. We qualitatively reproduce the trajectories and line-of-sight velocities of the possible streamers using a theoretical model approach. An ad-hoc model of a flared disk comprising a hot dust disk embedded in cold gas fits the H$_2$S emission, which revealed the molecular disk as crescent-shape with a prominent central absorption. Another fit to the central absorption spectrum suggests that the absorption is probably caused by different external cold layers from the envelope or the accretion streamers. Finally, the analysis of the rotation pattern of the different molecular transitions in the molecular disk, suggests that there is an inner zone devoid of molecular content.      
\end{abstract}


\section{Introduction} \label{sec:intro}

The presence of rotationally supported disks around low-mass young stellar objects (YSO) appears to be almost ubiquitous. The disks timescales are about a few million years, large enough to be considered planet-forming disks \citep{2001Haisch}. Recent ALMA observations have allowed a significant advance in the understanding of the properties of these disks and their evolution \citep[e.g.,][]{2020Andrews,2021Oberg}. 

Because their typically shorter timescales, larger distances, and large obscuration, the properties of the disks around massive stars, as well as their evolution, are less known. There are many large, $\sim 1000$~au, disk--like structures with clear velocities gradients \citep[e.g.][]{Sanchez13, Cesaroni17, Beuther17, 2020Tanaka, 2022Williams}. Smaller disks are, however, more difficult to find \citep[e.g.,][]{Ilee18,Maud19}, but they are usually very massive (2--5~\msun) and hot ($\gtrsim $400~K). The closest case of a disk around a massive star, Orion~I, has peculiar features such as its SiO maser emission and its association with an explosive outflow \citep{Plambeck16, Hirota17, Ginsburg19, Wright22}, only found in a handful of sources \citep{2022GuzmanCcolque} and it probably cannot be considered as a reference case. The rapid evolution of massive stars, makes the disk to be fully ionized in relatively short timescales \citep[e.g.,][]{2020JimenezS}.

The traditional view is that rotationally supported disks are fed by infalling, relatively flattened envelopes. In this scenario, the envelope can be understood as the expected pseudodisk in the classical theoretical picture of unstable magnetized cloud cores \citep{1993Galli}. However, recent observations show that accretion towards disks around low-mass YSOs is probably significantly asymmetric along all their evolutionary stages \citep{2019Alves,2019Yen,2019LeGouellec,2020Alves, 2020Pineda,2020Huang,2021Huang,2021Grant,2022Garufi,2022Murillo,2022Thieme,2022Pinte,2022Pineda}. This transforms the steady and more static view of the accretion process into a more dynamic and chaotic scenario.

The massive star driving the highly collimated jet associated with HH~80-81 \citep[e.g.,][]{1989Rodriguez, 1993Marti, 2012Masque, 2015Masque}, is one of these protostellar systems surrounded by a relatively compact, very massive and hot disk, associated with the millimeter source GGD27-MM1 \citep{2018Girart}. The emission can well be fitted by scale-up versions of accretion disk models around low-mass young stars \citep{2020AnezLopez}. This allowed to constrain the disk radius, $\simeq$170~au, the stellar mass, $\simeq$20~\msun, and the disk mass, $\simeq$5~\msun\ \citep[assuming a distance of 1.4~kpc, see][]{2020AnezLopez}. It has also allowed to theoretically derive the density and temperature radial and vertical profiles of the disk \citep{2020AnezLopez}. The dusty disk is surrounded by a bright molecular, disk-like structure with a radius of $\sim$1000~au, which is also perpendicular to the radio jet \citep{2003Gomez, 2011FernandezLopez1, 2011FernandezLopez2, 2012Carrasco, 2017Girart}. The presence of a rotating disk perpendicular to a highly collimated jet suggests that at least some massive stars form similarly to their lower mass counterparts, but with a more vigorous and in a significantly more energetic scale \citep{1998Heathcote, 2001Molinari, 2010Carrasco, 2017Kamenetzky, 2019Kamenetzky}.

In this work we present new continuum and molecular line observations taken with the Atacama Large (sub)Millimeter Array (ALMA) telescope toward the GGD27-MM1 protostellar system. A description of the observations is provided in Section \ref{sec:obs}. Section \ref{sec:results} describes the continuum and molecular diversity content of the system, and Section \ref{sec:analysis} presents a detailed analysis of its physical properties from studying the molecular emission in the disk, the envelope and an ensemble of possible streamers reported in the system for the first time. In Section \ref{sec:discussion} we discuss the results of the analysis and in Section \ref{sec:conclusions} we summarize the main conclusions of this contribution.

\section{Observations} \label{sec:obs}
We performed ALMA Cycle 4 observations (project 2015.1.00480.S)
taken at Band 7 in GGD\,27. These observations comprised two tracks carried out on 6 and 7 September 2016. Both used 39 antennas for a total on-source integration time of 60.7\,minutes. The baselines ranged between 15.1\,m and 2500\,m. The weather conditions reported at the ALMA site during the observations indicated a precipitable water column of 0.5\,mm and 0.6\,mm for the two tracks respectively, with about 1.1 airmass for both. The median T$_{sys}$ ranged from 80\,K to 120\,K at the observing frequency. The phase center was placed at ($\alpha$,$\delta$)$_{J2000}$=(\dechms{18}{19}{12}{101},\decdms{20}{47}{30}{984}), and the primary beam of the 12\,m antennas is $19\arcsec$.


The correlator setup included four spectral windows (two in each sideband) centered at 298.923\,GHz (spw25), 300.435\,GHz (spw27), 309.715\,GHz (spw31) and 312.853\,GHz (spw29). The spectral resolution was 0.5\kms for spw27 and spw31, and 1.0\kms for spw25 and spw29. The total continuum bandwidth was about 3.750\,GHz and 2.320\,GHz before and after spectral line removal, respectively. 


The data were calibrated following the standard pipeline procedures inside the CASA package \citep[version 4.7.0-1]{2007McMullin}. J1924-2914 was used as flux and bandpass calibrator, while J1832-2039 was used as phase calibrator. We set the flux of J1924-2914 at 298.923\,GHz to 3.431\,Jy, and we assume its spectral index to be -0.595\footnote{The flux and spectral index of this quasar were provided with the pipeline and extracted from the ALMA flux monitoring of quasars. The spectral index is defined here as $\alpha$, with $S_{\nu}\propto\nu^{\alpha}$.}. After setting J1924-2914 data amplitudes to this flux scale, and apply the correction to all the data, we obtained a flux for J1832-2039 of 303$\pm$4mJy at 298.923\,GHz for track\,1 and 312$\pm$2mJy for track\,2. 

We constructed a continuum line-free dataset using a Briggs weighting with robust=0.5. We made three self-calibration iterations which improved the signal-to-noise ratio of the continuum image. The final continuum image has an rms of 0.14\mjy and a synthesized beam of $0\farcs14\times0\farcs11$ (P.A.= $61\degr$). Then, the self-calibration solutions were applied to the spectral windows. In this work, we only analyzed a few spectral lines of the detected in these observations (see Table \ref{t:molecules}). The rest of the lines will be presented in a forthcoming contribution.

\begin{deluxetable*}{lccccccc}
\tablewidth{0pt}
\tablecolumns{8}
\tabletypesize{\scriptsize}
\tablecaption{Spectral lines}
\tablehead{
\colhead{Transition} & \colhead{$\nu_{rest}$} & \colhead{E$_{up}$} & \colhead{log$_{10}$(A$_{ij}$)}& \multicolumn{2}{c}{Synthesized Beam} & \colhead{$\delta v$} & \colhead{rms}  \\
\colhead{}  & \colhead{[GHz]} & \colhead{[K]} & \colhead{[log$_{10}$(s$^{-1}$)]} & \colhead{[$\arcsec\times\arcsec$]} & \colhead{[$\degr$]} & \colhead{[\kms]} & \colhead{[\mjy]}
}
\startdata
CH$_3$CN 17$_{0}$-16$_{0}$ & 312.687743 & 135.1 & -2.73710 & $0\farcs17\times0\farcs13$ & 77.3 & 0.9 & 1.9 \\
CH$_3$CN 17$_{1}$-16$_{1}$ & 312.681731 & 142.2 & -2.73867 & $0\farcs17\times0\farcs13$ & 77.3 & 0.9 & 1.9 \\
CH$_3$CN 17$_{2}$-16$_{2}$ & 312.663699 & 163.6 & -2.74327 & $0\farcs17\times0\farcs13$ & 77.3 & 0.9 & 1.9 \\
CH$_3$CN 17$_{3}$-16$_{3}$ & 312.633653 & 199.4 & -2.75105 & $0\farcs17\times0\farcs13$ & 77.3 & 0.9 & 1.9 \\
CH$_3$CN 17$_{4}$-16$_{4}$ & 312.591607 & 249.3 & -2.76222 & $0\farcs17\times0\farcs13$ & 77.3 & 0.9 & 1.9 \\
CH$_3$CN 17$_{5}$-16$_{5}$ & 312.537576 & 313.6 & -2.77699 & $0\farcs17\times0\farcs13$ & 77.3 & 0.9 & 1.9 \\
CH$_3$CN 17$_{6}$-16$_{6}$ & 312.471584 & 392.1 & -2.79579 & $0\farcs17\times0\farcs13$ & 77.3 & 0.9 & 1.9 \\
CH$_3$CN 17$_{7}$-16$_{7}$ & 312.393656 & 484.8 & -2.81905 & $0\farcs17\times0\farcs13$ & 77.3 & 0.9 & 1.9 \\
CH$_3$CN 17$_{8}$-16$_{8}$ & 312.303825 & 591.7 & -2.84738 & $0\farcs17\times0\farcs13$ & 77.3 & 0.9 & 1.9 \\
CH$_3$CN 17$_{9}$-16$_{9}$ & 312.202126 & 712.8 & -2.88193 & $0\farcs17\times0\farcs13$ & 77.3 & 0.9 & 1.9 \\\hline
SO$_2$ 9$_{2,8}$-8$_{1,7}$ & 298.576307 & 51.0 & -3.84092 & $0\farcs19\times0\farcs14$ & 88.2 & 1.0 & 1.5 \\
SO$_2$ 12$_{6,6}$-13$_{5,9}$ & 312.258421 & 160.0 & -4.38674 & $0\farcs16\times0\farcs13$ & 77.0 & 1.0 & 1.6 \\
SO$_2$ 19$_{3,17}$-19$_{2,18}$ & 299.316816 & 197.0 & -3.69102 & $0\farcs17\times0\farcs14$ & 78.5 & 1.0 & 1.4 \\
SO$_2$ 32$_{3,29}$-32$_{2,30}$ & 300.273416 & 518.7 & -3.59400 & $0\farcs18\times0\farcs13$ & 79.1 & 0.5 &  1.7 \\
SO$_2$ 33$_{10,24}$-34$_{9,25}$ & 310.017205 & 760.5 & -4.21948 & $0\farcs16\times0\farcs13$ & 76.8 & 0.5 & 1.8 \\\hline
CH$_3$OH 3$_{1,2}$-2$_{0,2}$ & 310.192994 & 35.0 & -4.05554 & $0\farcs17\times0\farcs13$ & 76.8 & 0.5 & 1.8 \\
H$_2$CO 4$_{1,3}$-3$_{1,2}$ & 300.836635 & 47.9 & -3.14420 & $0\farcs18\times0\farcs13$ & 80.0 & 0.5 & 1.8 \\
H$_2$S 3$_{3,0}$-3$_{2,1}$ & 300.505560 & 168.9 & -3.98895 & $0\farcs18\times0\farcs13$ & 79.2 & 0.5 & 1.9 \\
HC$_3$N 33-32 & 300.159647 & 244.9 & -2.66753 & $0\farcs18\times0\farcs13$ & 79.1 & 0.5 & 1.7 \\
\enddata 
\tablecomments{The quoted rest frequencies have been extracted from the Molecular Spectroscopy database of the Jet Propulsion Laboratory, JPL, using the on-line port Splatalogue.}
\label{t:molecules}
\end{deluxetable*}

\section{Results}\label{sec:results}
\subsection{The GGD27-MM1 disk and envelope system: continuum emission}\label{sec:disk_continuum}
\begin{figure}
  \includegraphics[width=\linewidth]{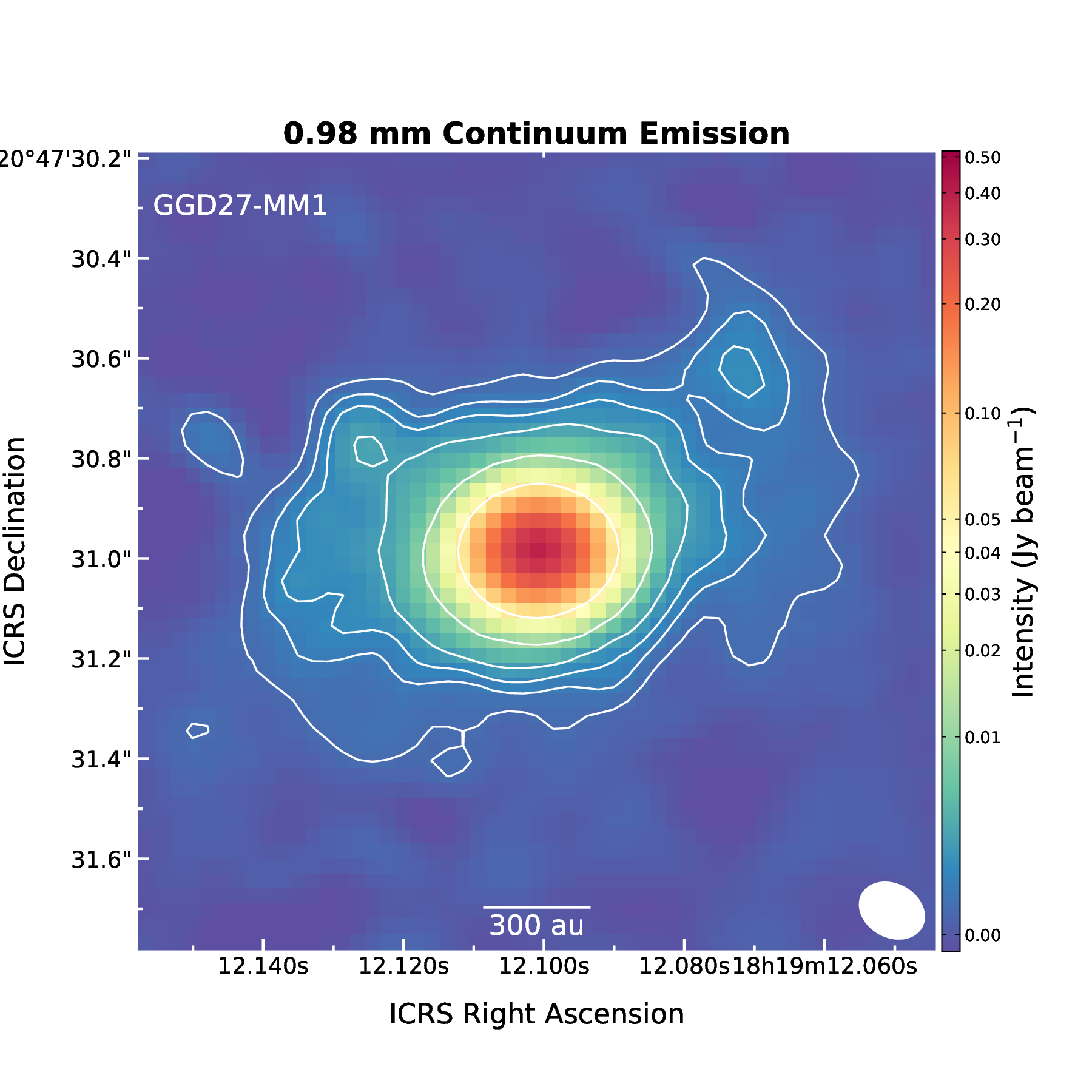}
\caption{ALMA 0.98\,mm dust continuum emission toward GGD27-MM1. Contours display the emission at levels 5, 10, 15, 25, 75, 375 times the rms noise level of the image, $\sim0.14$\mjy. The synthesized beam is shown in the bottom right. 
}
\label{f:cont} 
\end{figure}

Figure \ref{f:cont} shows the dust continuum emission at 0.98\,mm (306.1\,GHz) toward GGD27-MM1, as observed with ALMA. 
The emission is dominated by a compact component, slightly elongated roughly along the East-West direction, surrounded by a tenuous emission, also elongated in the same direction, that extends in total about $1''$ (1400\,au at the source distance). Within the tenuous component, there are two peaks of emission toward the Northwest and Northeast with respect to the continuum peak. 
These two peaks are not present in the images build up using only long baselines \citep{2018Girart}.
From \cite{2018Girart} and \cite{2020AnezLopez} we know that there is an unresolved (with a radius of $\lesssim$6~au) continuum source at the very center of the system with a spectral index of $\simeq$1.2 \citep[see Section 7.1 of ][]{2020AnezLopez}.
The nature of this source is still unknown. For instance, it could be due to an extremely compact ionized HII region, or to the basis of the thermal radio-jet. By extrapolating its contribution at the observing frequency and using a spectral index of 1.2, we estimate its flux intensity to be 23\mjy.
We modelled the 0.98~mm continuum emission assuming that the following components have a 2D-Gaussian profile: (i) an unresolved source, (ii) a disk, (iii) the envelope, (iv) an unresolved NE component, and (v) a NW compact component. 
We used the peak intensity and position of each component as initial guess values. For the first three components, we used the intensity and position of the continuum peak. For the central source (i), we fixed the peak intensity with a value of 23\mjy and a size consistent with an unresolved point-like source. We also assumed that the NE compact component is unresolved. We fixed the position of the compact and the NW components. The rest of parameters for all the components were treated as free parameters. We used the CASA task {\em imfit} to do the fitting. Table~\ref{t:contfit}  shows the results. A visualization of the fit can be found in Appendix \ref{sec:appendix_continuum}.
We note that the values of the fit for the envelope and of the disk may not be unique and there may be some degeneracy, specially in the flux density.
The disk's radius, 133~mas (about 190~au) is slightly smaller than the value derived by \citet{2018Girart} and \citet{2019Busquet}, 170~mas (240~au), from higher angular resolution 1.14 mm ALMA observations. Interestingly, the value obtained by fitting the 1.14~mm emission using the $\alpha$-viscosity disk model \citep{2020AnezLopez} was 170~au. 
The values of the position angle, $\simeq$113\degr, and of the minor-to-major axis ratio, $\simeq0.68$ (i.e., a line-of-sight disk inclination of $\simeq$42\degr),  are similar to those previously reported.
The envelope is also elongated approximately in the same direction as the disk (the difference is small: $\simeq 5$\degr). Both structures also have the same axis ratio.
The NW dusty component appears to be elongated roughly along the North-South direction, and has a major axis length of $0\farcs26$ (360~au).  
The total flux from all the components in MM1 \ad{(compact  object, disk, and envelope) is 816~mJy, while the flux from the residual map at the location of the sources is 3~mJy}. If we compare this value with the value obtained with the Submillimeter Array at 1.36~mm at an angular resolution of $\sim 0\farcs5$ (with this angular resolution, all three components would be barely resolved), 441$\pm$6~mJy \citet{2011FernandezLopez1}, then the overall spectral index is $\sim 1.9$. This spectral index is in agreement with the previous measurements toward this source \citet{2011FernandezLopez1}. 

Table \ref{t:masses} contains information about the mass and column density of the different dusty components. 
As a rough approximation, we assumed that the dust emission in the envelope is optically thin and isothermal, despite we know that the temperature progressively decreases with distance to the center (see section \ref{sec:tk}). We also used a gas-to-dust ratio of 100, a dust opacity of 1.0\,cm$^2$\,g$^{-1}$ at 1.3\,mm appropriate for dust with thin ice mantles \citep{1994Ossenkopf}, which we extrapolate to the observing wavelength using an opacity spectral index with an exponent $\beta=1.6$, typical of Interstellar Medium grains \citep{2006Draine}, and the 306\,GHz flux extracted in this work. Finally, we considered a range of dust temperatures for the envelope; we assumed that the temperature of the envelope may be between 110\,K and 150\,K \citep[see section \ref{sec:tk} and][]{2011FernandezLopez1}. Hence, we derived an envelope mass of $\sim0.22$\msun and a column density $9-12\times10^{22}$\cmd. Likewise, we also estimate the mass of the NE and NW components (about 0.005\msun and 0.01\msun, respectively). For their temperature we assume they are at a similar temperature as the rest of the envelope. However, the NE component is at a position where the CH$_3$CN analysis (see Section \ref{sec:analysis}) show two separate gas temperature values: a warm $\sim120$~K and a hotter $\sim200$~K temperature. We used a range of temperatures (110-250~K) to derive a plausible mass for this component. 
Also, we refrain from derive the mass and density of the disk, since it is optically thick and its dust millimeter emission may comprise an unknown contribution due to scattering. \cite{2020AnezLopez} estimated the disk mass to be around 5\msun.

\begin{deluxetable*}{lccccccc}
\tablewidth{0pt}
\tablecolumns{8}
\tabletypesize{\scriptsize}
\tablecaption{MM1 continuum emission}
\tablehead{
\colhead{Component} &
\multicolumn{2}{c}{Position}  &
\colhead{Peak Intensity} & 
\colhead{Flux Density} & 
\colhead{Deconvolved Size}   \\
\colhead{}  & 
\colhead{RA (s) [18$^h$19$^m$]} & 
\colhead{DEC ($\arcsec$) [$-20\degr$ $47\arcmin$]} & 
\colhead{(\mjy)} & 
\colhead{(mJy)} & 
\colhead{(mas $\times$ mas, \arcdeg)}  
}
\startdata
Compact & 12.101\tablenotemark{$*$} & 30.95\tablenotemark{$*$}
        & 23.0\tablenotemark{$\dagger$} & 23.0\tablenotemark{$\dagger$}
        & unresolved\tablenotemark{$\ddagger$} \\
Disk    & 12.10079$\pm$0.000004 & 30.98585$\pm$0.00004 
        & 353.6$\pm$0.3 & 656.6$\pm$0.7 
        & 133.2$\pm$0.3$\times$89.6$\pm$0.3, 113.1$\pm$0.3  \\
Envelope& 12.1020$\pm$0.0003 & 30.966$\pm$0.003 
        & 9.6$\pm$0.2 &  127$\pm$2
        & 564$\pm$10$\times$341$\pm$7, 108$\pm$2  \\
NE      & 12.1258$\pm$0.0003 & 30.765$\pm$0.003 
        & 3.5$\pm$0.2 & 3.5$\pm$0.2 
        & unresolved\tablenotemark{$\ddagger$} \\
NW      & 12.071\tablenotemark{$*$} & 30.62\tablenotemark{$*$}
        & 1.8$\pm$0.2 & 6.0$\pm$0.8 
        & 261$\pm$2$\times$124$\pm$4, 29$\pm$9  \\
\enddata 
\tablecomments{The positional uncertainties quoted are just statistical. The absolute astrometric error of the observations is $0\farcs013$.}
\tablenotetext{*}{Peak position fixed.}
\tablenotetext{\dagger}{Flux fixed. The value has been scaled from 263~GHz \citep{2018Girart,2020AnezLopez}, assuming a spectral index of 1.2.}
\tablenotetext{\ddagger}{Assumed to be unresolved.}
\label{t:contfit}
\end{deluxetable*}

\begin{deluxetable}{lcccc}
\tablewidth{0pt}
\tablecolumns{5}
\tabletypesize{\scriptsize}
\tablecaption{Dust Masses}
\tablehead{
\colhead{Component} &
\colhead{R$_d$} &
\colhead{T$_d$} & 
\colhead{Mass} & 
\colhead{N$_{H_2}$}  \\
\colhead{}  & 
\colhead{(au)}  &
\colhead{(K)}  & 
\colhead{(\msun)}  & 
\colhead{($10^{22}$\cmd)}
}
\startdata
Disk    & 190 & 300-600  & $\simeq$5\tablenotemark{$*$} & \nodata\tablenotemark{$*$}  \\
Envelope& 790 & 110-150  & 0.19-0.26 & 8.9-12.4  \\
NE      & $\lesssim170$ & 110-250\tablenotemark{$\dagger$} & 0.003-0.007 & 2.9-6.7 \\
NW      & 250 & 110-150 & 0.008-0.012 & 3.7-5.2 \\
\enddata 
\tablenotetext{*}{Value derived from fits to CH$_3$CN spectra (see Section \ref{sec:tk} and Figure \ref{f:fits}}
\tablenotetext{\dagger}{Values estimated from the fitting of the continuum emission using a $\alpha$-viscosity disk model \citep{2020AnezLopez}.}
\label{t:masses}
\end{deluxetable}


\subsection{The GGD27-MM1 disk and envelope system: molecular content}\label{sec:disk_molecular}
\begin{figure*}
\vspace{-3.cm}
\minipage{0.5\linewidth}
\includegraphics[width=\linewidth]{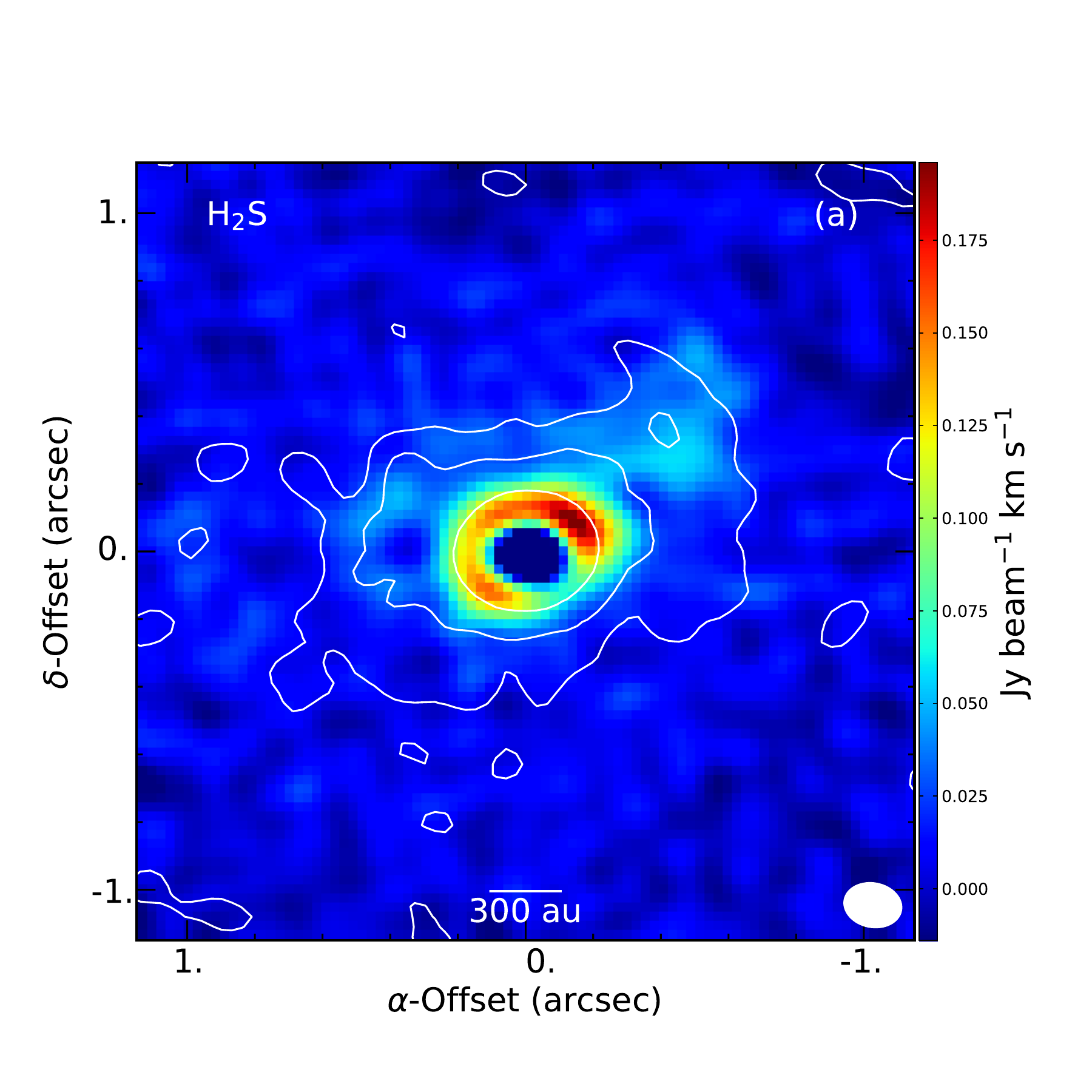}
\endminipage\hfill
\minipage{0.5\linewidth}
\includegraphics[width=\linewidth]{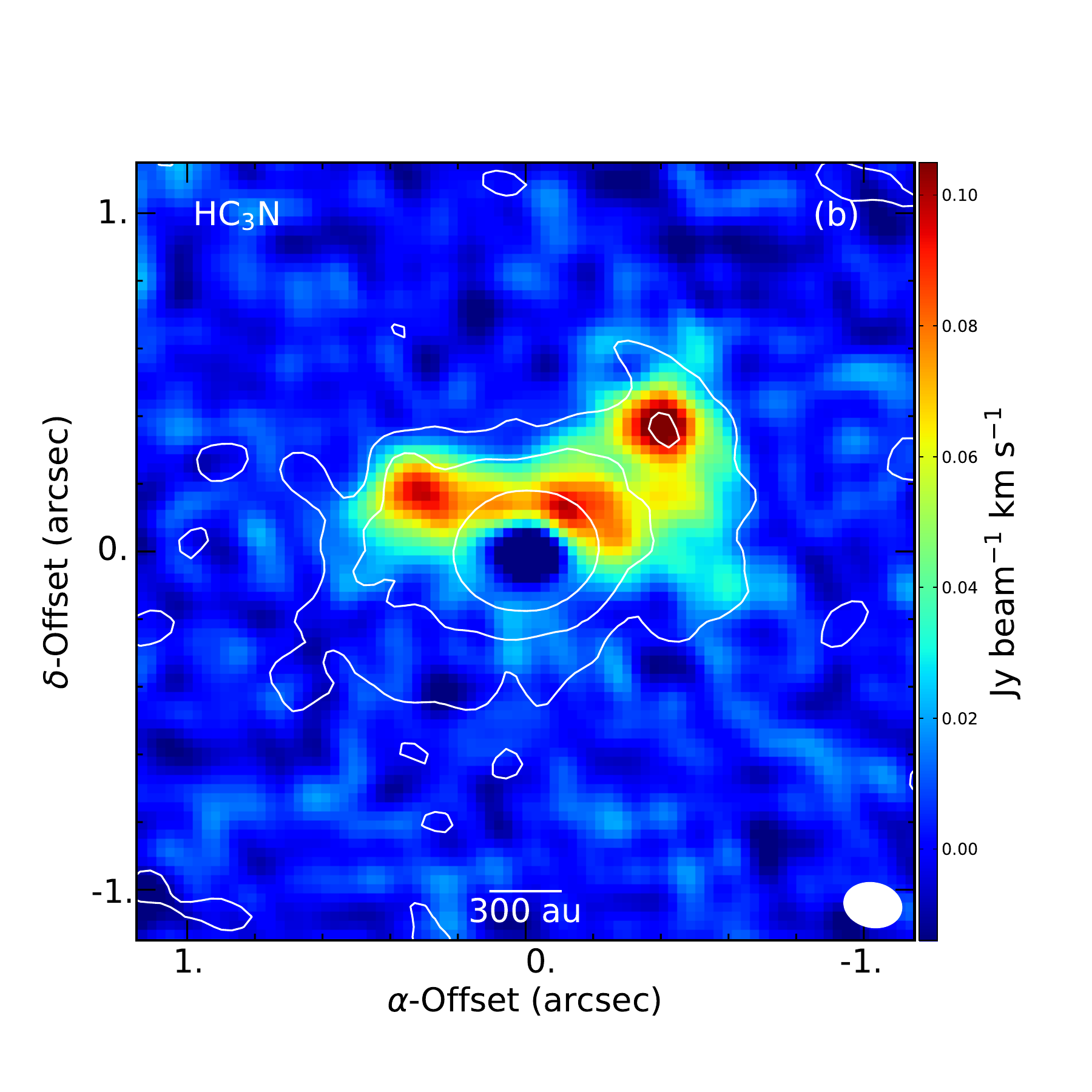}
\endminipage\vspace{-0.5cm}\\
\minipage{0.5\linewidth}
\includegraphics[width=\linewidth]{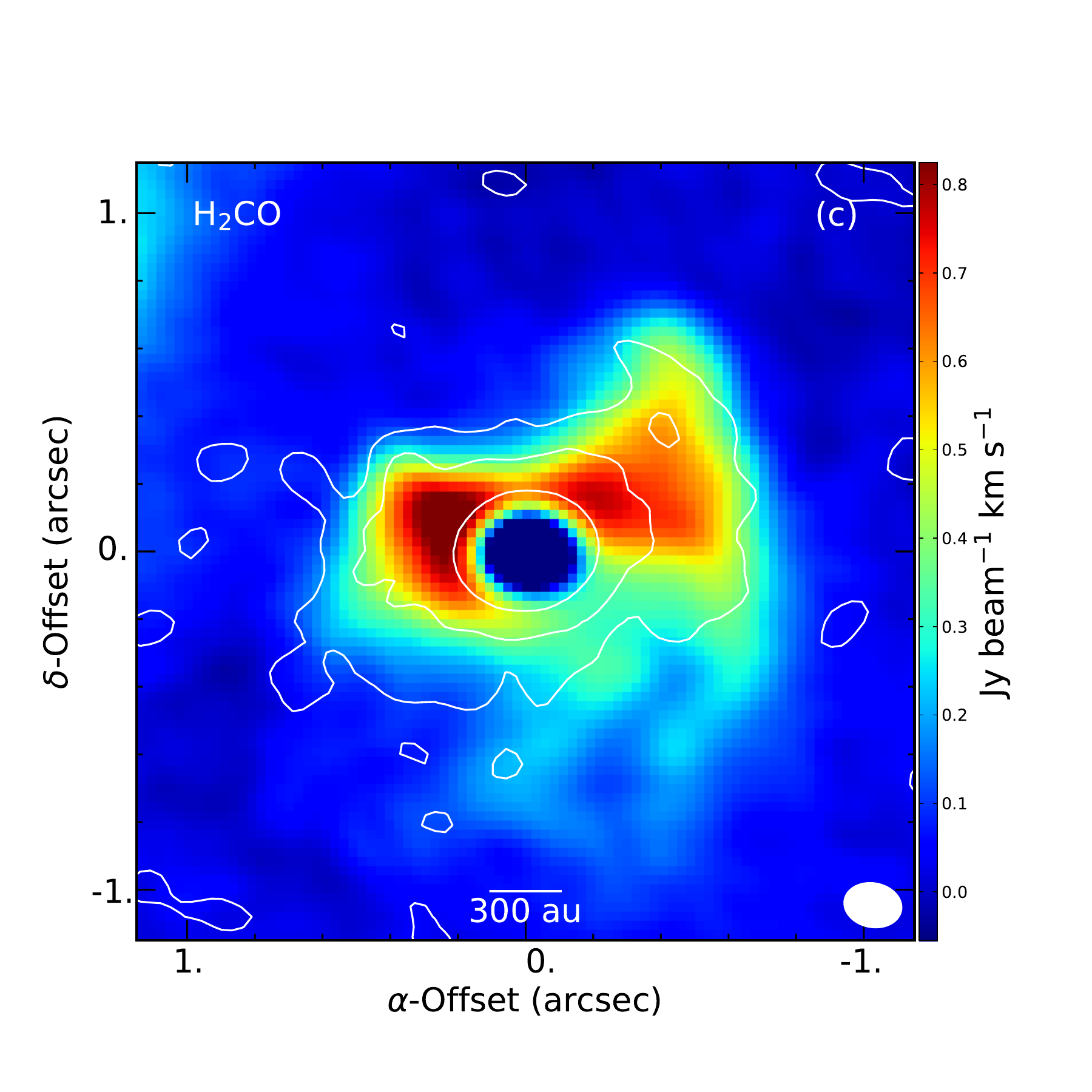}
\endminipage\hfill
\minipage{0.5\linewidth}
\includegraphics[width=\linewidth]{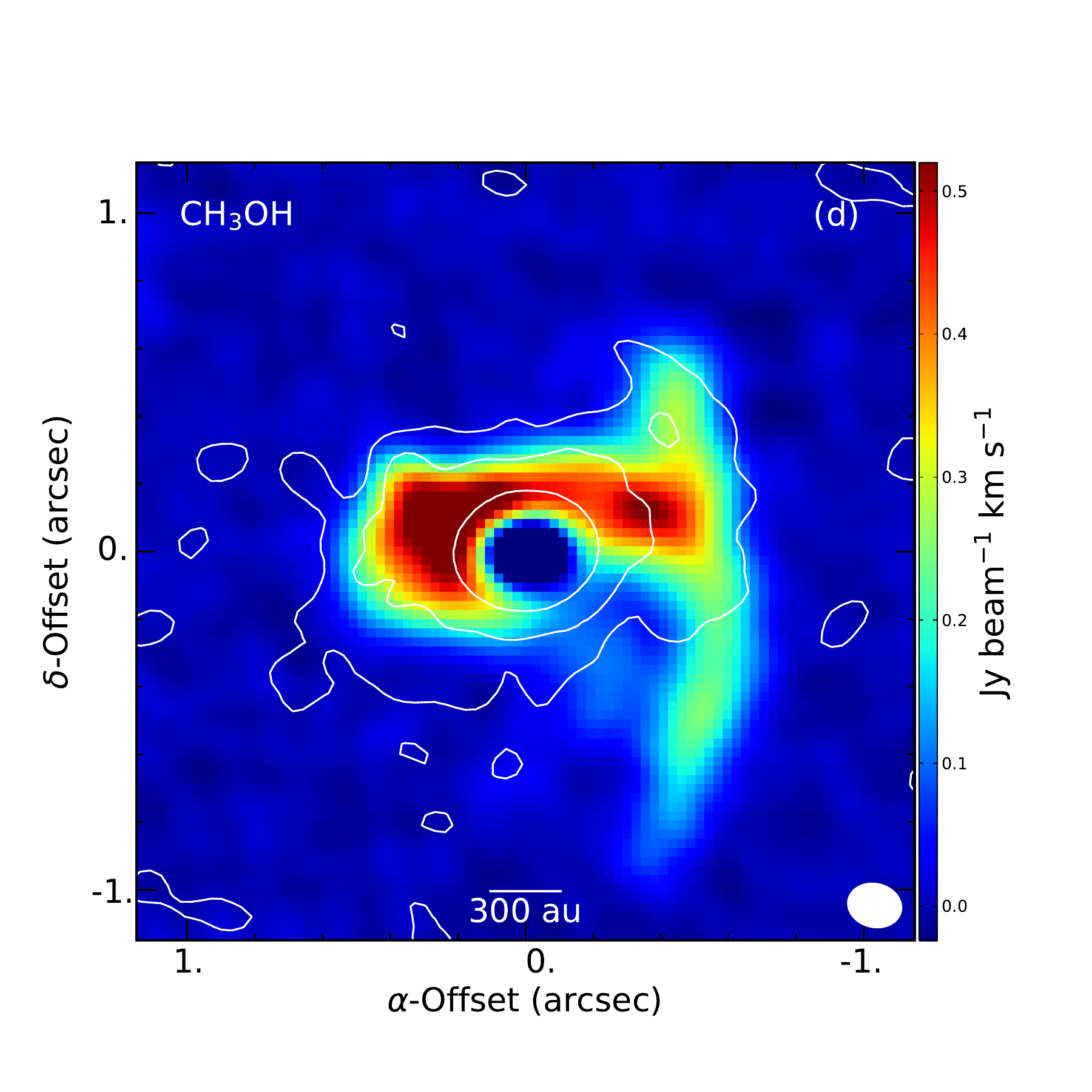}
\endminipage\vspace{-0.5cm}\\ 
\minipage{0.5\linewidth}
\includegraphics[width=\linewidth]{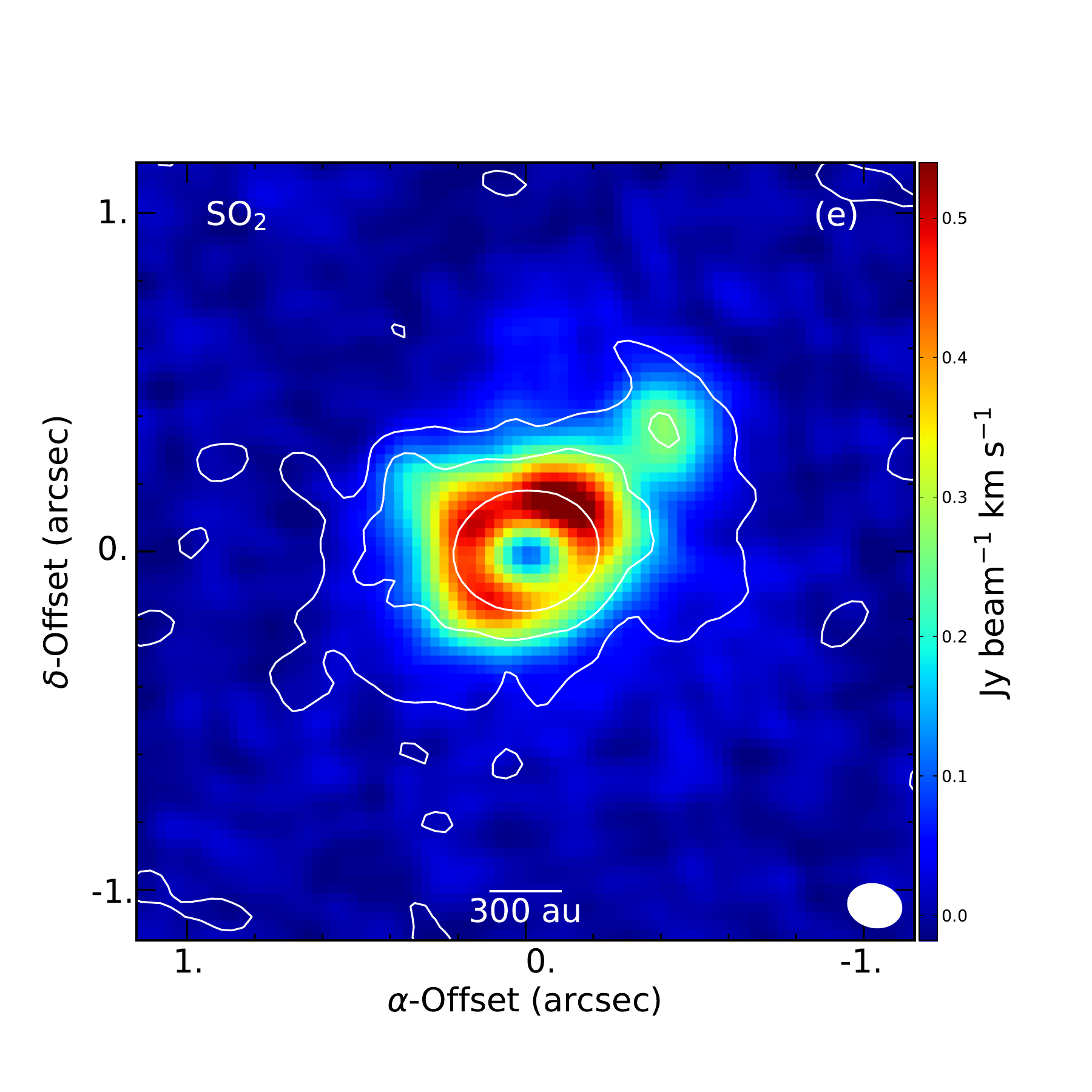}
\endminipage\hfill
\minipage{0.5\linewidth}
\includegraphics[width=\linewidth]{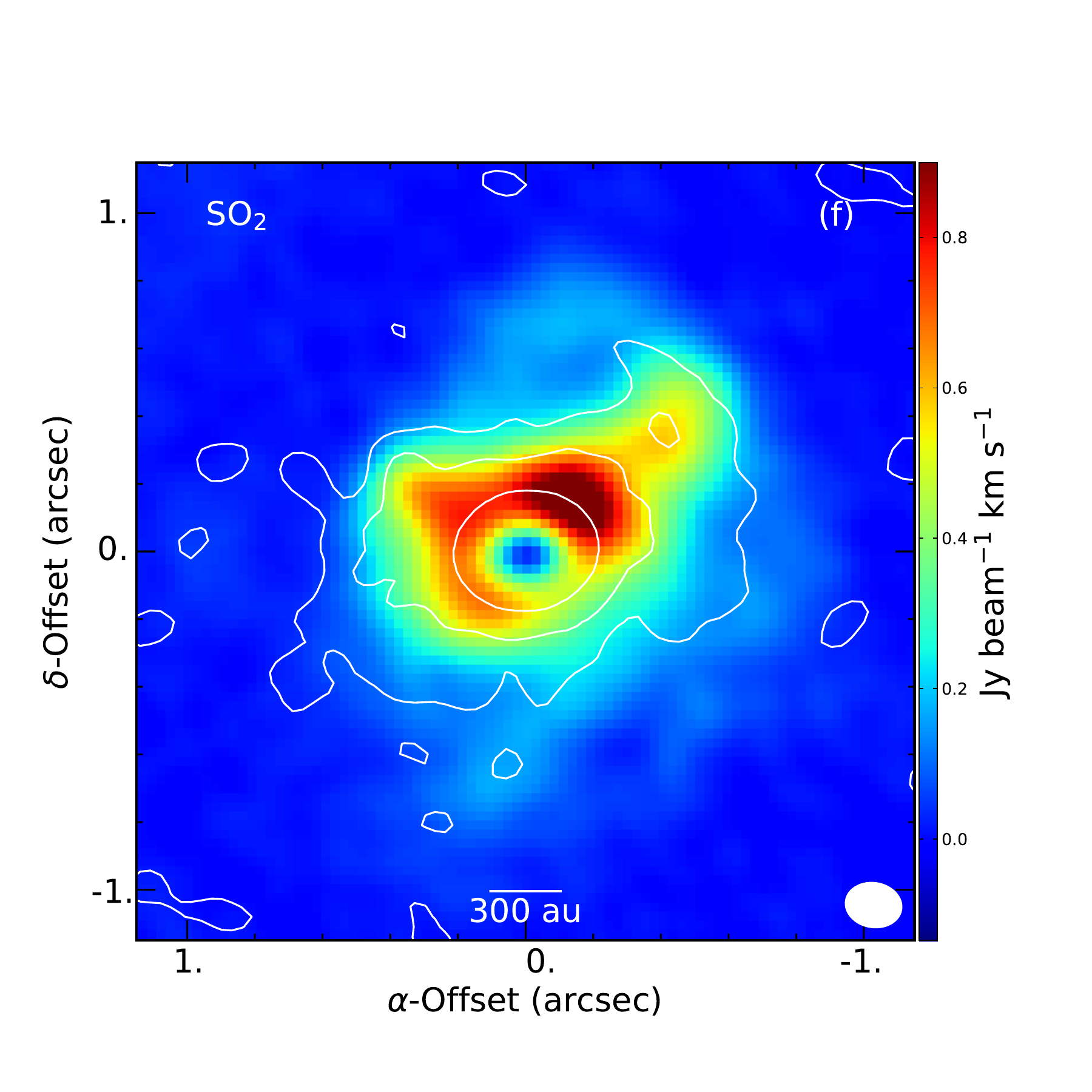}
\endminipage
\caption{Following rows from top to bottom and panels from left to right, this figure presents the H$_2$S\,(3$_{3,0}$-3$_{2,1}$), HC$_3$N\,(33-32), H$_2$CO\,(4$_{1,3}$-3$_{1,2}$), CH$_3$OH\,(3$_{1,2}$-2$_{0,2}$), SO$_2$\,(12$_{6,6}$-13$_{5,9}$) and SO$_2$\,(9$_{2,8}$-8$_{1,7}$) moment 0 images. Contours represent the continuum emission at 3, 15 and 100 times the rms noise level of 140\mujy. Synthesized beams appear in the bottom right corner.}
\label{f:moms0} 
\end{figure*}

\begin{figure}
\includegraphics[width=\linewidth]{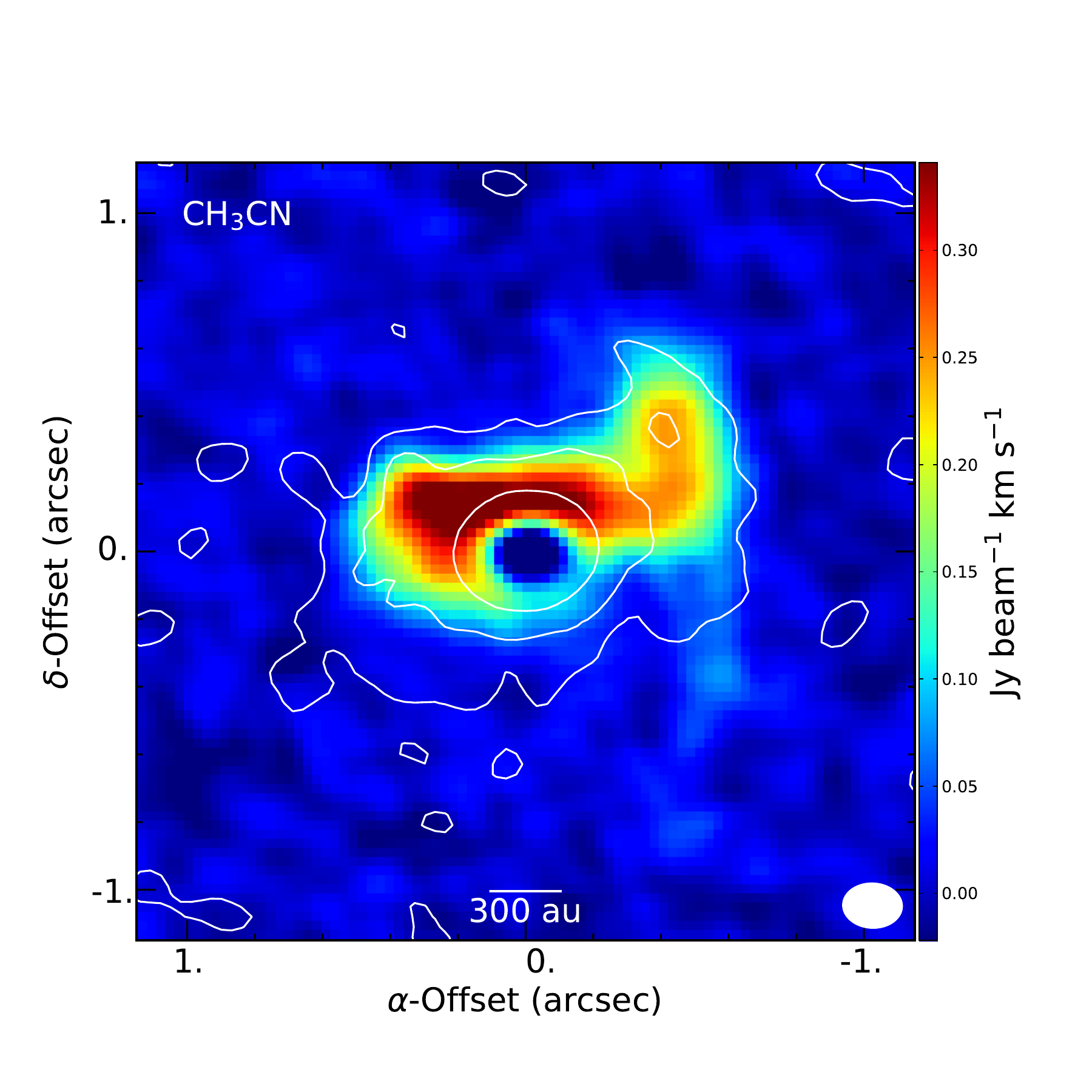}
\caption{CH$_3$CN\,J$=17_3-16_3$ moment 0 image with contours presenting the continuum emission as in Figure \ref{f:moms0}. The synthesized beam appears in the bottom right corner. The K=3 is the strongest of the K-ladder transitions observed.}
\label{f:ch3cn_moms0} 
\end{figure}

Figures \ref{f:moms0} and \ref{f:ch3cn_moms0} show a comprehensive view of some of the more common molecular species detected toward the disk of GGD27-MM1 in the present Band\,7 ALMA observations. They present velocity total integrated emission images (i.e., moment zero images) providing a glimpse of the spatial distribution diversity of the molecular emission. Complete velocity cube images, with the details on the gas kinematics, can be found in the Appendix \ref{sec:appendix_cubes}.

Panel (a) in Figure \ref{f:moms0} shows the H$_2$S\,(3$_{3,0}$-3$_{2,1}$) line emission, which preferentially traces the molecular component of the disk structure (average radius of about $0\farcs18$, i.e. 250\,au). This molecular component embraces the more compact dust continuum emission from the disk. A remarkable feature of the molecular disk\footnote{From now on we use the terms dust disk and molecular disk to designate the dust component of the disk \citep[traced by the continuum emission carefully analyzed and modeled in][]{2020AnezLopez} and the molecular component of the disk \citep[traced by several emission lines, and already reported by e.g.,][]{2011FernandezLopez2,2018Girart}.} is the weak intensity of its southern rim, while revealing a bright northwest crescent shape and southeast bright spot. These asymmetries are not present in the continuum emission, though. The H$_2$S line is also characterized by a strong absorption toward the inner part of the disk. Interestingly, the absorption peaks at different velocities (see Section \ref{sec:absorption} for more details). The H$_2$S velocity cube (Figure \ref{f:h2scube}) makes a nice example of a rotating annular disk, spreading from +5.0\kms to 19.5\kms in LSR velocity. In Section \ref{sec:flared}, we analyze the possibility that the asymmetries in the molecular disk and the inner absorption are originated in a system comprising a hot flat continuum disk, immersed in a flared molecular disk structure, which contains colder gas inside.

HC$_3$N~(33-32) emission (Figure \ref{f:moms0}, panel b) comprises three peaks northward of the inner part of the dust disk. Two of these peaks are prominent protuberances northwest ($0\farcs54=760$\,au, from the protostar location) and northeast ($0\farcs39=550$\,au from the protostar) of the dust disk. These protuberances spatially coincide with the NE and NW dust components identified in Section \ref{sec:disk_continuum}. The emission from the rotating molecular disk is very weak for this line (mainly seen at high-velocities, Figure \ref{f:hc3ncube}) suggesting a different origin for most of this line emission. Besides, the emission from the three main locations does not match the pattern of rotation of the disk, when compared with the emission from H$_2$S (Figure \ref{f:moms0}, panel a). They appear mostly at the systemic and more redshifted velocities ($>12.1$\kms). We note as well that, interestingly, the central absorption peaks at different radial velocities (see Section \ref{sec:absorption}).

The H$_2$CO\,(4$_{1,3}$-3$_{1,2}$) line (Figure \ref{f:moms0}, panel c) shows more extended emission than the rest of the lines detected here spreading further beyond the boundaries of the extended dust envelope emission. Its moment zero has a prominent absorption hole at the center of the disk, surrounded by emission overlapping the continuum emission from the envelope. It shows the northwest and northeast protuberances aforementioned, plus two weak curved arcs starting at the western side of the disk with tails extending to the southeast. 

In panel (d) (Figure \ref{f:moms0}), the CH$_3$OH\,(3$_{1,2}$-2$_{0,2}$) emission shows a prominent peak of emission east of the protostellar position. Contrarily to the disk rotation pattern (east/west are blueshifted/redshifted, respectively), this peak is contributed mainly by redshifted emission (velocities larger than 12.1\kms, Figure \ref{f:ch3ohcube}). There is also a $1\farcs2$ (850\,au) curved arc starting from the western side of the disk with a tail pointing south. A second, much fainter $0\farcs7$ (950\,au) arc-like structure, is seen south of the disk. This is more clearly seen in the velocity cube (Figure \ref{f:ch3cncube}). The width of these arcs  is  marginally resolved with deconvolved sizes ranging between $\sim0\farcs12$ and $\sim0\farcs17$ (i.e., about 200\,au). In addition, the CH$_3$OH emission traces the northwest protuberance seen in the other lines. 

The two bottom panels (e) and (f) of Figure \ref{f:moms0} show the SO$_2$ integrated emission of two lines with different excitation temperatures. The emission from SO$_2$\,(12$_{6,6}$-13$_{5,9}$) is clearly more compact than that from SO$_2$\,(9$_{2,8}$-8$_{1,7}$). The former line (E$_{up}=160$~K, panel e) traces the molecular disk, which is more extended than the compact dust disk. It also peaks northwest and northeast from the disk at the locations of the aforementioned NE and NW dust components. The main emission in this line delineates a annular-like structure, broken up at the southern rim (similarly to the H$_2$S emission), tracing the molecular disk. The SO$_2$\,(9$_{2,8}$-8$_{1,7}$) line (E$_{up}=51$~K, panel f) shows the broken annular-shaped molecular disk structure as well, but it has more extended emission spreading over a wider region. A curved arc-like structure runs from the western side of the disk with a tail pointing southeast; a gaseous loop can be noticed northward as well. The central absorption is deeper in the SO$_2$\,(9$_{2,8}$-8$_{1,7}$) line than in the higher-excitation line SO$_2$\,(12$_{6,6}$-13$_{5,9}$).

In Figure \ref{f:ch3cn_moms0} we show the moment zero image of the CH$_3$CN\,J$=17_3-16_3$ line. We have detected several transitions of the CH$_3$CN\,J$=17-16$ ladder (Table \ref{t:molecules}); we use this molecule in Section \ref{sec:tk} to estimate the physical conditions of the disk and envelope toward GGD27-MM1. The spatial distribution of the emission from this molecule is similar to that of the H$_2$CO (Figure \ref{f:moms0}, panel (c)). It is associated with the molecular disk, the absorption hole at its center, its immediate envelope material, the northern protuberances and one of the curved arcs extending south from the disk's western side (like the one seen in the CH$_3$OH emission).

In a nutshell, we detect emission from a molecular diversity in GGD27-MM1 (sulfured, carbonaceous and oxygenated species). We identify some characteristic features closely related with the compact continuum dust disk (central absorption, molecular disk annular structure), and the extended envelope, which comprises the northern protuberances (NE and NW dust components) and the curved arcs. We find that different molecules preferentially exist in different regions. For instance, the sulfur-bearing H$_2$S appears related mostly with the molecular disk, but the also sulfureted SO$_2$ low-excitation lines trace the extended envelope and curved arcs as well. Likewise, the oxygenated CH$_3$OH shares some similitude (e.g., northern protuberances and curved arc) with the carbonaceous CH$_3$CN. Finally, the HC$_3$N and CH$_3$OH lines are weak in the molecular disk, contrarily to the rest of the chemical species; these lines preferentially trace gas structures in the vicinity of the molecular disk and also coincident with the NW and NE components, with velocities that do not belong to the rotation pattern of the molecular disk. The two southern curved arcs spatially coincide in CH$_3$OH and CH$_3$CN. The most prominent has a subtle difference in its curvature when seen in H$_2$CO. However, the low-excitation transitions of SO$_2$ trace preferentially emission from the easternmost arc with a slightly different trajectory.

\section{Analysis}\label{sec:analysis}
\subsection{T$_{ex}$ and column density from the CH$_3$CN analysis}\label{sec:tk}
\begin{figure}
  \includegraphics[width=\linewidth]{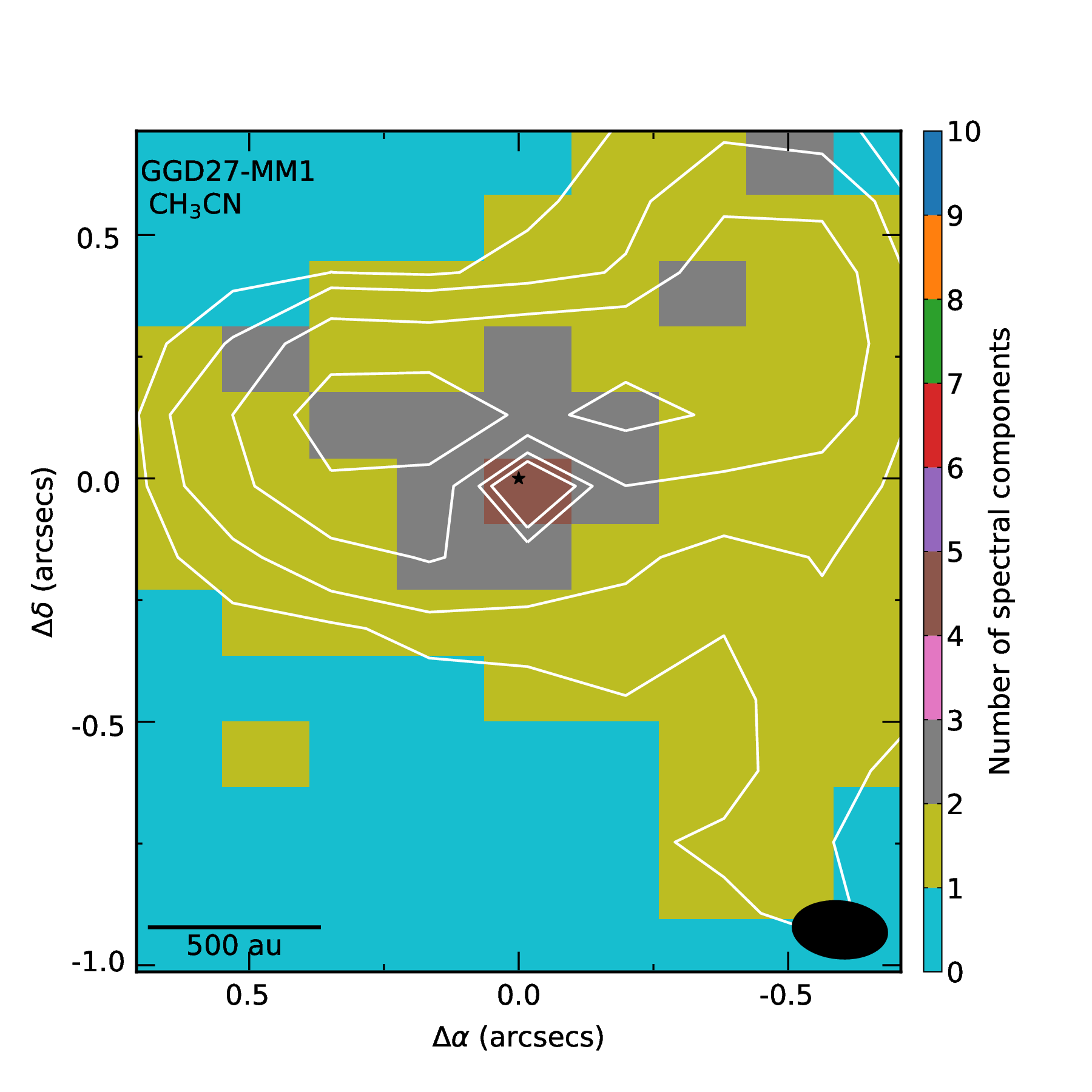}
\caption{Number of velocity components found in the spectra of the CH$_3$CN~(17-16) ladder emission toward GGD27-MM1. Contours show the integrated intensity from the CH$_3$CN~K=3. Pixel size roughly coincides with the beam size.}
\label{f:ncomps} 
\end{figure}

This section presents the analysis of the CH$_3$CN\,(17-16) ladder detected with ALMA toward the GGD27-MM1 disk. We derived excitation temperatures, column densities and radial velocities of the gas using a manual, pixel-by-pixel spectral fitting using CASSIS\footnote{Analysis carried out with the CASSIS software \citep[http://cassis.irap.omp.eu; ][http://adsabs.harvard.edu/abs/2015sf2a.conf..313V]{2015Vastel} and the JPL and VAMDC databases. CASSIS has been developed by IRAP-UPS/CNRS.} (see Appendix \ref{sec:appendix_cassis} for details on the fitting procedure and fitting examples in Fig. \ref{f:example_fits}).
Before starting the fitting process, we spatially binned the ALMA data to build an image in which the pixels have the approximate size of the synthesized beam ($0\farcs17\times0\farcs13$). This allowed to analyze truly spatially independent spectra. Second, we identified the number of different spectral components in the spectrum of every pixel (Figure \ref{f:ncomps}). The complex spectrum from the central pixel (coincident with the position of the continuum peak) contains at least four different spectral components. Surrounding this position, there is a annular-like set of pixels with two spectral components. There are two filament-like structures starting from this annulus and going northeast and northwest of the central position showing two spectral components, as well. The rest of the CH$_3$CN emission shows more simple spectra, presenting a single component.   

\begin{figure*}
  \includegraphics[width=\linewidth]{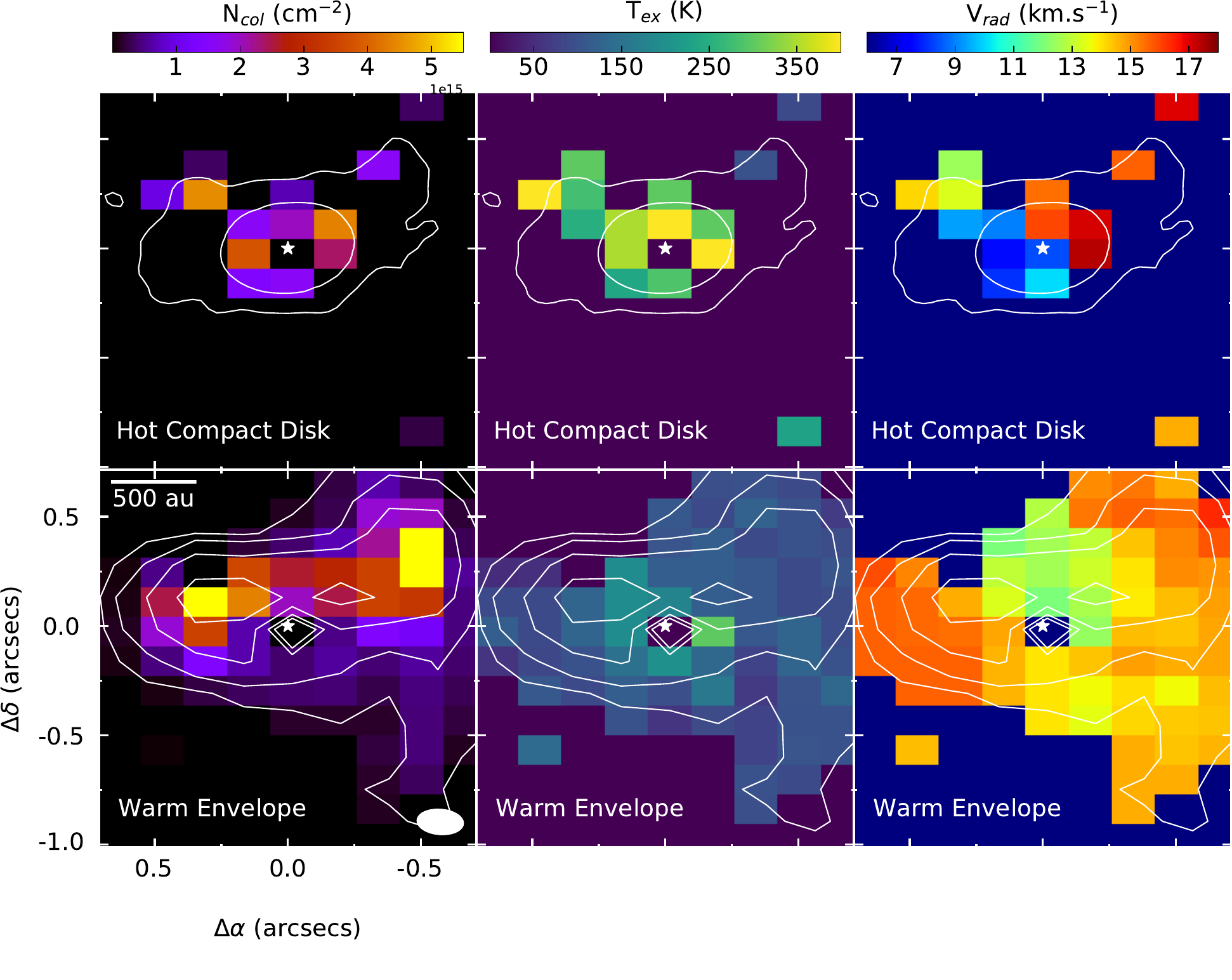}
\caption{CH$_3$CN maps obtained from the fit to the compact hot molecular disk (upper panels), and the warm envelope (bottom panels). From left to right are the column density, N$_{CH3CN}$, in units of cm$^{-2}$, the excitation temperature, T$_{ex}$, in units of K, and the LSR velocity, V$_{rad}$, in units of \kms.}
\label{f:fits} 
\end{figure*}

After simultaneously fitting every spectral component at each pixel we could distinguish two different temperature-velocity-spatial structures toward GGD27-MM1: (i) a hot and compact rotating structure, which we identify with the molecular disk, and (ii) a warm, extended and more quiescent structure, which we identify with the molecular envelope. From this, we built two maps for the derived excitation temperature, CH$_3$CN column density and radial velocity, separating the fitted solutions using a threshold of $T_{ex}=200$\,K, which clearly separates both structures (Figure \ref{f:fits}). 
 
An attempt to fit the CH$_3$CN absorption spectrum of the central pixel using a model with three components shows a good agreement with the observations. The fit and its results are shown in Section \ref{sec:absorption} (see also Figure \ref{f:example_fits}).

The top row of Figure \ref{f:fits} shows the results for the hot compact molecular disk, which has $T_{ex}>200$\,K, $N_{CH_3CN}\sim2-4\times10^{15}$\cmd, and a clear velocity gradient along the disk major axis compatible with rotation (east is redshifted, west is blueshifted). In addition to the disk, the figure reveals two filament-like structures at velocities redshifted from the V$_{LSR}$ velocity. The northeast filament particularly disagrees with the rotation pattern of the molecular disk, suggesting that these structures do not belong to the disk or envelope (which is colder), although may be physically connected. It is thereby plausible that they are part of two independent streamer-like structures. 

The bottom row of Figure \ref{f:fits} shows the gas from the warm envelope, which has $100$\,K$<T_{ex}<200$\,K, $N_{CH_3CN}\sim0.5-5\times10^{15}$\cmd, and velocities $13-15$\kms close to the systemic, redshifted from the 12.1\kms center of the rotating pattern of the disk, and not showing clear velocity gradients. The envelope shows a denser rim at the northern part of the disk, with two prominent peaks at both east and west sides. These peaks are slightly redshifted from the rest of the envelope's gas, which is also a bit more blueshifted north of the continuum peak. The temperature seems more uniform, slightly increasing toward the disk center.

\begin{figure}
  \includegraphics[width=\linewidth]{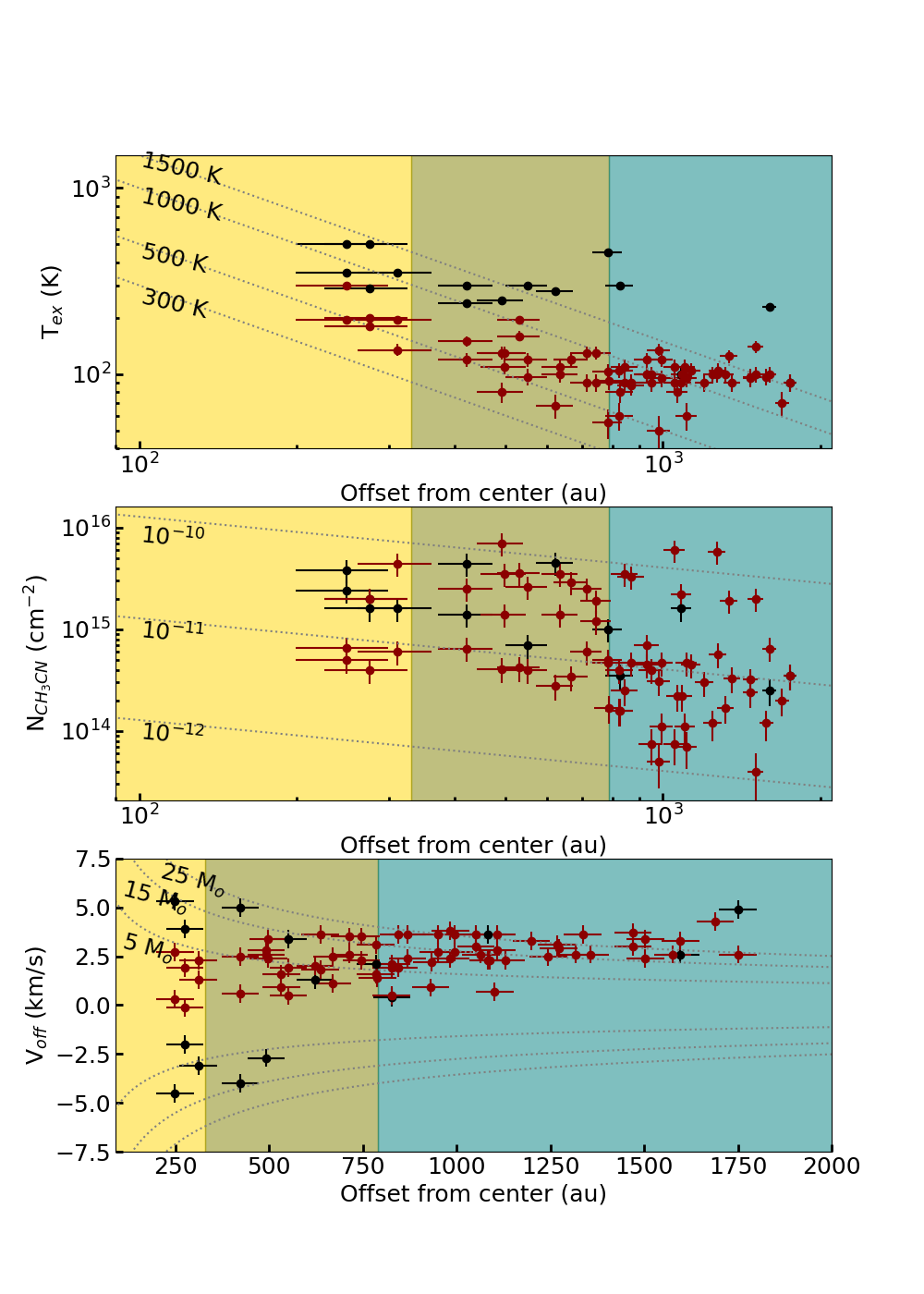}
\caption{Radial behavior of the excitation temperature, column density and offset radial velocity from the hot molecular disk component (black symbols) and the warm envelope component (red symbols). The spatial offsets are estimated assuming a disk/envelope model with a position angle of $113\degr$ and an inclination of $49\degr$ ($0\degr$ for a face-on disk). The hot compact disk extends up to $R\simeq360$~au. $T_{ex}\propto T_0 / R$ (dotted curves for different values of T$_0$). $N_{CH3CN}= (500\cdot X_{CH3CN}/(\mu_{mH}))\cdot\sqrt{100au/R}$ (dotted curves for various abundances $X_{CH3CN}$). V$_{off}$ velocity relative to the systemic velocity (12.1\kms) vs deprojected spatial offsets (the system inclination is taken into account). Dotted lines show Keplerian rotation curves for} different central masses. Offset error bars are 50\,au, $\sim1/4$ of the beam size.
\label{f:radials} 
\end{figure}

Figure \ref{f:radials} presents annular averages of the excitation temperature, the column density, and the LSR line-of-sight velocity of the CH$_3$CN. The averages were calculated assuming a disk model with a position angle of $113\degr$ and an inclination of $49\degr$ \citep{2018Girart,2020AnezLopez}. Black/red symbols belong to the spectra from the hot disk/warm envelope components. The plots are divided in three zones (yellow, olive and blue) separated at 330\,au and 790\,au, corresponding with the centrifugal radius\footnote{We follow the same procedure as \cite{2014Sakai} identifying the highest and lowest velocity peaks of the H$_2$S emission in a position-velocity diagram along the major axis of the disk. Their average position is $0\farcs12\pm0\farcs5$ ($166\pm70$\,au), which is the radius of the centrifugal barrier, likewise half of the centrifugal radius ($\sim330$\,au).} that we estimate from the H$_2$S emission, following the procedure described in \cite{2014Sakai}, and the radius of the envelope (as measured in the continuum emission, see Table \ref{t:masses}). 

The CH$_3$CN excitation temperature (Figure \ref{f:radials}, top panel) decreases with distance from the star up to $790$~au. It shows a clear segregation between the hotter and warmer data points (i.e., between disk and envelope). To stress this segregation we set a temperature threshold at 200\,K and use symbols with different colors for the hot and warm points (black and red, respectively). Several power-laws (dotted lines) of the form $T\propto T_{100} (100/R)$ are plotted for values of T$_{100}$ between 300~K and 1500~K, where T$_{100}$ is the excitation temperature at 100\,au. The inverse linear trend is assumed following the derivation in \cite{2020AnezLopez}, who used a radiative code model to simulate the continuum emission of a flared disk of 170~au radius. If the gas is thermalized, the excitation temperature of the hot disk and warm envelope can be a proxy for dust temperature. The best model for the continuum disk emission resulted in  a power-law with T$_{100}=300$~K, but the molecular data do not have the spatial resolution to determine the temperature at 100~au. However, for radii $<790$\,au (i.e., area within the envelope), the black dots show a $\sim1/R$ trend with T$_{100}\gtrsim750$~K. Likewise, the red dots (warm envelope) apparently show a similar trend with a T$_{100}\gtrsim300$~K. This indicates that the gas temperature would be clearly higher than the dust temperature. Beyond the envelope outer boundary, the temperature of the warm component (red dots) stays at $\sim100$\,K, while the hot component (black dots) reveals the presence of two hot spots.

The middle panel of Figure \ref{f:radials} shows the CH$_3$CN column density. There is no clear segregation between the disk and envelope data points. For distances $<790$\,au, the densities lie in a plateau, mostly between $3\times10^{14}$\cmd and $8\times10^{15}$\cmd; beyond that radius, densities are lower, ranging between $10^{14}$\cmd and $10^{15}$\cmd. A clear trend of decreasing density with distance from the protostar is hampered due to some outliers; we overlapped the result trends derived in the continuum disk modeling by \cite{2020AnezLopez} as guidelines for comparison. We used different values for the CH$_3$CN abundance, which seems to be between $10^{-10}$ and $10^{-11}$ (see Section \ref{sec:discussion}).

The bottom panel of Figure \ref{f:radials} displays the CH$_3$CN radial profile of the V$_{LSR}$ velocity. As expected the hot disk (black dots) shows a rotation pattern consistent with the overlapped Keplerian curves for central masses 5, 15 and 25\msun. The scarcity of sampling toward the inner region (lack of angular resolution), the lack of high-velocity gas, and the uncertainties on the measurements (lack of spectral resolution), hamper a finer determination of the central mass. Besides a more detailed analysis of the disk rotation shown by the H$_2$S emission resulted in no improvement in the accuracy of the mass determination, compared with the value obtained by \cite{2020AnezLopez}. The warm envelope (red dots) shows a different trend. It lies at low velocities (less than $\sim 4$\kms), redshifted, and have a smooth increasing tendency further from the protostar. Close to the protostar there is envelope emission at least at two different velocities: one close to 0\kms and another at about 2\kms with respect to the systemic velocity of the rotating disk.       

\subsection{Flared Disk Model} \label{sec:flared}
Motivated by the broken annular-shape (or crescent) of the H$_2$S $3_{3,0}-3_{2,1}$ integrated emission image toward GGD27-MM1 molecular disk, this Section presents a toy-model that can explain it. 
Alternative ways to create a broken annular-shaped structure with an asymmetric brightness distribution can be attributed to the formation of substructures within the disk. Indeed, lopsided disks and crescent shapes have been seen to be originated by vortices in the dust traps of low-mass transitional disks or clumps infalling into the disk \citep[e.g.,][]{2021vanderMarel,2022Demidova}. These type of disk asymmetries have been detected preferentially in the dust continuum emission. 

\begin{figure}
  \includegraphics[width=\linewidth]{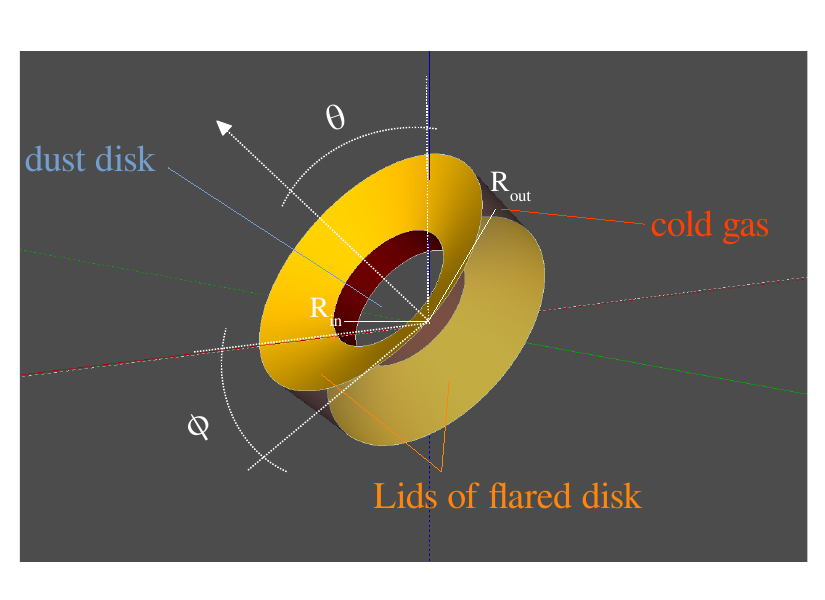}    
\caption{Flared disk model for the H$_2$S emission comprising three components: a compact dust disk (indicated but not shown in the sketch), a conical truncated flared molecular disk, whose lids are seen in emission, and an optically thick cold gas filling the volume between the conical lids.}
\label{f:flared_sketch} 
\end{figure}

\begin{figure*}
\minipage{0.33\linewidth}
\includegraphics[width=\linewidth]{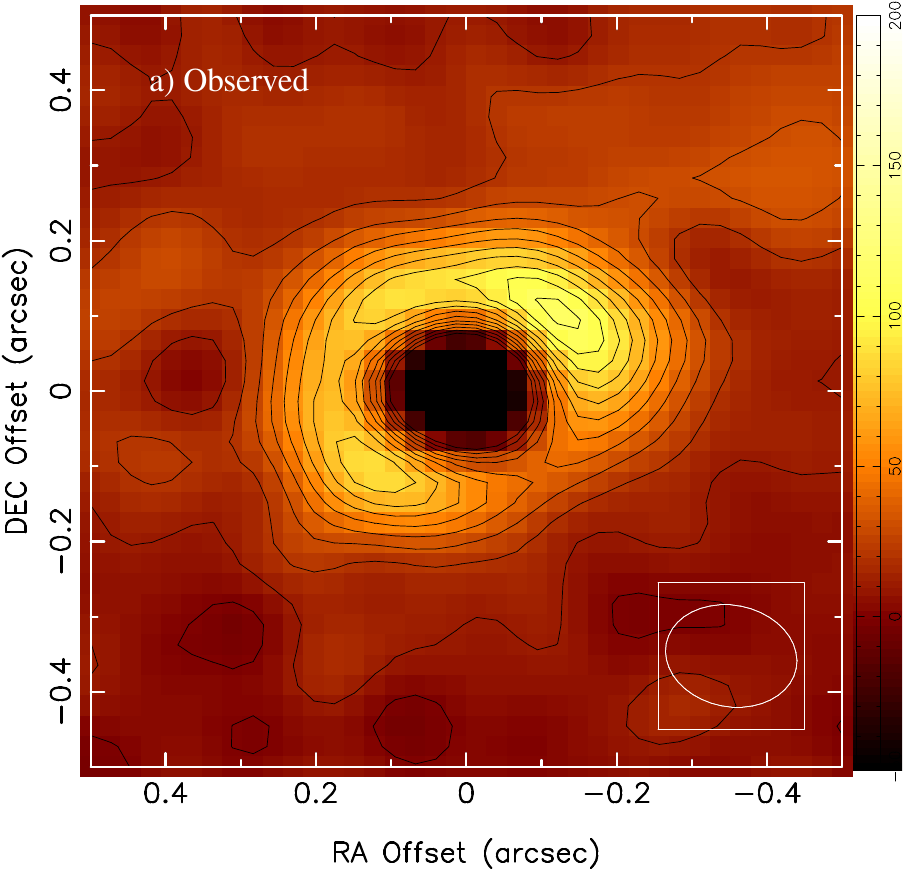}
\endminipage\hfill
\minipage{0.33\linewidth}
\includegraphics[width=\linewidth]{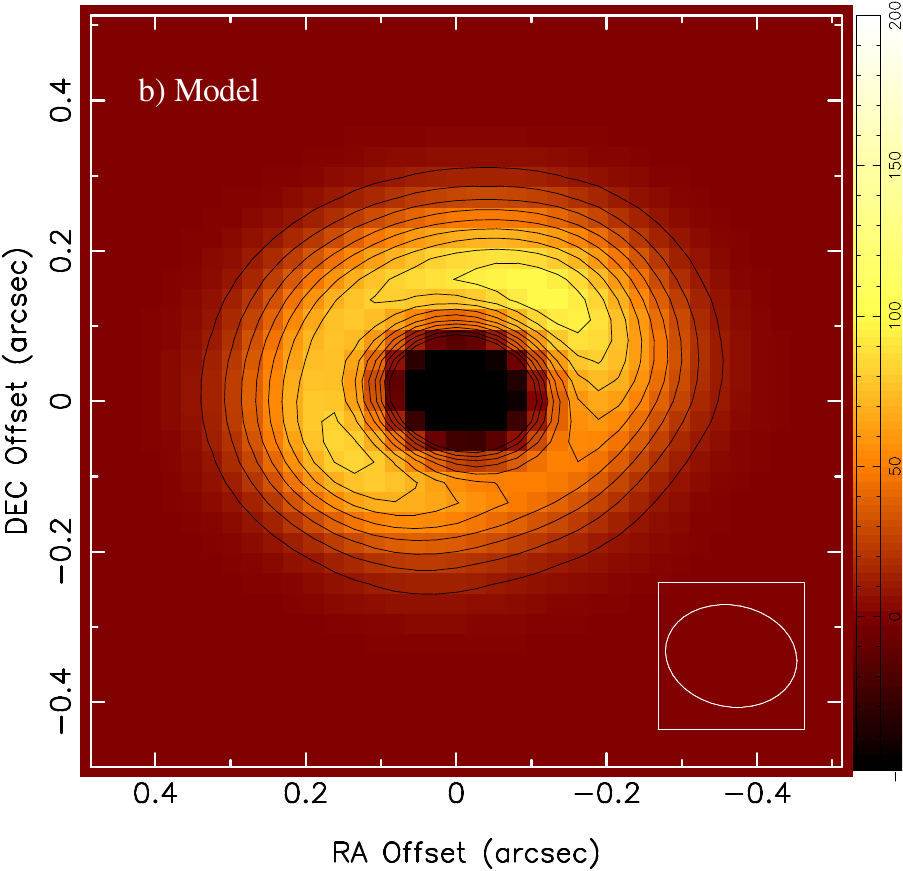}
\endminipage\hfill
\minipage{0.33\linewidth}
\includegraphics[width=\linewidth]{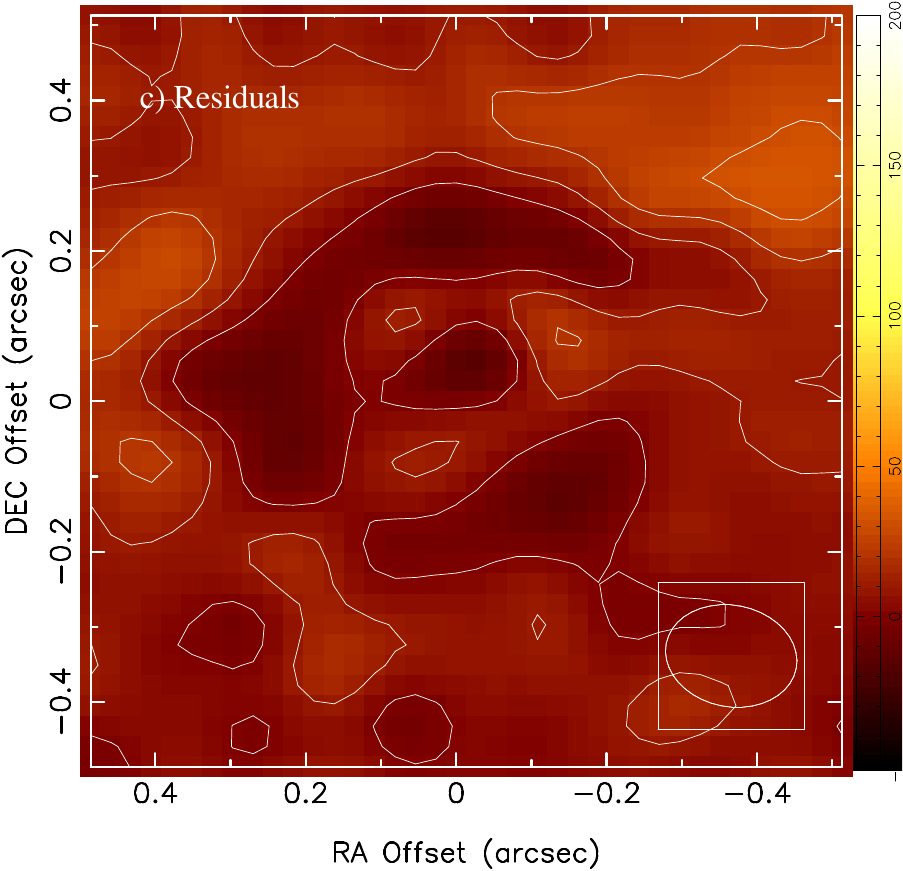}
\endminipage
\caption{Flared disk model for the H$_2$S emission. From left to right, the images show the observed moment\,0 integrated intensity image, the model obtained (see Appendix \ref{sec:appendix_flared_disk}), and the residual image after the fit. 
The three images have the same color scale and contours, at multiples of 10\,K\,\kms. Their rms noise value is 5\,K\,\kms.
The beam is shown at the bottom right corner of each image.}
\label{f:flared_model} 
\end{figure*}

The simple, phenomenological, model has three different components:
\begin{itemize}
\item A central dust disk with radius R$_{dust}$ devoid of line emission, seen in absorption after subtracting the continuum emission.
\item A flared molecular disk illuminated by the central protostar. The flared disk
extends from an inner radius $R_{inn}$ to an outer radius $R_{out}$.
\item Cold gas that fills the volume between the upper and lower lids of the flared disk with a non-zero optical depth, which, in this model, is the responsible for the absorption of the continuum emission of the dust disk. Note, however, that in Section 4.4 we find that the H$_2$S absorption is probably caused by absorption of foreground layers rather an internal cold gas, since it does not show the expected (Keplerian) rotation pattern. Nevertheless, here we chose this cold gas filling the disk lids for our toy model.
\end{itemize} 
The model has several geometrical and physical parameters that are described in detail in the Appendix \ref{sec:appendix_flared_disk}. 
The disk is described by its center coordinates, the position angle, the inclination with respect to the line of sight, the radius and the intensity of the dust emission. 
The intensity of the dust emission is taken as a power law of the radius, $I_{dust}\propto r^{-1}$, \citep{2020AnezLopez}.
The dust disk is assumed to be optically thick and geometrically flat. 
The flared molecular disk comprises two lids with an upper and lower conical faces (Figure \ref{f:flared_sketch}). 
This disk is described by an inner and outer radii, a flare angle, and the intensity of the two faces.
The center, position angle and inclination of the molecular disk are the same of those of the dust disk. 
Finally, we included cold gas filling the volume between the faces of the molecular disk. This gas is responsible for the absorption of the continuum emission of the disk and is parameterized through an intensity and an optical depth.

The results of the best fits of the phenomenological model to the integrated intensity H$_2$S data toward GGD27-MM1 are presented in Table \ref{t:flare_fit} and in Figure \ref{f:flared_model}. Note that, for the fits, we did not consider the velocity structure of the emission and we just used the integrated intensity.
The model that works best (i.e., minimizes the residuals) is one with an inner compact dust disk of $60\pm14$\,au radius. This size is smaller than the one that results from the radiative transfer modeling of the continuum emission and the size of the continuum emission. This may indicate that only the central part of the disk (the hottest part) is being absorbed by the cold gas and that there is a strong drop in the dust temperature within 100\,au from the central protostar. The molecular disk extends from 140\,au to 360\,au in radius (which agrees well with measurements from the moment\,0 image along the major axis) and has a flare angle (measured from the mid-plane to one of the conical lids) of $11\pm4\degr$. 
The gas between the lids of the disk is cold and optically thick, while the warm faces of the gas disk would be heated by the protostellar radiation. The dust emission from the disk correspond to a brightness temperature, about 4 times that of the gas disk lids, and more than 30 times larger than that of the cold gas.

The flared disk model fits qualitatively the central absorption, the dim of the southern part of the H$_2$S emission, and the crescent-like features of its northern part, leaving residuals between 2 and 3 times the noise level. It allows to obtain the flare angle and the width of the molecular disk. 


\begin{deluxetable}{llcl}
\tablewidth{0pt}
\tablecolumns{4}
\tabletypesize{\scriptsize}
\tablecaption{Best fit results from the flared disk model
\label{t:flare_fit}
}
\tablehead{
{Param.} & {Units} & \colhead{Value} & {Description} 
}
\startdata
$x_0$      &(mas)        & $3.2\pm3.7$    & Center coordinate \\
$y_0$      &(mas)        & $-0.4\pm3.8$   & Center coordinate \\
$\theta$   &(deg)        & 112.9          & Disk-axis position angle (fixed) \\
$i$        &(deg)        & 40.8           & Inclination (fixed; 0 if face-on) \\
$\phi$     &(deg)        & $11.0\pm3.6$   & Flare angle (from mid-plane to lid) \\
$\R{dust}$ &(au)         & $60\pm14$      & Dust-disk radius \\
$\R{in}$   &(au)         & $141\pm17$     & Disk inner radius \\
$\R{out}$  &(au)         & $357\pm9$      & Disk outer radius \\
$\I{dust}$ &(K\kms) & $498\pm80$     & Dust-disk intensity at 100\,au \\
$\I{face}$ &(K\kms) & $125\pm14$     & Flare-disk intensity \\
$\I{cold}$ &(K\kms) & $<16$          & Cold-gas intensity \\
$\tau_0$   &             & $>18$          & Cold-gas opacity \\
Fit rms    &(K\kms) & 12.3              
\enddata 
\end{deluxetable}

\subsection{Possible Accretion Streamers}\label{sec:streamers}

In Section \ref{sec:disk_molecular} we showed that the spatial distribution of the emission of various molecules (H$_2$CO, CH$_3$OH, SO$_2$, CH$_3$CN) have elongated and curved structures, connecting the outer envelope with the envelope/disk system. A plausible explanation is that these structures form a complex system of streamers that drive material from the outer parts of the original cloud surrounding GGD27-MM1 onto the inner envelope, or even toward the equatorial plane of the disk. These streamers could even transform into rotating spiral arm-like structures when reaching the envelope. The streamers can comprise gas from the original cloud that may have been dragged and pushed by the thermal jet, and is now infalling onto the disk. Alternatively, the structures could be the shreds of gas associated with a fly-by passage through the disk of widely separated companions, or the tidal streams from a close-encounter \citep{2022Cuello}. However, these latter scenarios have no current observational support (i.e., no companions have been confirmed so far associated to GGD27-MM1).  Motivated by the recent discovery of accretion streamers in other circumstellar disks \cite[e.g., ][]{2014Yen,2019Yen,2020Alves,2020Pineda,2020Hsieh,2021Huang,2022Thieme}, we analyze this hypothesis by fitting a model to the gas emission of the streamers more clearly identified.     

We assumed that the material of the observed gas streamers comes from the original accreting envelope. If the material of the envelope has a certain amount of angular momentum and zero total energy, it should follow a parabolic trajectory in which the focus corresponds to the origin of the coordinates system (\citealt{1976Ulrich}, hereafter Ulrich's model). These trajectories projected onto the plane of the sky can form similar structures to the observed streamers \citep{2014Yen}.

Ulrich's model considers that the accreting envelope is a rotating cloud in gravitational collapse, where the fluid particles, initially at rest, have a distribution given by a rigid body rotation, falling to the center and conserving the specific angular momentum. The collapse is assumed to be pressureless and thus the orbits are ballistic. When the fluid particles approach to the equatorial plane, they collide with their symmetric counterparts, where they could form an equatorial disk around the central object.

More recently, \cite{2009Mendoza} (hereafter Mendoza's model) proposed an analytic accretion envelope where the radius of the rotating cloud or accreting envelope is finite and the fluid particles can have an initial radial velocity (i.e., not a line-of-sight velocity but a velocity component in the direction to the central protostar). Also, the orbits are contained in a plane and are described by conic sections. More specifically, the trajectories of the fluid particles are given by

\begin{equation}
    r=\frac{\sin^2\theta_0}{1-e\cos\varphi},
    \label{eq:rm}
\end{equation}
where $r$, $\theta$ and $\phi$ are the spherical coordinates, and the radius $r$ is expressed in units of the equatorial disk radius (from now on the inner radius, $r_{in}$) where the material streams to, $\theta_0$ is the initial polar angle of the orbit of the fluid element at the beginning of the collapse toward the center, and $\varphi$ is the azimuthal angle over the plane orbit.
Finally, $e$ is the eccentricity of the orbit given $r$ by

\begin{equation} 
    e=\sqrt{1+\varepsilon \sin^2\theta_0}.
    \label{eq:exc}
\end{equation}
The variable $\varepsilon$ is the dimensionless specific energy. Under the assumptions of Mendoza's model, the energy is constant 
along each particular trajectory. This energy can be written as

\begin{equation}
    \varepsilon=\nu^2+\mu^2\sin^2\theta_0-2\mu,
    \label{eq:energy}
\end{equation}

where the parameter $\nu$ is the ratio between the initial radial velocity and the Keplerian velocity, and the parameter $\mu$ is the ratio between the inner radius $r_{in}$ and the radius of the rotating cloud or accreting envelope $r_0$ ($\mu=r_{in}/r_0$). At the border of the cloud, $r=r_0=1/\mu$. If we use this condition in the equation (\ref{eq:rm}), we obtain the following condition for $\varphi_0$

\begin{equation}
\cos\varphi_0=\frac{1}{e}\left(1-\mu\sin^2\theta_0\right).
\label{eq:varphi0}
\end{equation}
We can define this angle as the initial position of the particle trajectory in the orbital plane.

The trajectory of the fluid particles is defined by the angles $\varphi$, $\theta$, $\theta_0$, $\phi$, and $\phi_0$. Where $\theta$ and $\phi$ are the polar and azimuthal coordinates, respectively. The variables $\theta_0$ and $\phi_0$ are the polar and azimuthal initial angles, respectively. These angles are related between them as

\begin{equation}
    \cos\left(\varphi-\varphi_0\right)=\frac{\cos\theta}{\cos\theta_0},\,\,\cos\left(\phi-\phi_0\right)=\frac{\tan\theta_0}{\tan\theta}.
    \label{eq:angles}
\end{equation}

In terms of the angles defined in the above equation, the radius and velocity field of the accretion envelope in  Mendoza's model can be written as

\begin{equation}
    r=\frac{\sin^2\theta_0}{1-e\cos\xi},
    \label{eq:rmf}
\end{equation}
\begin{equation}
    v_r=-\frac{e\sin\xi\sin\theta_0}{r\left(1-e\cos\xi\right)},
    \label{eq:vr}
\end{equation}
\begin{equation}
    v_\theta=\frac{\sin\theta_0}{r\sin\theta}\left(\cos^2\theta_0-\cos^2\theta\right)^{1/2},
    \label{eq:vtheta}
\end{equation}
and

\begin{equation}
    v_\phi=\frac{\sin^2\theta_0}{r\sin\theta}.
    \label{eq:vphi}
\end{equation}
Where the angle $\xi$ is given by

\begin{equation}
    \xi=\cos^{-1}\left(\frac{\cos\theta}{\cos\theta_0}\right)+\phi_0.
    \label{eq:xi}
\end{equation}

If we fix $\mu=0$ and $\nu=0$, the radius and velocity field of Mendoza's model converge to the solutions of Ulrich's model, a more detail comparison between both models is presented in Appendix \ref{sec:appendix_model}. 

\begin{figure*}
  \includegraphics[width=\linewidth]{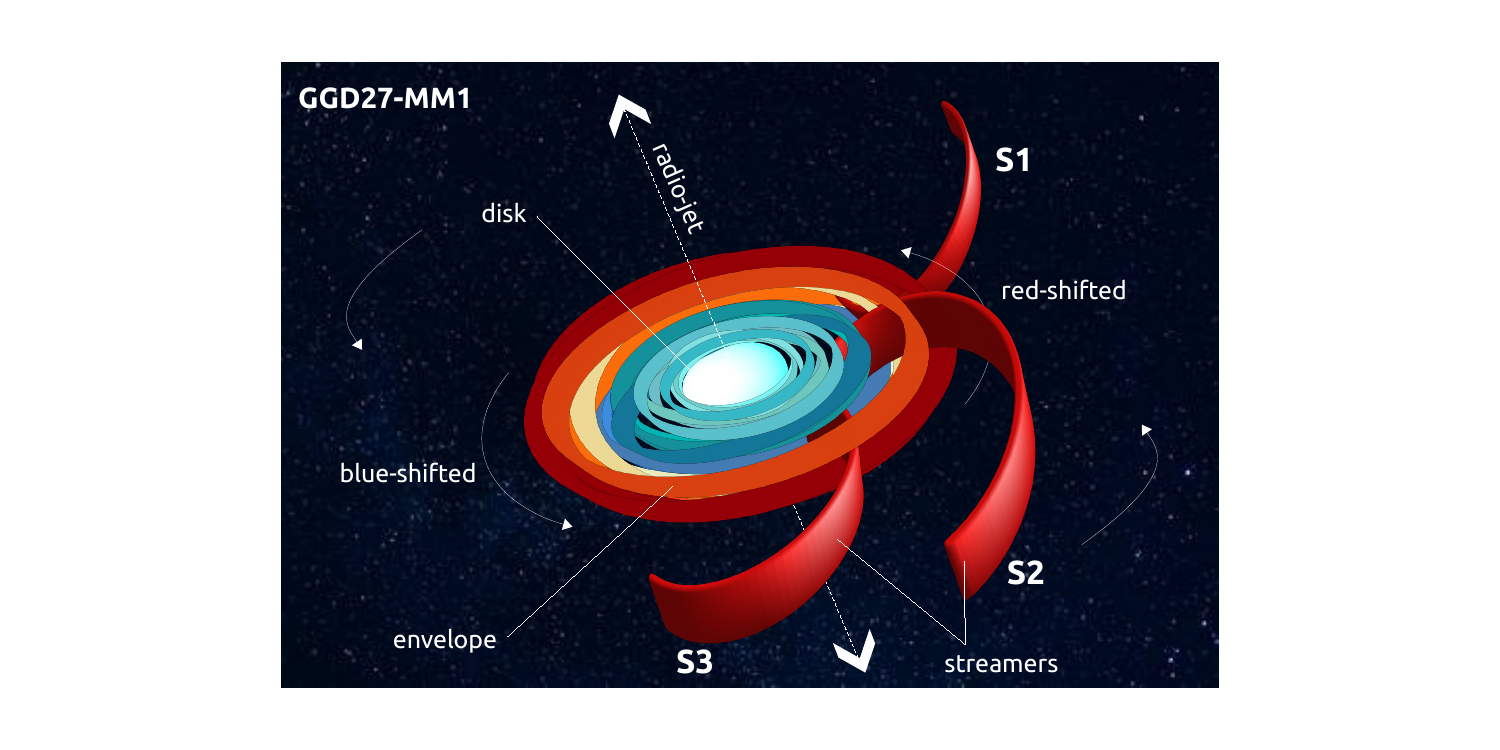}
\caption{Sketch showing the disk, envelope and the location of three possible accretion streamers: S1, S2 and S3. The figure indicates the approximate orientation of the thermal radio-jet HH~80-81, of the disk and the system rotation. The rough representation is not to scale.}
\label{f:sketch} 
\end{figure*}

For the purpose of testing the accretion streamer scenario in GGD27-MM1, we identify some curved arcs analyzing the CH$_3$OH\,(3$_{1,2}$-2$_{0,2}$) emission. We named these structures S1, S2 and S3 (Figure \ref{f:sketch}). These structures can also be identified in the H$_2$CO(4$_{1,3}$-3$_{1,2}$) moment~0 image (Figure \ref{f:moms0}). For each of these structures we selected gas condensations of the CH$_3$OH emission by carefully inspecting its channel maps (Figure \ref{f:ch3ohcube}). We considered condensations with intensity peaks of at least 5-$\sigma$. We additionally avoided the area close to the molecular disk, where emission from several structures is blended at the resolution of the present ALMA observations. S1 is related to the northwest protuberance which seems to connect with the northwest part of the disk (Figure \ref{f:moms0} and channel maps from 14\kms to 17\kms in Figure \ref{f:ch3ohcube}); S2 is a $\sim1\arcsec$ curved arc with the tail pointing south and connecting with the west side of the disk (Figures \ref{f:moms0} and \ref{f:ch3ohcube}); S3 corresponds with the second fainter arc with the tail pointing southeast that connects with the southwest part of the disk (Figure \ref{f:moms0} and channel maps from 11.5\kms to 14.0\kms in Figure \ref{f:ch3ohcube}). Despite the method of identifying streamer's condensations can be refined with higher angular resolution observations, the idea of the analysis is just to check the plausibility of the accretion streamer scenario.
 
In order to explain the S1, S2, and S3 streamers we use both the Ulrich's model and the Mendoza's model. We adopted a stellar mass of $M_*=20M_\odot$ \citep{2020AnezLopez}, an inclination angle of the disk and the envelope system with respect to the plane of the sky of $42^\circ$ (Table \ref{t:flare_fit}) and a position angle of $113^\circ$. Also, we consider two different inner radii $r_{in}$ for the fits. First, the molecular disk outer radius of 357\,au (see Table \ref{t:flare_fit}); second, the envelope radius of 790\,au (see Table \ref{t:contfit}). The angles $\theta_0$ and $\phi_0$ are left as free parameters in both models. For Mendoza's model $\mu$ and $\nu$ are left as free parameters too.

\begin{deluxetable*}{cccccccc}
\tablewidth{0pt}
\tablecolumns{8}
\tabletypesize{\scriptsize}
\tablecaption{Values of the different parameters used for the best fits in the two models considered and the three different streamers.}
\label{t:parametersmodels}
\tablehead{
\colhead{Radius} &
\colhead{Streamer} & 
\multicolumn{2}{c}{Ulrich's model} & 
\multicolumn{4}{c}{Mendoza's model}  \\
\colhead{(au)}  & 
\colhead{}  & 
\colhead{$\theta_0$ ($\degr$)}  & 
\colhead{$\phi_0$ ($\degr$)} &
\colhead{$\mu$}  & 
\colhead{$\nu$}  & 
\colhead{$\theta_0$ ($\degr$)}  & 
\colhead{$\phi_0$ ($\degr$)} 
}
\startdata
357 & S1 &  52.55 & 108.68 & 0.35 & 0.13 & 121.19 & 94.804 \\
    & S2 &  73.59 & 283.16 & 0.56 & 0.97 & 144.30 & 325.64 \\
    & S3 & 128.32 & 251.44 & 7.9$\times$10$^{-2}$ & 3.8$\times$10$^{-2}$ & 116.02 & 256.09 \\
\hline
790 & S1 &  63.43 & 133.79 & 0.27 & 0.44 & 58.12 & 152.04 \\
    & S2 &  76.99 & 254.21 & 0.54 & 5.8$\times$10$^{-3}$ & 72.21 & 289.60 \\
    & S3 & 117.84 & 226.65 & 0.46 & 3.1$\times$10$^{-2}$ & 56.06 & 236.76 \\
\enddata 
\end{deluxetable*}

The best set of parameters fitting the data are shown in Table \ref{t:parametersmodels} for both models. These parameters were estimated using the so called \textit{asexual genetic algorithm} (AGA) developed by \citet{2009Canto}.

\begin{figure*}
\includegraphics[width=\linewidth]{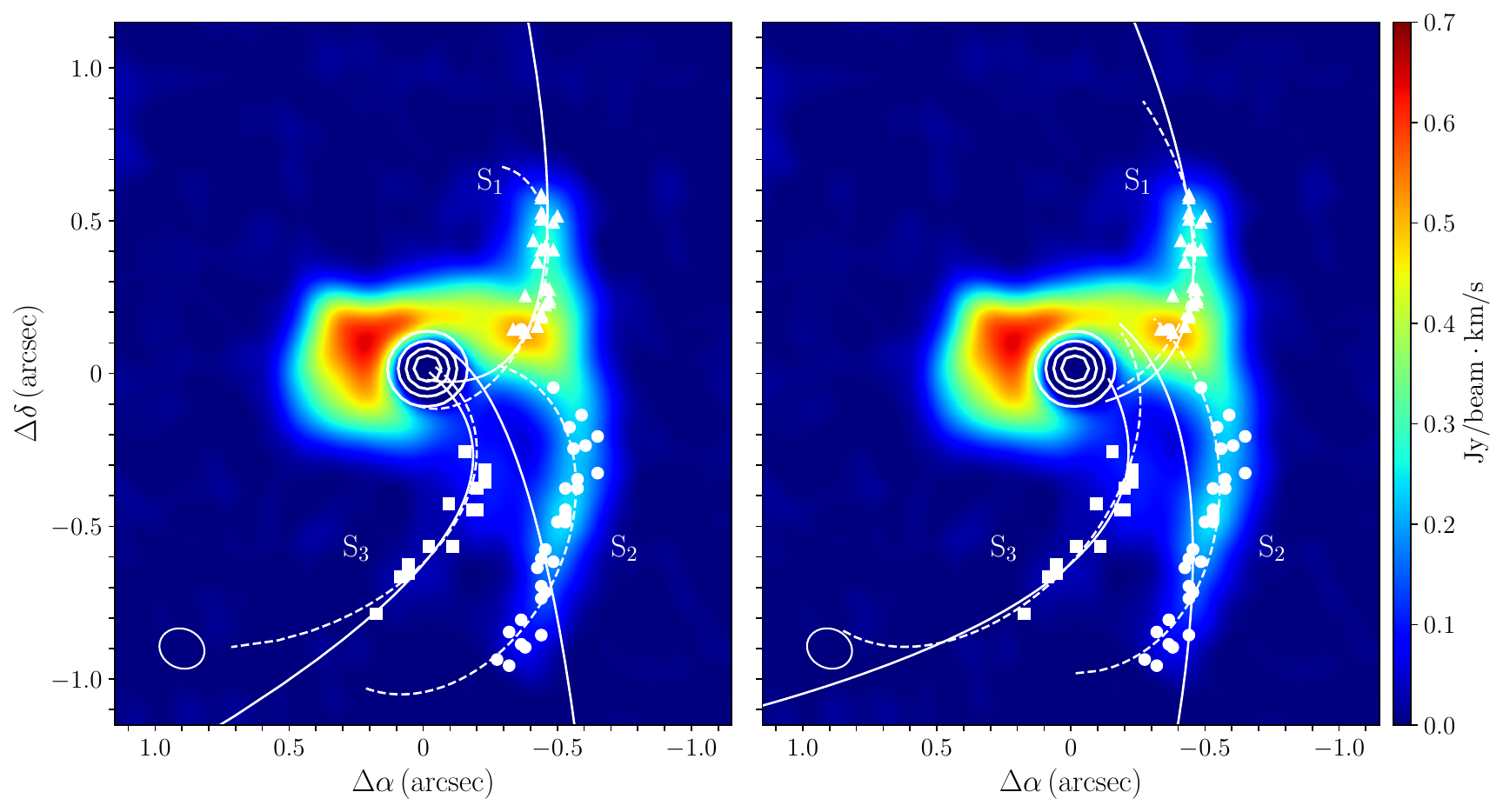}
\caption{CH$_3$OH moment 0 image with contours representing the strongest continuum emission. In both panels, the solid lines represent Ulrich's model while the dashed lines show Mendoza's models (see text in section \ref{sec:streamers}). Left panel: models using the inner equatorial radius, $r_{in}$, as the molecular disk radius (357\,au). Right panel: same but using $r_{in}$ as the envelope radius (790\,au). The angles $\phi_0$ and $\theta_0$ and the parameters $\mu$ and $\nu$ of the fits are shown in Table \ref{t:parametersmodels}. The triangles, circles, and squares mark the position of the condensations identified in the CH$_3$OH velocity cube.}
\label{f:streams1} 
\end{figure*}

Figure \ref{f:streams1} shows two panels with the CH$_3$OH integrated intensity emission (moment 0) in color scale. In these plots, the streamers denoted by S1, S2, and S3 (see Figure \ref{f:sketch}) can be observed. The white triangles, circles, and squares show the position of the CH$_3$OH condensations identified in the velocity cube in channel by channel basis. The solid lines correspond to our best fits manually found using parabolic orbits (Ulrich's model), while the dashed lines represent the conic orbits (Mendoza's model). The left panel shows the models for $r_{in}$ equal to the molecular disk radius, while the right panel presents the models for $r_{in}$ equal to the envelope radius. Both models fit reasonably well the observational data and could explain the trajectories of the streamers. However, Ulrich's model cannot reproduce the curvature of the streamer S2. In addition, the curvature of the  streamer S1 is contrary to S2 and S3. This could mean that S1 is rotating in the opposite direction. Alternatively, S1 could be associated with material that has already reached the periastron and has crossed the plane of the accretion disk.

\begin{figure*}
\includegraphics[width=\linewidth]{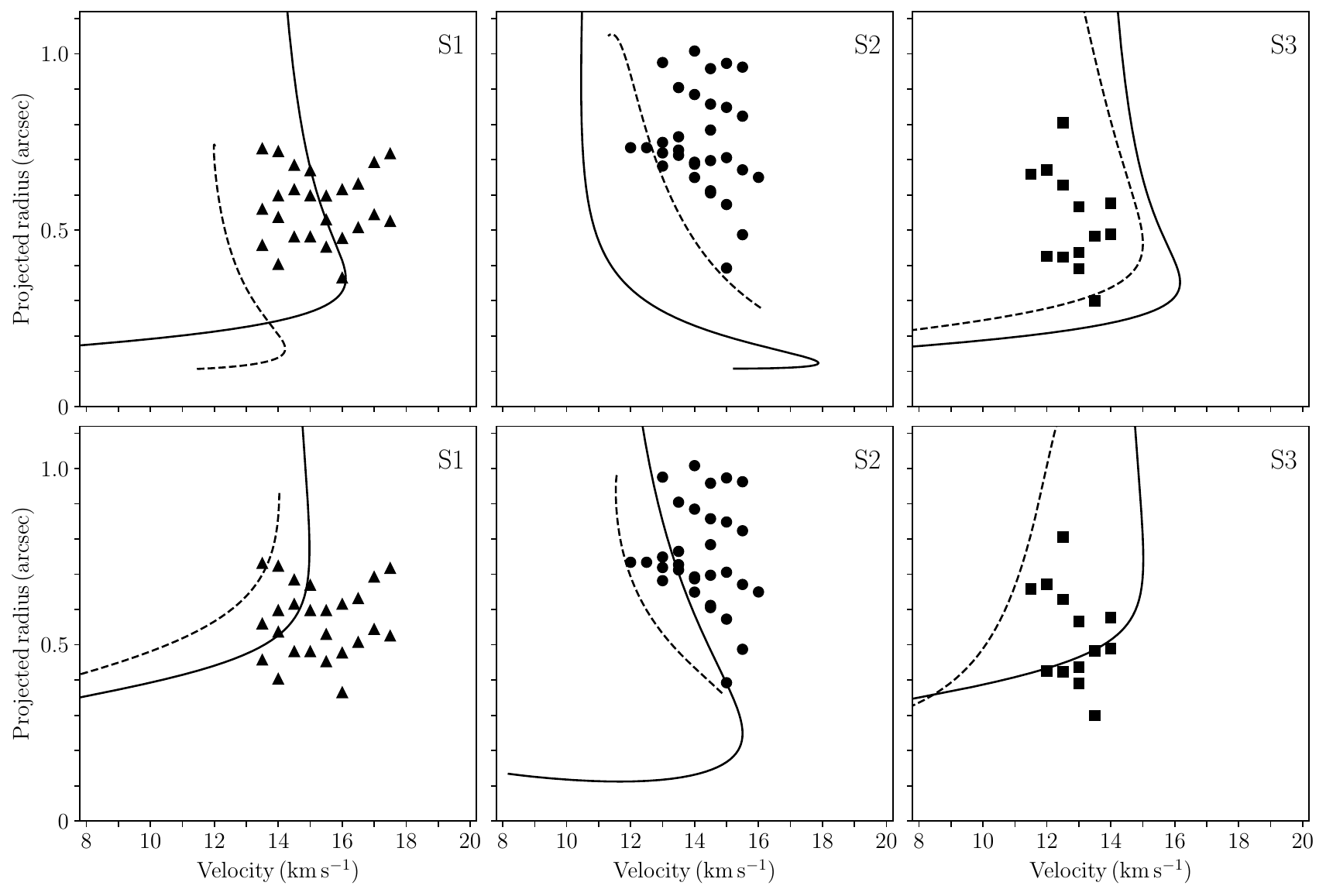}
    \caption{The CH$_3$OH position velocity diagrams  along the streamers S1, S2, and S3 (i.e., observed line-of-sight velocity versus projected distance to the protostar location) for our best fits to the data using the Ulrich's (solid lines) and Mendoza's (dashed lines) models. Note that the system velocity is 12.1\kms. Top panels: using $r_{in}$ as the molecular disk radius. Bottom panels: using $r_{in}$ as the envelope radius. The triangles, circles, and squares correspond to the S1, S2 and S3 streamers, respectively.}
    \label{f:pvdiagrams}
\end{figure*}

The position-velocity diagrams along the streamers S1, S2, and S3, are shown in Figure \ref{f:pvdiagrams}. The solid and dashed lines represent our best fits using Ulrich's model and Mendoza's model, respectively (same as in Fig. \ref{f:streams1}). The top panels present the models where $r_{in}$ equals to the molecular disk radius, and the bottom panels show the models where $r_{in}$ equals to the envelope radius. The triangles, circles, and squares correspond to S1, S2 and S3 condensations (same as in Fig. \ref{f:streams1}). The panel in this Figure show that both models can roughly match the velocities in the streamers. Both models show a lack of detailed coincidence, that could be motivated by coarse angular and spectral resolutions and the manual selection of the putative streamer trajectories. The velocities obtained using the AGA algorithm should be considered as a rough approximation to the observational line-of-sight velocities, since the simple models used do not consider different physical phenomena such as the thermal pressure, the magnetic pressure, or the expansion of MM1's outflow.

\begin{figure*}
\includegraphics[width=\linewidth]{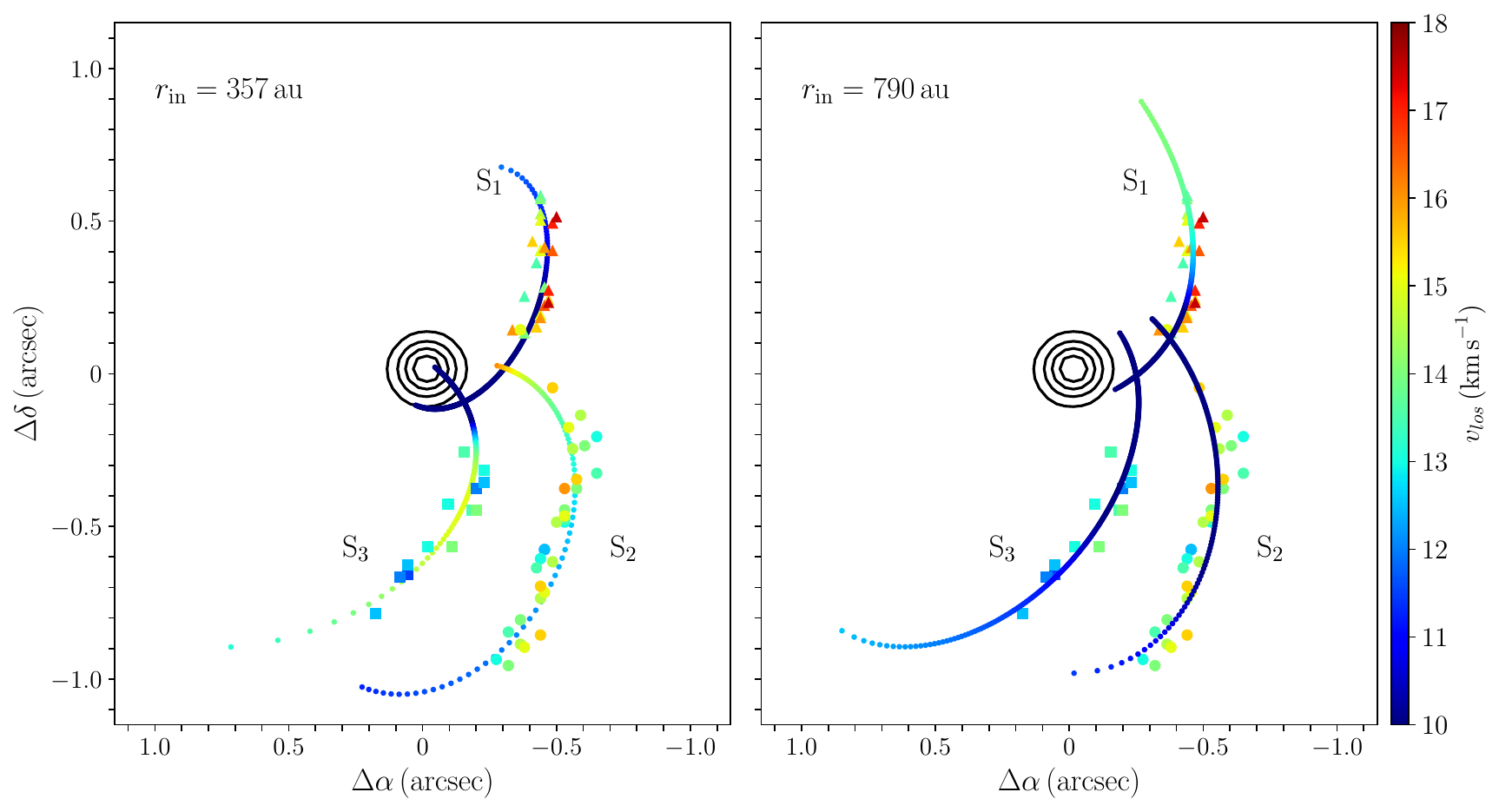}
\caption{Trajectories of the S1, S2 and S3 streamers on top of the contours marking the strongest continuum emission. The line-of-sight velocity of the streamers for our best fit to the CH$_3$OH data using Mendoza's model is also depicted (color scale). Left panel: using $r_{in}$ as the molecular disk radius. Right panel: using $r_{in}$ as the envelope radius. The triangles, circles, and squares correspond to condensations of the S1, S2, and S3 streamers and the colors follow the same velocity color scale as the model trajectories.}
\label{f:rvelocity} 
\end{figure*}

Figure \ref{f:rvelocity} presents the line-of-sight velocity for our best fit trajectories found using Mendoza's model in the three streamers. We chose to use Mendoza's model because Ulrich's model cannot fit the curved trajectories of the streamers. In this Figure we made a comparison between the line-of-sight velocities obtained from Mendoza's model and those measured in the CH$_3$OH condensations. The left panel corresponds to the model where $r_{in}$ is the molecular disk radius and the right panel represents the model where $r_{in}$ is the envelope radius. a Few contours of the continuum emission are shown for reference purposes.


\subsection{Molecular absorption in the inner region of the disk}\label{sec:absorption}

\begin{figure}
  \includegraphics[width=\linewidth]{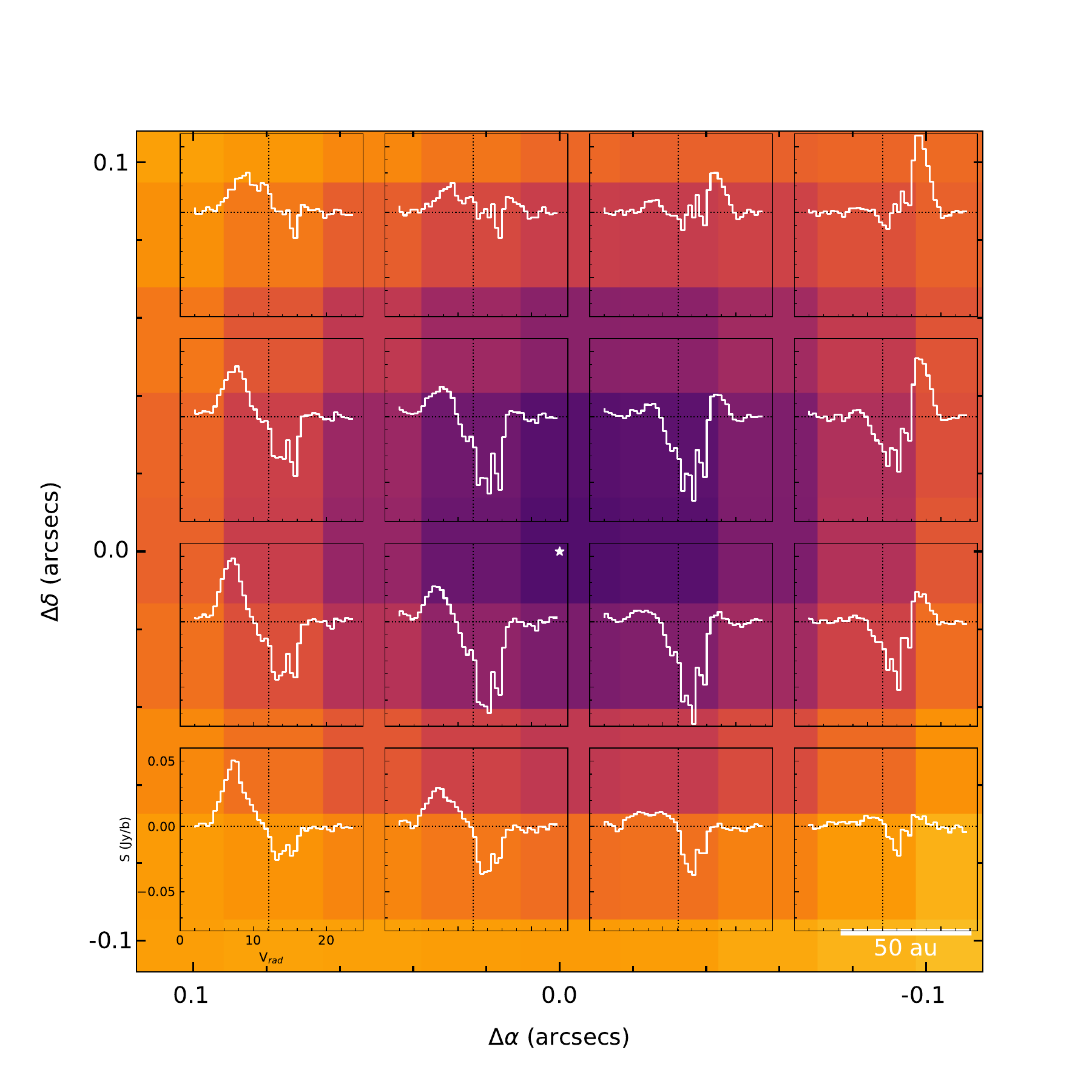}
\caption{H$_2$S integrated intensity image (moment 0) of the inner region of the GGD27-MM1 disk. The color scale shows the negative absorption toward the center position.
The figure shows H$_2$S\,3$_{3,0}$-3$_{2,1}$ spectra extracted and overlay every quarter of a beam, approximately. The star marks the position of the peak of the millimeter continuum emission.}
\label{f:cgspec} 
\end{figure}

Figure \ref{f:cgspec} shows the absorption feature in the spectra toward the center of the disk ($0\farcs2\times0\farcs2$ box, or 280\,au$\times$280\,au at the source distance) in the H$_2$S\,3$_{3,0}$-3$_{2,1}$ transition (Table \ref{t:molecules}). This figure presents one spectrum approximately every quarter of the synthesized beam. The emission in these spectra nicely shows the well known rotation pattern of the disk --a southeast (blueshifted) northwest (redshifted) gradient-- but the most noticeable feature is the broad absorption toward most of the positions (between 10.5\kms and 16.5\kms). The H$_2$S absorption peaks do not follow a pattern of rotation analogous to that of the emission; they do not trace a velocity gradient along the inner disk positions. This can indicate that the absorption is produced in colder and more quiescent foreground layers, not sharing the disk kinematics, or in the cold flared structure from the southern rim, roughly aligned with the disk's minor axis, which may not display significant rotation features. The absorption features mainly peak at redshifted velocities and they are more prominent southward of the continuum peak, marked with a star symbol. Moreover, there seems to be three main absorption peaks at velocities 13.0\kms, 14.5\kms and 16.0\kms. 

\begin{figure}
\minipage{\linewidth}
\includegraphics[width=\linewidth]{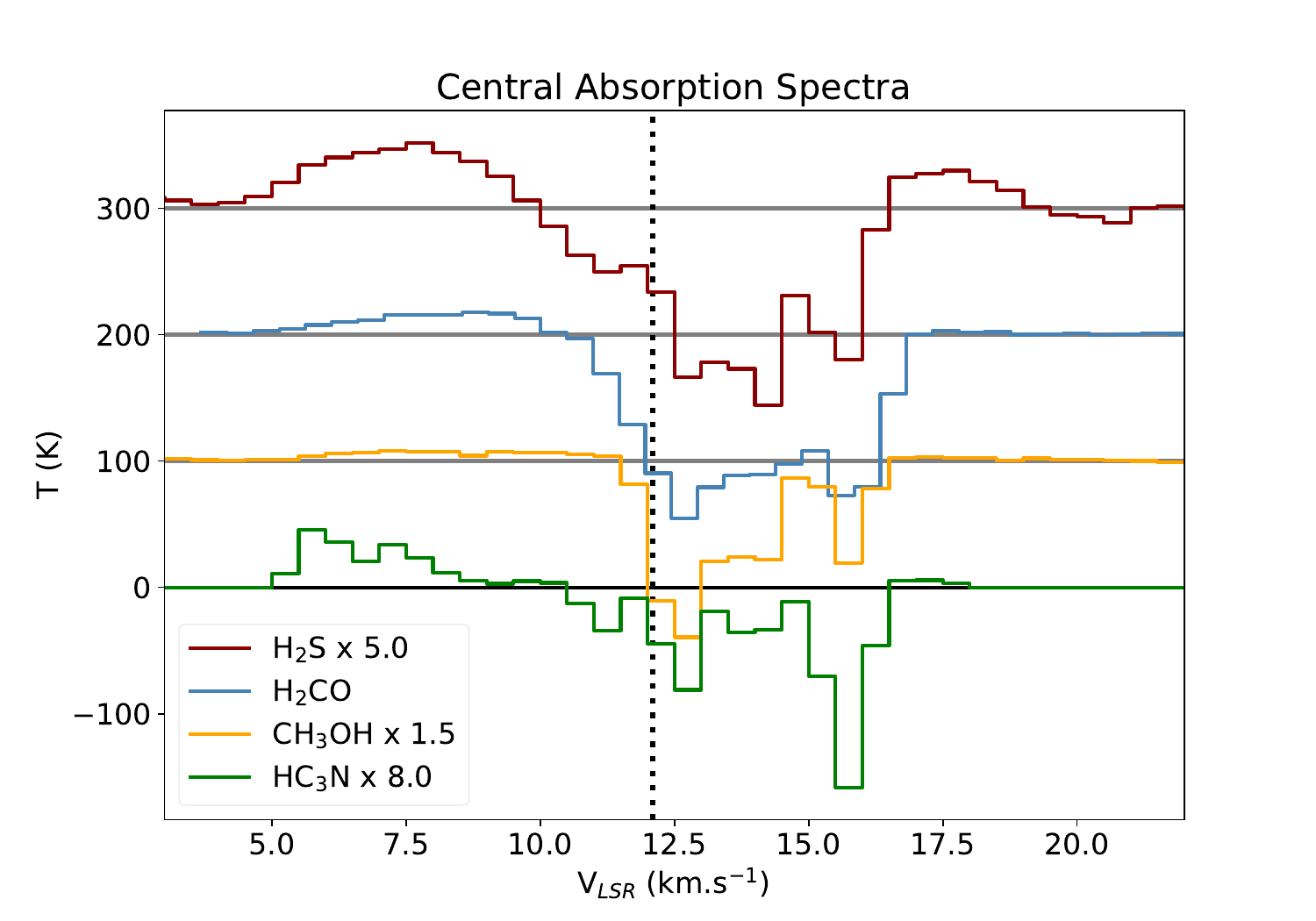}
\endminipage\\
\minipage{\linewidth}
\includegraphics[width=\linewidth]{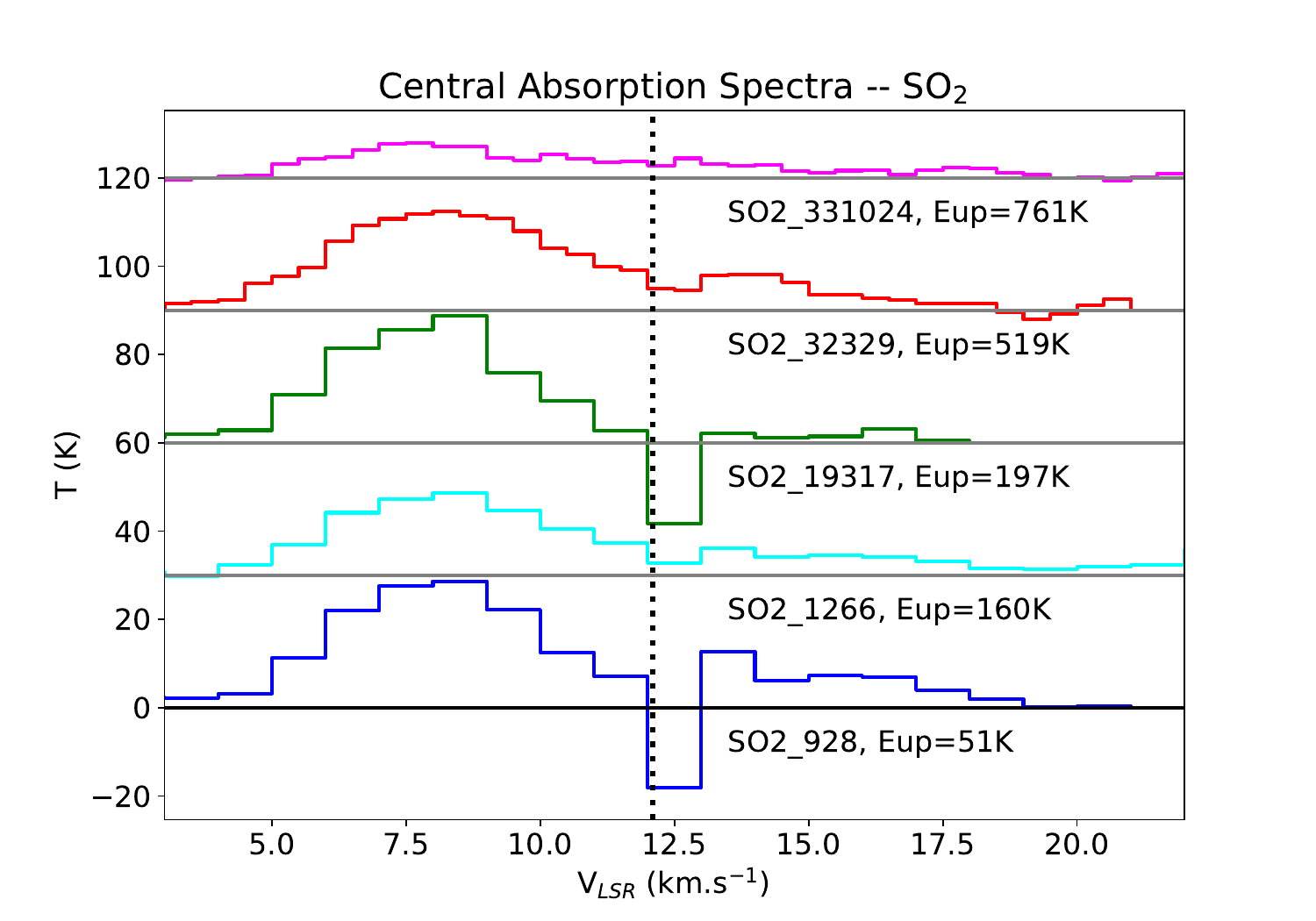}
\endminipage\\
\minipage{\linewidth}
\includegraphics[width=\linewidth]{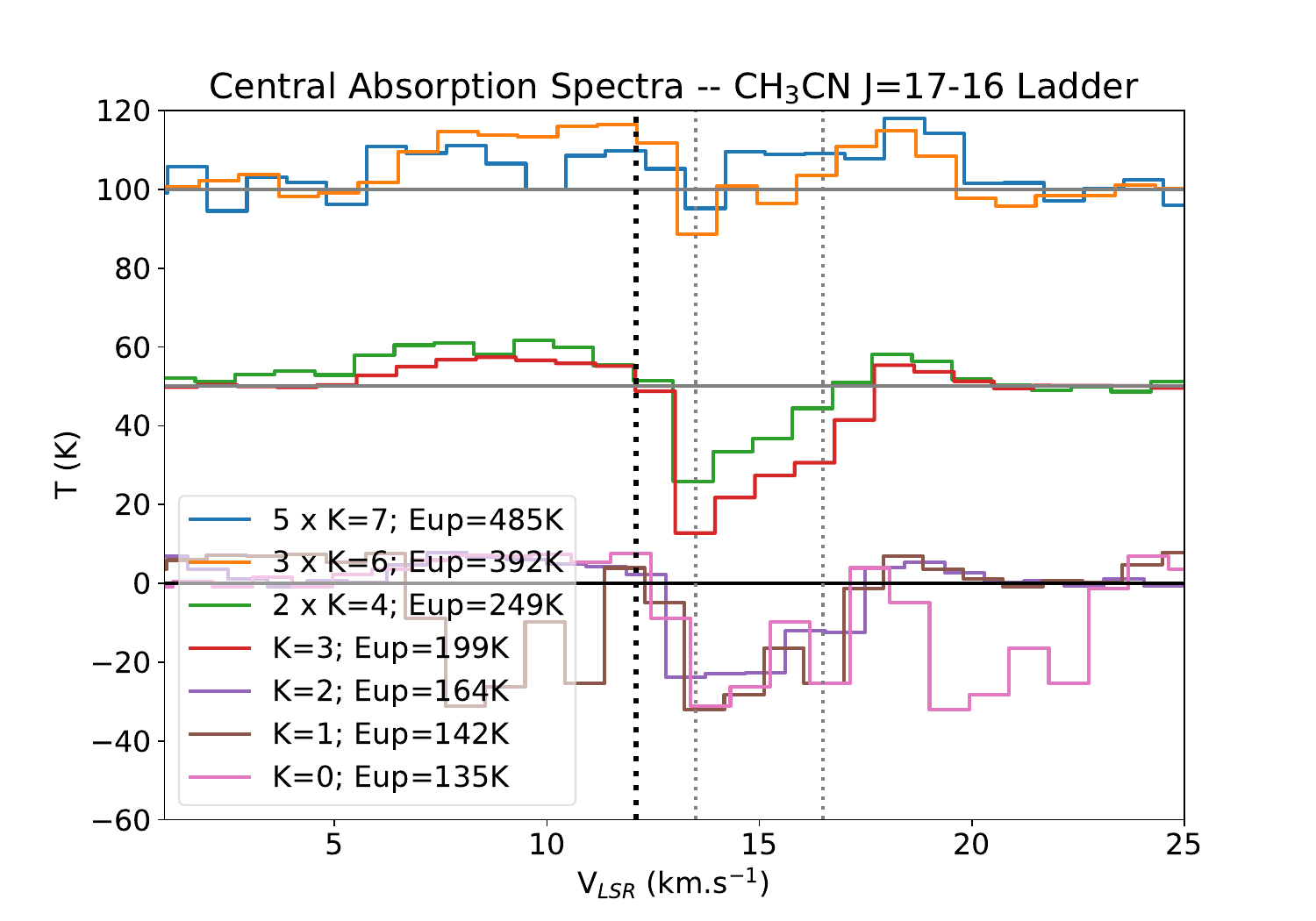}
\endminipage
\caption{Spectra of different molecular transitions integrated over the central absorption region. The dotted black line in all the panels marks the 12.1\kms systemic velocity. Top: H$_2$S, H$_2$CO, CH$_3$OH and HC$_3$N lines. Middle: SO$_2$ lines with E$_{up}$ ranging 51\,K and 761\,K. Bottom: several lines of the CH$_3$CN\,J=17-16 ladder. To help the readability some spectra are shifted in the y axis, and some are also multiplied by an integer factor indicated in the legend. The two grey dotted lines mark the velocity of the two main absorption peaks at 13.5\kms and 16.5\kms, see text.}
\label{f:absorption} 
\end{figure}

We integrated the spectra of several molecular transitions found in the ALMA observations over the central beam of the GGD27-MM1 disk (Figure \ref{f:absorption}). Some of these lines trace mainly the gas of the hot molecular disk (e.g., H$_2$S, SO$_2$), some others trace the colder envelope as well (e.g., CH$_3$CN ladder, H$_2$CO, SO$_2$). As can be seen in the three panels of Figure \ref{f:absorption} (one for the CH$_3$CN ladder, one for the SO$_2$ lines and one for the other absorbed transitions), most of the lines show redshifted absorption in the same velocity range as H$_2$S. However, different species show a different number of absorption peaks. While the H$_2$S line shows three peaks, the CH$_3$CN, CH$_3$OH and H$_2$CO lines show two peaks at 13.0\kms and 16.0\kms, and the SO$_2$ lines show one absorption peak at $\sim12.5$\kms.  

Figure \ref{f:absorption} (middle panel) shows that the two SO$_2$ lines with absorption have upper-energy levels (E$_{up}$) lower than 200\,K. The only line which do not follow this trend is SO$_2$[12$_{6,6}$-13$_{5,9}$]. 

For the CH$_3$CN lines of the J$=17-16$ K-ladder, the ALMA data show absorption profiles with two peaks at 13.5\kms and 16.5\kms. The absorption at 13.5\kms is noticeable from line K$=0$ to K$=7$ (E$_{u}=250$\,K and 485\,K, respectively). The K$=5$ line is contaminated by other molecular transitions and does not show a clear profile. Interestingly, the absorption at 16.5\kms is only significant up to K$=4$. The emission of the ladder is detected up to the K=9 line (with E$_u=713$\,K). The differences in the spectra from these two absorption peaks could indicate the presence of two different kinematic components, with different physical properties.

Finally, at the central pixel, the fit of the CH$_3$CN K-ladder is more complicated than for the rest. The line profile shows two redshifted absorption peaks and two emission peaks: one blueshifted and other redshifted. Given the degeneracy in the space of parameters resulting of using that many spectral components, we choose to fit the profile using a simple approach comprising three interacting layers in CASSIS: (i) a background emission layer with uniform temperature corresponding with the dust continuum emission from the compact disk, (ii) a layer corresponding to the gaseous component of the hot disk, and (iii) a foreground layer corresponding to the warm envelope. As a proof of concept, we fixed the background temperature to a uniform value close to that reported in \cite{2020AnezLopez} (300\,K at 100\,au from the central star). However, fixing this parameter has an important effect on the excitation temperatures of the other two layers (the disk of gas and the envelope). Hence, we have repeated the fit using temperatures from 250\,K to 450\,K in steps of 50\,K. The fits are qualitatively similar and the fit residuals are minimum between 250\,K and 300\,K. The results of the spectrum fit (Table \ref{t:centralpix}, Fig. \ref{f:example_fits3}), qualitatively agree with the warm envelope and hot disk picture outlined in this work, but the simplicity of the model leave large uncertainties that ask for more information at the center of the system.  Higher angular and spectral resolution observations of photodissociated gas tracers may give further evidence for a more detailed characterization.  Despite of this, one relevant takeaway is that the full width at zero intensity of the emission line is smaller than 14.5\kms for any line reported here (i.e., they show emission at maximum from +5.0\kms to 19.5\kms). This would imply to probe the rotating molecular gas at a radius of 340\,au from the central 20\msun protostar (if the mass of the protostar is 10\msun, the radius would be 170\,au). 

\begin{deluxetable}{ccccc}
\tablewidth{0pt}
\tablecolumns{5}
\tabletypesize{\scriptsize}
\tablecaption{CH$_3$CN central pixel spectral fitting}
\tablehead{
\colhead{Spectral Component} & \colhead{N$_{CH_3CN}$} & \colhead{T$_{ex}$} & \colhead{V} & \colhead{$\Delta$V} \\
\colhead{}  & \colhead{(\cmd)} & \colhead{(K)} & \colhead{(\kms)} & \colhead{(kms)}
}
\startdata
Dust Disk\tablenotemark{$\dagger$} & \nodata & 300 & \nodata & \nodata \\
Gas Disk & 7$\pm1\times10^{16}$ & 480$\pm190$ & 9.8$\pm0.1$ & 5.8$\pm0.7$ \\
Envelope & 3.5$\pm0.3\times10^{15}$ & 150$\pm40$ & 13.7$\pm0.5$ & 3.0$\pm0.5$ \\
\enddata 
\tablenotetext{\dagger}{Optically thick background emission source at a fixed an uniform temperature assumed to be between 250\,K and 450\,K. Note that this background temperature has a big uncertainty and needs to be finetuned for a more detailed analysis. The uncertainties of the free parameters are estimated taking into account the fits at the two background temperature extremes.}
\label{t:centralpix}
\end{deluxetable}

\subsection{Accretion Rate}\label{sec:accretion}
The redshifted absorption in the spectra of molecules tracing the envelope and/or disk material surrounding a protostellar system can be interpreted as an inverse P-Cygni line profile, originated by active accretion of material. In the case of GGD27-MM1, the absorption is seen at the center of the system, within the inner 190\,au, but the ALMA data also reveals two other signatures related to accretion processes: the redshifted quiescent material from the warm extended envelope and the accretion streamers (Sections \ref{sec:tk} and \ref{sec:streamers}). On one hand, we now assume that the redshifted absorption is due to gas moving inwards toward the center. On the other hand, we assume that there is ongoing accretion from the extended envelope into the hot gaseous disk component. Following these hypotheses, we can apply an analogous procedure to that described in \cite{2006Beltran} and \cite{2022Beltran} to estimate the mass accretion rate at two highlighted radii: (i) the centrifugal radius, 330\,au, which roughly coincides with the outer radius of the H$_2$S gas disk ($\sim360$\,au) and (ii) the dust disk radius, of 190\,au. These radii were picked to make estimates of the (i) infalling rate from the extended envelope onto the hot disk, and (ii) of the accretion rate from the external parts of the hot disk (mostly molecular) onto the dust disk. We use the expression:
\begin{equation}
  \dot{M}_{acc}=\Omega~R^2~m_{H_2}~n_{H_2}~v_{acc} \quad, 
\end{equation}
where $m_{H_2}=\mu\cdot m_H$ is the mass of the hydrogen molecule \citep[we use $\mu=2.8$ as the mean mass per H$_2$ molecule, see][]{2008Kauffmann}, $R$ is the radius at which the rate is calculated, $n_{H_2}$ is the H$_2$ column density at radius $R$, and $v_{acc}$ is the infalling accretion velocity. We approximate \begin{equation}
 n_{H_2}(R)=\frac{N(R)}{H(R)}=\frac{N(R_0)}{H(R_0)}\left(\frac{R}{R_0}\right)^{-1.5} \quad ,   
\end{equation}
where $H(R)$ is the height of the disk at radius $R$, which we assume is proportional to $R$. Likewise, the column density $N$ goes as $R^{-0.5}$, as derived from the modeling by \cite{2020AnezLopez}. Using expressions 11 and 12, the mass accretion rate is:
\begin{equation}
   \dot{M}_{acc}=\Omega~R^2~m_{H_2}~\frac{N(R_0)}{H(R_0)}~\left(\frac{R}{R_0}\right)^{-1.5}~v_{acc} \quad. 
\end{equation}
We derive the infall velocity measuring the velocity of the two main peaks in the CH$_3$CN line profiles of the $J=17-16$ ladder (at 13.5\kms and 16.5\kms, Fig. \ref{f:absorption}); therefore, we estimate $v_{acc}$ to be between 1.4\kms and 4.4\kms (2.0\kms and 6.2\kms, corrected by projection, in case that the accretion is through the disk midplane). For the column density of the disk we assume the relationship derived via radiative transfer modeling of the 1.14\,mm continuum emission \cite[see equation 3 of][]{2020AnezLopez}. This results in $N=7.8\times10^{25}$\cmd at radius $R_0=190$\,au. Extending the density trend up to 330\,au gives a density of $N=5.9\times10^{25}$\cmd. For an isotropic infall, the mass accretion rate will be in the range $0.03-0.09$\msunyr at 190\,au, which is 3 orders of magnitude larger than the accretion onto the protostar estimated by \cite{2020AnezLopez} ($7\times10^{-5}$\msunyr). However, we can consider that the bulk of the accretion process is not spherical but goes through a smaller solid angle, $\Omega$.
As a raw estimate for this factor we take into account the vertical profile of the dust disk estimated in \cite{2020AnezLopez}. They estimated that the hydrostatic scale height of the disk would be 13\,au at a radial distance of 190\,au from the center. This implies a disk semi-opening angle of $4\degr$. From the model of the H$_2$S disk emission the total flare angle of the disk would be $22\degr$ (see Section \ref{sec:flared}). This implies a mass accretion rate one order of magnitude smaller, ranging $7\times10^{-4}-1.24\times10^{-2}$\msunyr at 190\,au, and a larger $5.2\times10^{-3}-1.64\times10^{-2}$\msunyr at 330\,au (for this latter distance from the protostar, we use only the large flare value of $22\degr$).  

\section{Discussion} \label{sec:discussion}
The ALMA data presented in this work reveals that the GGD27-MM1 protostellar system comprises four different structures. (i) At small scales, surrounding the protostellar object(s), there is a dust continuum disk, with a radius  estimated in 170-190\,au; (ii) encompassing the dust there is a rotating hot gaseous disk, seen in emission between 140\,au and 360\,au; (iii) a more spread envelope, roughly 1600\,au in diameter, detected in dust emission but also in several molecular components; (iv) at least three molecular accretion streamers with lengths between 850--1700\,au apparently bridging external material onto the disk structure.

\subsection{The compact continuum disk}
Regarding the kinematics of the compact dust disk, the central emission/absorption detected in several species does not show the larger velocity spread expected in the inner part of a rotating structure. For the H$_2$S line, the radial velocity ranges from +5.0\kms to +19.5\kms in the LSR, suggesting a rotation velocity of $\sim7.25$\kms at the innermost radius of its emission. However, for a 20\msun central star, the expected Keplerian velocity at, for example, 20\,au, would be of 30\kms. If the central mass estimate is accurate enough, this implies that the molecular content detected in the GGD27-MM1 disk reaches an innermost radius of $\sim 340$\,au, which is approximately twice the 141\,au of the inner radius derived from our phenomenological molecular model (see Section \ref{t:flare_fit}). To reconcile these two radii, the central mass should be of about 10\msun, or the measured rotation motions are sub-Keplerian. 
The lack of molecular emission at high velocities could explain the problems constraining the dynamical mass of the system and, more generally, the problems probing the keplerianity of high-mass protostellar disks and pseudodisks. 
A possible explanation for the lack of molecular emission inward of this radius \ad{at large velocities}, is that there is not much molecular content left embedding the compact continuum disk, maybe due to photoevaporation and/or radiation pressure. The radiation pressure effect is less likely since it is various orders of magnitude lower in molecules than in dust grains. Therefore, the lack of abundant molecular material in the inner part of the system, is more likely due to photodissociation and/or photoevaporation. Interestingly, while this inner region is apparently devoid of \ad{molecular gas at high velocities}, the radiation has not destroyed the dust grains at the disk's midplane yet. Actually, the dust is optically thick at the inner part, and could hide the molecular remnants within the $\tau=1$ layer, completely absorbing their emission. The emission/absorption detected at the central position seems to come from gas from the inner radius of the molecular disk component and from the continuum radiation absorbed by the tenuous and widespread envelope.

In Section \ref{sec:accretion} we estimated the rate at which the dust disk is replenished with material from the hot molecular disk in $\sim5\times10^{-4}$\msunyr. This rate implies to increase the mass of the disk in 1\msun in $\sim2000$\,yr, which is currently estimated to be of 5\msun \citep{2020AnezLopez}. This rate is one order of magnitude larger than the accretion rate onto the protostar estimated by \cite{2020AnezLopez} ($7\times10^{-5}$\msunyr), meaning that the infalling material is bottle-necked onto the disk and may stay in the compact disk for $\sim15000$\,yr, orbiting the central object(s). This can explain the formation  of a massive disk, which may have time to form gravitational instabilities.
The accretion rate onto the compact disk would be lower if the solid angle of the accretion (as seen from the central protostar) is lower than the assumed here. One possible way is by feeding the compact dust disk asymmetrically (e.g., through streamers). However, the current angular resolution of the molecular data is not enough to discern if the detected streamers carry material below $\sim360$\,au from the protostar(s).     

\subsection{Central absorption}
Figure \ref{f:absorption} shows how different molecules have absorption peaks at slightly different velocities at the center of the  GGD27-MM1 system. The absorption is not detected at large velocities expected in the inner part of the rotating compact disk, indicating that it happens likely in colder and quiescent gas layers associated with the envelope. 
Redshifted absorption toward circumstellar disks or massive hot cores have been interpreted as signposts of infall motions from actively accreting layers with colder gas than the continuum \citep[e.g.,][]{2013Zapata,2018Beltran,2022Beltran}. \ad{An alternative explanation, is that line intensities could be suppresed due to a continuum over-subtraction, caused by the CH$_3$CN lines being marginally optically thick at the disk center. Regardless, in} the case of GGD27-MM1, the continuum at the inner disk is absorbed by the colder gas surrounding it. Interestingly, some of the accreting streamers shown in previous Section \ref{sec:streamers} (and other possible streamers not identified here, but whose presence is hinted from the molecular gas distribution of some species such as H$_2$CO or SO$_2$ and CH$_3$OH) end at the redshifted velocities of the absorption peaks. Assuming that the accreting streamers cover the central part of the disk, causing the detected absorption, the material may be just hovering the disk surface.


To estimate the mean column density in the inner 210\,au of the GGD27-MM1 protostellar system, we take, as an informed value, the CH$_3$CN column density extracted from our fit of the $J=17-16$ ladder at the central pixel (Table \ref{t:centralpix}): $7.0\times10^{16}$\cmd. We can compare this value with an average of the H$_2$ column density in this region, estimated using the density profiles given by equation 3 in \cite{2020AnezLopez}: $3.1\times10^{26}$\cmd. Hence, we derive an abundance of CH$_3$CN with respect to H$_2$ in the range $2\times10^{-10}$. This is a value about 2 orders of magnitude lower than the fractional abundances estimated toward several hot molecular cores \citep[$\sim 10^{-8}$ ,e.g.,][]{2014HernandezHernandez,2020Liu}, and about one order of magnitude smaller to that found in the Orion hot core \citep[$10^{-10}-10^{-9}$][]{1994Wilner,2014Bell}. If this means that the abundance of the gas decreases at the center, it can explain why the inner regions of the disk do not display the Keplerian pattern of rotation expected at higher velocities (or why CH$_3$CN transitions with higher energy levels or even vibrational transitions were not detected at these high-velocities). As commented before, the reasons behind this low abundance are unknown and we can only hypothesize about processes such as photoionization and/or radiation pressure. \ad{There is also the possibility of a continuum over-subtraction due to radial variations of temperature and opacity of the lines and the continuum.} Further experiments probing atomic and/or infrared ro-vibrational transitions from the simplest molecules --such as CO-- that may have escaped destruction, are needed to give more insight in this respect. \ad{Also, a good knowledge of the radial profile of the dust temperature and opacity, through detailed multi-wavelength observations, can help to test the continuum over-subtraction effect.}

\subsection{Hot molecular disk}



The current data hints at the possibility that several accretion streamers channel material that infalls onto the molecular disk (e.g., Figs. \ref{f:moms0} and \ref{f:streams1}). We estimate a large accretion/infall rate ranging between $10^{-3}$\msunyr and $10^{-2}$\msunyr. For this estimate we consider that the infalling solid angle is the flared region of $22\degr$ surrounding the midplane of the disk. If the presence of gas streamers means that the infall is channeled through them, the solid angle from the protostar depends on the section of the streamers at the distance of the deposition and the number of the streamers. The current data do not allow us to clearly resolve the width of the streamers. Hence assuming a $0\farcs15$ width (the geometrical size of the beam), the solid angle for one streamer, decreases a factor of 3--10 relative the solid angle of the flared disk. This factor is compensated by the number of streamers. Therefore, with the present data constrains, we can assume that the infall of material from the outskirts onto the disk may proceed very quickly.

\subsection{Warm gas envelope}
The warm envelope is seen at quiescent velocities ($<4$\kms), that are consistently redshifted with respect to the disk central velocity of 12.1\kms derived in the kinematic analysis of the disk rotating pattern \citep[e.g.,][]{2011FernandezLopez2,2017Girart}.  If the material from the envelope is gravitationally bound\footnote{The escape velocity at 790\,au, the radius of the envelope, is 4.7-6.7\kms for a central mass of 10-20\msun.} in the potential well of GGD27-MM1, then the fact that we only see redshifted emission can be explained by the disk being at the background edge of the original dense pocket. This pocket of material may be infalling onto the central system. Another redshifted-only envelope has been seen in RU\,Lup \citep{2020Huang}, and for example, the HH\,30 protostellar system is known to be at the edge of its dense cloud \citep[][]{2018Louvet}.


\subsection{Possible Accretion Streamers feeding GGD27-MM1}\label{sec:disc_streamers}
Accretion streamers have lately been reported to channel material into low-mass young stellar systems from all evolutionary stages \citep[see][and references therein for a current review]{2022Pineda}. The structures found in GGD27-MM1 match the expected shape, length and kinematics shared by most of these streamers (as showed by our fitting), suggesting that the accretion process from large scales continue asymmetrically beyond the formation of a massive protostar and its massive disk. In principle, an asymmetric infall of material could be regarded as a slower accretion process (that is, considering two scenarios with the same flux of material through a fixed solid angle, infall through a few small cones will provide a slower accretion rate --less accretion-- than isotropic infall), but if instead of the isotropic process, accretion is carried out via a relatively small flare angle (i.e., the infalling solid angle is just a portion of the whole spherical surface), and if there are several wide enough accretion streamers, the subtended angle of the accretion could be even larger than in the case of a \textit{flared} isotropic accretion.

The analysis in Section \ref{sec:streamers} shows the plausibility that the filament-like structures extending outwards from the molecular disk/envelope of the GGD27-MM1 system could be accretion streamers. Both the morphology and the kinematics are qualitatively fitted by our modeling (Figs. \ref{f:streams1}-\ref{f:rvelocity}). The model using Mendoza's paradigm of three streamers carrying material down to the envelope radius agrees specially well with the spatial distribution of the data. On the contrary, Ulrich's model does not work as well, rejecting the more static scenario proposed by this second model. Note, however, that from the line-of-sight velocities (Fig. \ref{f:pvdiagrams}), it is not obvious which model is better. The other relevant result is that using Mendoza's model, the current data do not allow to discern the endpoint for the accretion streamers. That is, the accretion streamers could well transport material onto both the outskirts of the envelope or the molecular disk. With the current angular resolution and sensitivity, other fitting solutions are possible (e.g., using different end radii for the streamers), so the present fitting is shown here as a proof of concept for the accretion streamer hypothesis. 

In addition, we have only used the three streamers that are most apparent in the data of the different molecules. However, the molecular protuberances to the north of the molecular disk and to the northeast of the envelope, along with the persistent redshifted emission toward the east (contrarily to the rotation disk pattern) could indicate that other accretion paths exist. 

Finally, let us note that, even though the hypothesis of accretion streamers seems to be very likely in this system, alternative explanations for these structures cannot be ruled out. A fly-by event or a close encounter with another object could display tails and spiral arms in large scales too \citep{2018Kurtovic,2020Zapata,2022Cuello}. A further analysis studying a wider field of view surrounding GGD27-MM1 is currently ongoing, and its results will be showed in a forthcoming publication. 

\section{Conclusions} \label{sec:conclusions}
In this contribution we have presented new ALMA 0.98\,mm observations with angular resolution $\sim0\farcs12$ toward the high-mass protostellar system GGD27-MM1. The main results and final ideas of the paper are:
\begin{itemize}
    \item The continuum emission is well reproduced by a model including an unresolved central ionized source, a 190\,au disk, a tenuous 0.2\msun envelope and two compact NW and NE components.  
    \item We mapped the emission of different SO$_2$ transitions, and CH$_3$OH, H$_2$CO, H$_2$S, HC$_3$N and CH$_3$N lines. The different lines trace the molecular rotating disk, the envelope and several arc-like structures, similar to the accretion streamers seen to feed material onto young stellar systems.
    \item The H$_2$S line traces the molecular disk, seen as a crescent with a dim in the southern rim. We modeled its emission through an ad-hoc model comprising a compact dust disk, whose emission is absorbed by the cold gas between the emitting lids of a flared molecular disk. The flared angle of the disk, extending between 140 and 360\,au, is derived to be $22\degr$. 
    \item We analyzed the emission from the CH$_3$CN~(17-16) ladder with CASSIS, producing maps and radial profiles of the temperature, the column density and the velocity for two different kinematical components: (i) the hot and rotating molecular disk, and (ii) the warm and quiescent redshifted envelope. We discuss that the redshifted envelope may be quiescently infalling onto the molecular disk, which may be located at the background end of its natal molecular pocket.
    \item Regarding the possible presence of accretion streamers, a model of infalling gas with an initial non-zero velocity toward the protostar (the so-called Mendoza's model) seems to qualitatively fit the trajectories and line-of-sight velocities of three of these streamers. We, nevertheless, include an estimate of an isotropic accretion infall rate onto the disk through the flared angle of $\sim5\times10^{-4}$\msunyr at a 190\,au radius, which is one order of magnitude larger than the accretion rate onto the protostar estimated by \cite{2020AnezLopez}. This may explain the formation of such massive disk (5\msun) in GGD27-MM1.
    \item The absorption toward the center of the system is detected at low velocities and, hence, it does not trace the disk rotation. It peaks at two or three different redshifted velocities depending on the transition, suggesting that may be related to different cold layers of gas. By fitting the CH$_3$CN spectrum, we found that material from the envelope is most likely producing the main absorption feature.
    \item A relevant implication from the lack of emission/absorption detected at high velocities is the possible existence of a central zone, roughly coincident with the extent of the dusty disk, devoid of \ad{rapidly rotating} molecular gas or hidden within the optically thick disk.
\end{itemize}


\begin{acknowledgements}
We thank the anonymous referee for a thorough and very constructive review that help to improve the consistency and quality of the paper. M.F.L. thanks the hospitality of the I.C.E. and the University of Barcelona; also the support from grant LACEGAL. J.M.G., R.E. and G.B. acknowledge support from the PID2020-11710GB-I00 grant funded by MCIN/AEI/10.13039/501100011033. S.C. acknowledges support from UNAM-PAPIIT IN103318 and IN104521 grants and from CONACyT, M\'exico. This work is also partially supported by the program Unidad de Excelencia Mar\'ia de Maeztu CEX2020-001058-M.

\end{acknowledgements}
\facility{ALMA}
\software{DS9 \citep{2003Joye}, CASA \citep[v5.6.2][]{2007McMullin}, Astropy \citep{2013Astropy}, Karma \citep{1995Gooch}, MAP\footnote{https://sites.google.com/fqa.ub.edu/robert-estalella-home-page/map?authuser=0}}

\newpage
\bibliography{biblio}

\appendix
\section{Continuum fit}\label{sec:appendix_continuum}
\begin{figure*}
\minipage{0.33\linewidth}
\includegraphics[width=\linewidth]{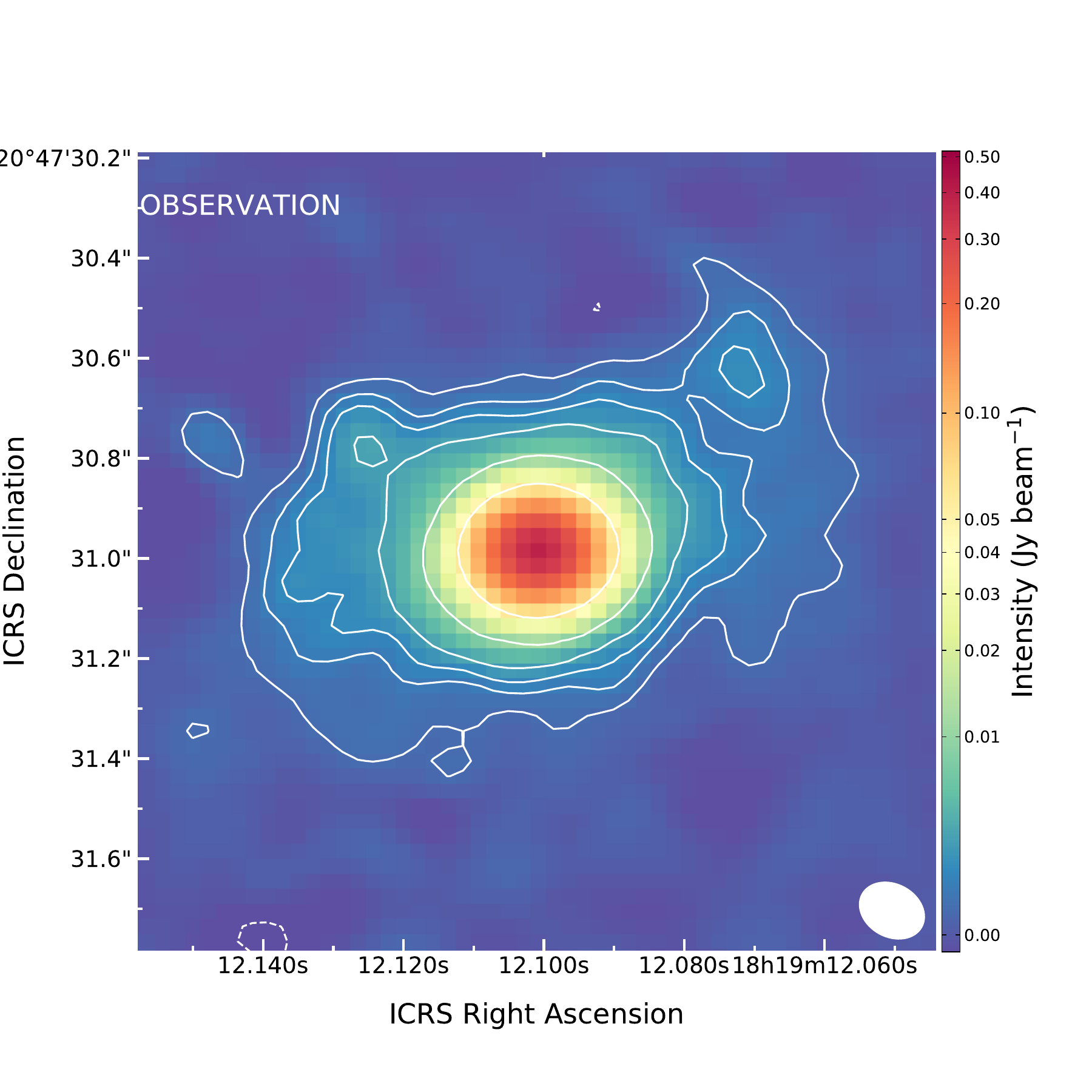}
\endminipage\hfill
\minipage{0.33\linewidth}
\includegraphics[width=\linewidth]{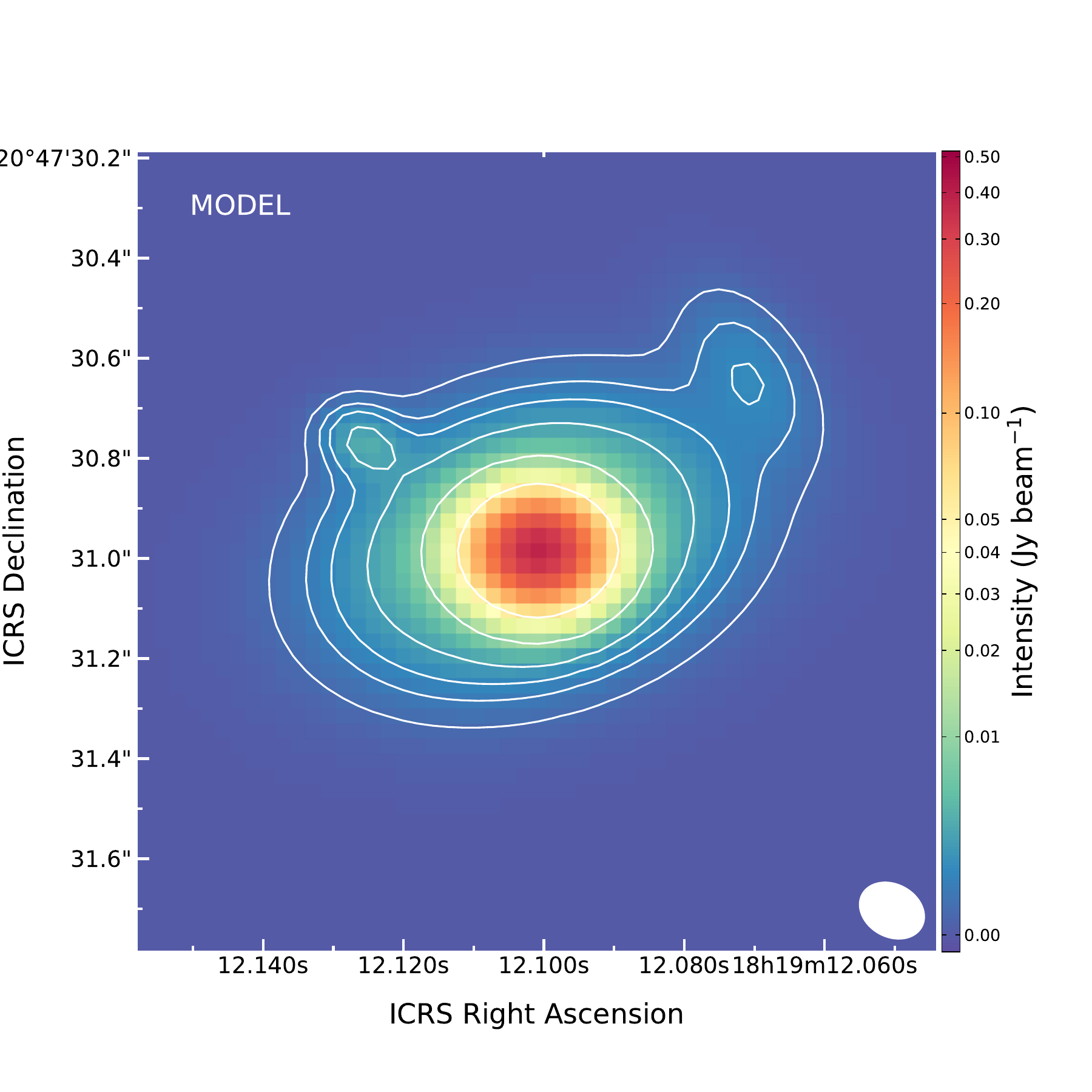}
\endminipage\hfill
\minipage{0.33\linewidth}
\includegraphics[width=\linewidth]{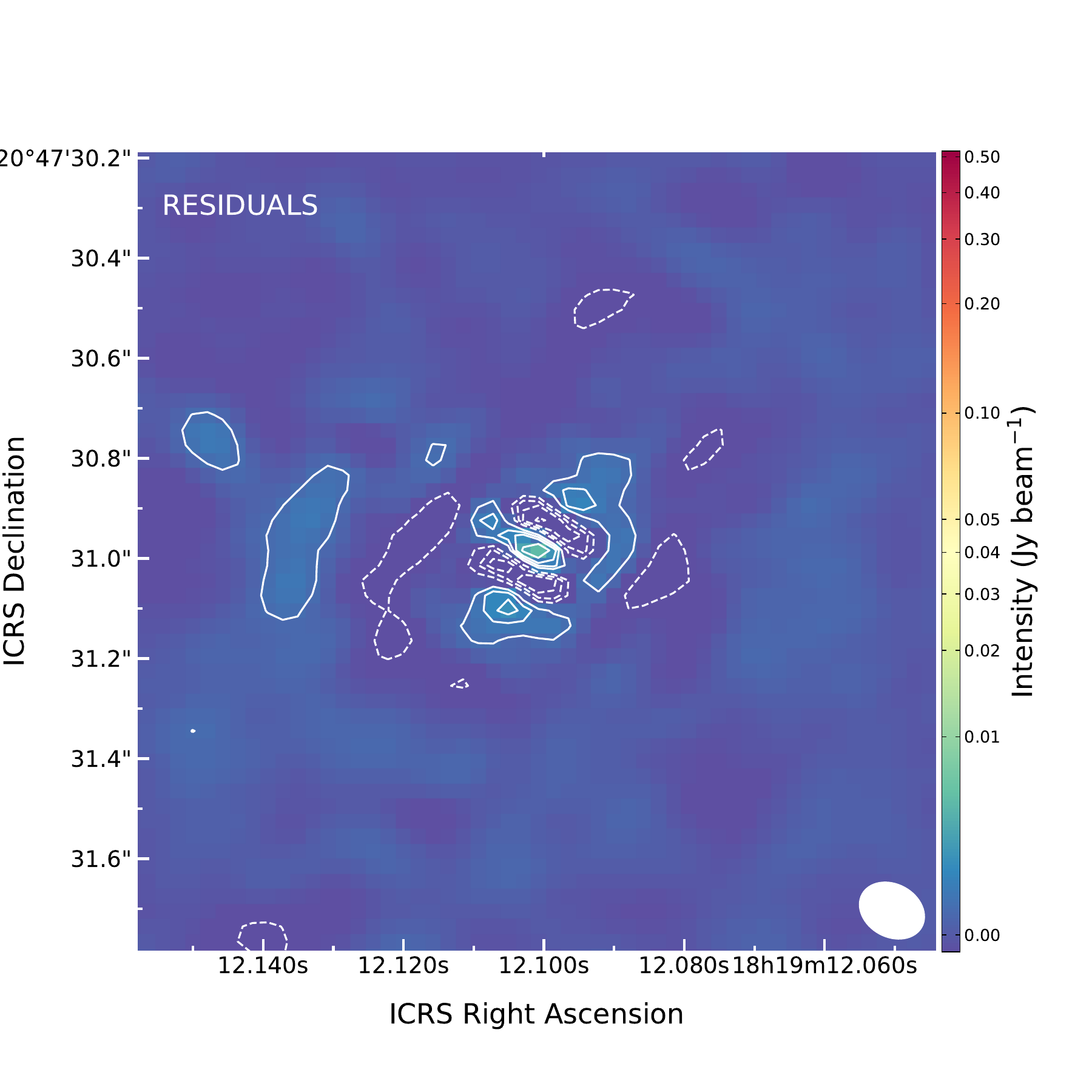}
\endminipage
\caption{Continuum model for the continuum emission. From left to right, the images show the observed image, the model obtained, and the residual image after the fit. 
The three images have the same color scale and contours, at levels -25, -10, -5, 5, 10, 15, 25, 75, 375 times the rms noise level of the image, $\sim0.14$\mjy.}
\label{f:continuum_model} 
\end{figure*}

This Section includes a visual comparison of the observed continuum emission and the model used to fit it in Section \ref{sec:disk_continuum}. A  residual image between both is also presented in Figure \ref{f:continuum_model}. The fit of the central source still leaves residuals above 20$\sigma$, while removing the bulk of the central $>2500\sigma$ emission.

\section{Velocity cubes}\label{sec:appendix_cubes}
This Section contains velocity cube images with the emission of the lines presented in this paper: H$_2$S\,(3$_{3,0}$-3$_{2,1}$), HC$_3$N\,(33-32), H$_2$CO\,(4$_{1,3}$-3$_{1,2}$), CH$_3$OH\,(3$_{1,2}$-2$_{0,2}$), SO$_2$\,(9$_{2,8}$-8$_{1,7}$), and CH$_3$CN~(17$_3$-16$_3$). The field of view is a zoom up to show the details of the disk and envelope toward GGD27-MM1. Enlarged versions of these figures may be presented in a forthcoming contribution. All of these figures show the ALMA Band 7 continuum emission overlapped with contours at 5, 15 and 100 times the rms noise level of the continuum emission (140\mujy). Details of each observed transition can be found in Table \ref{t:molecules}.   

\begin{figure*}
  \includegraphics[width=\linewidth]{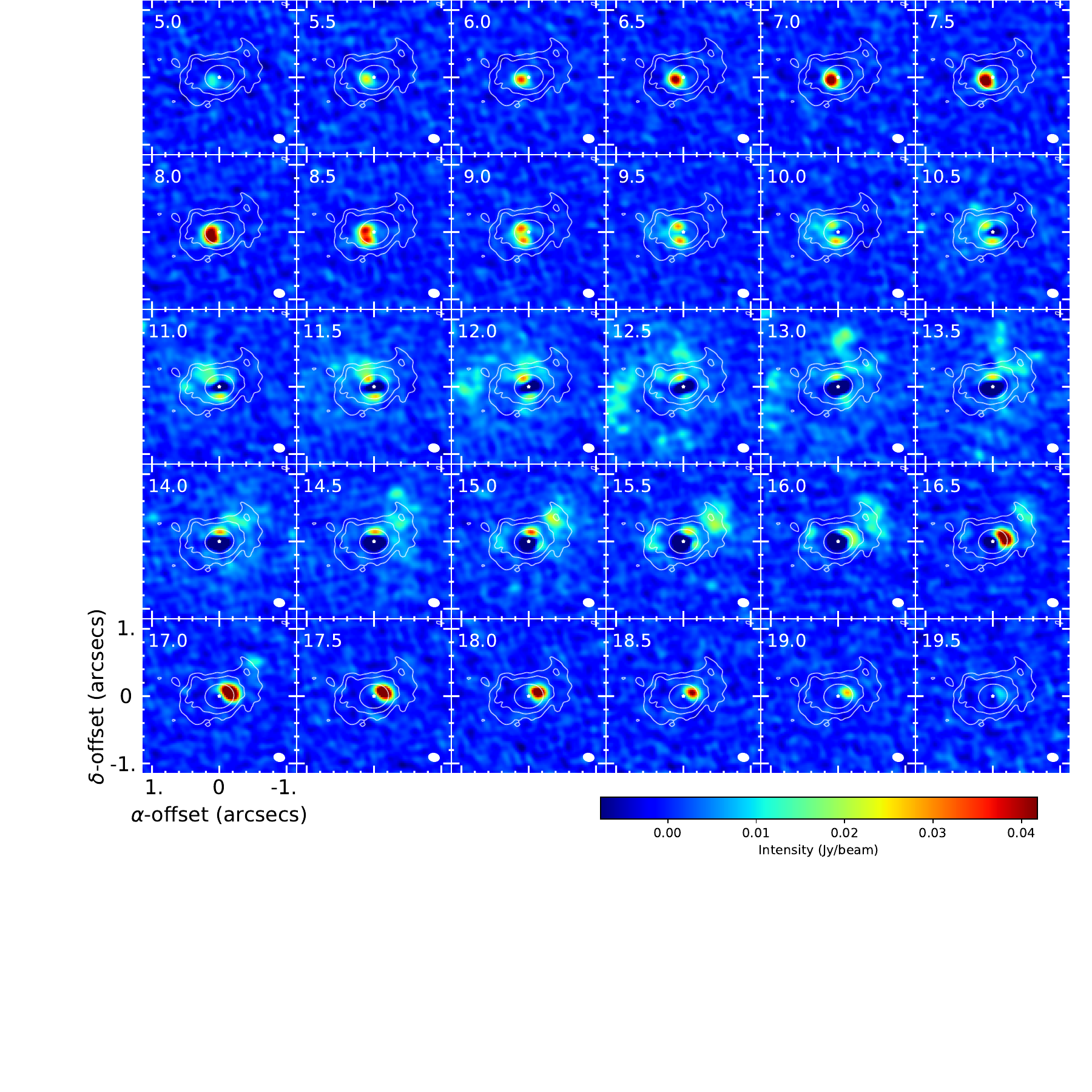}
\caption{Zoom up velocity cube image of the H$_2$S\,(3$_{3,0}$-3$_{2,1}$) emission toward GGD27-MM1. Velocity is labeled in the top left corner and the synthesized beam appears in the bottom right corner of each channel. Band 6 continuum emission contours are overlapped at levels 5, 15 and 100 times the rms.}
\label{f:h2scube} 
\end{figure*}

\begin{figure*}
  \includegraphics[width=\linewidth]{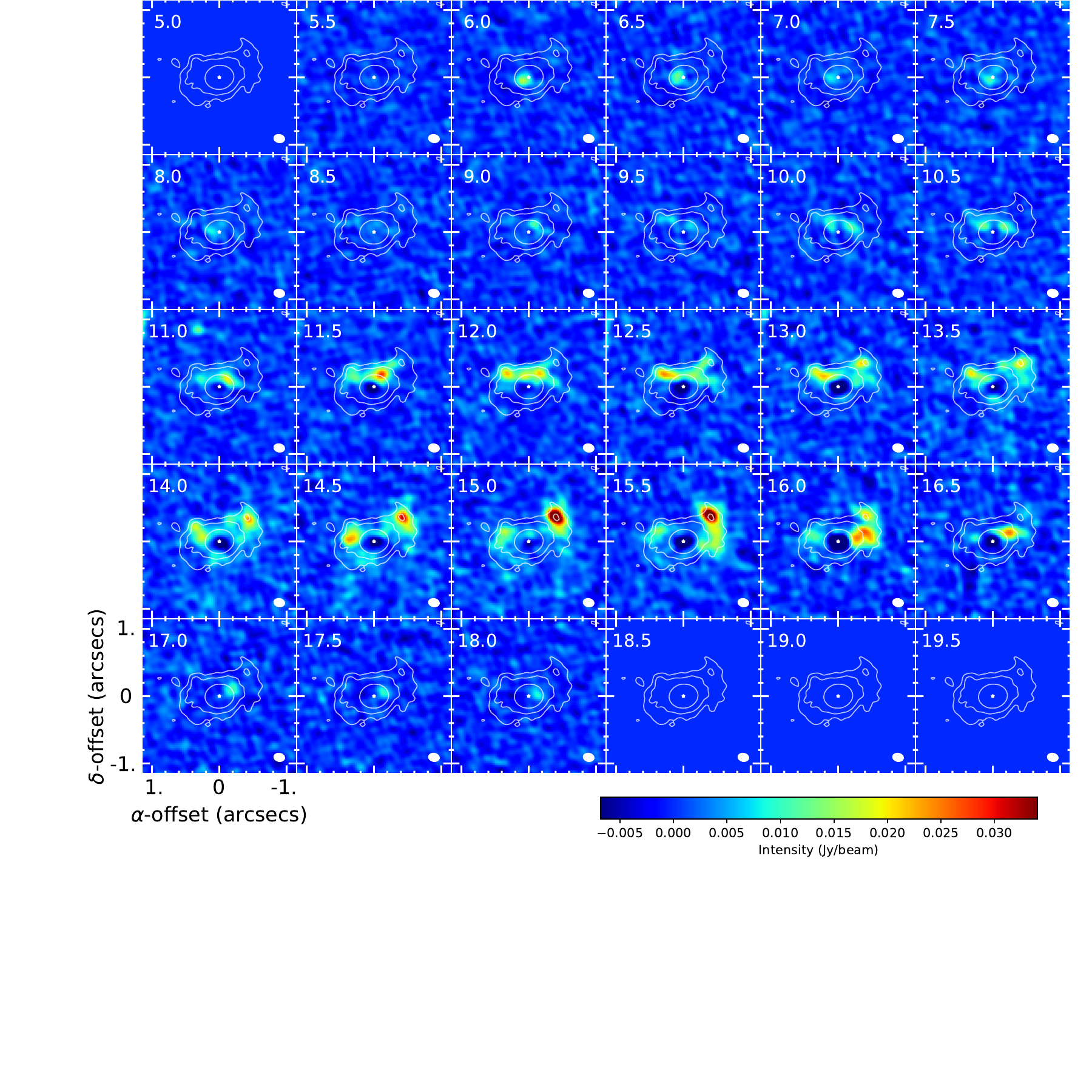}
\caption{Zoom up velocity cube image of the HC$_3$N\,(33-32) emission toward GGD27-MM1. Velocity is labeled in the top left corner and the synthesized beam appears in the bottom right corner of each channel. Band 6 continuum emission contours are overlapped at levels 5, 15 and 100 times the rms.}
\label{f:hc3ncube} 
\end{figure*}

\begin{figure*}
  \includegraphics[width=\linewidth]{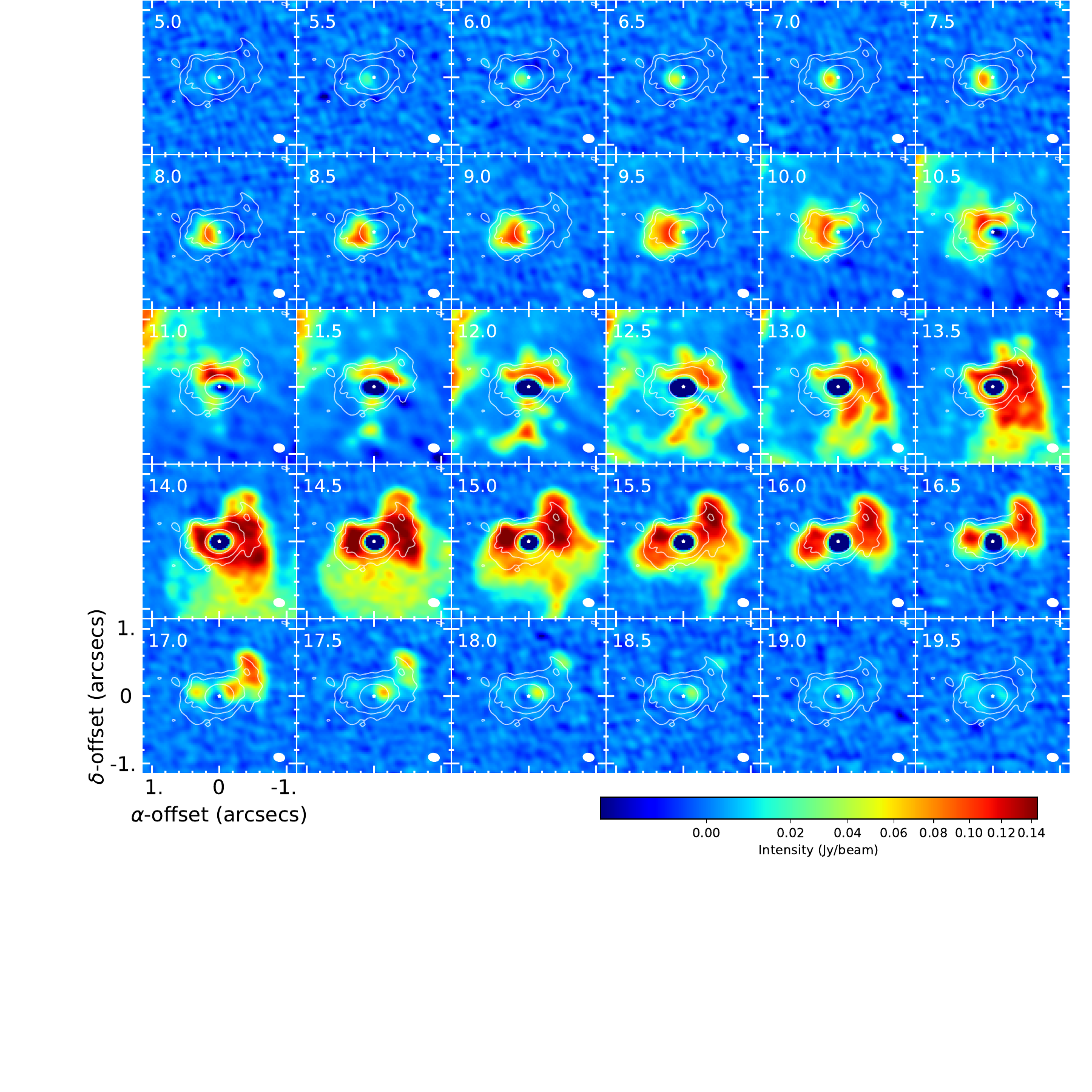}
\caption{Zoom up velocity cube image of the H$_2$CO~(4$_{1,3}$-3$_{1,2}$) emission toward GGD27-MM1. Velocity is labeled in the top left corner and the synthesized beam appears in the bottom right corner of each channel. Band 6 continuum emission contours are overlapped at levels 5, 15 and 100 times the rms.}
\label{f:h2cocube} 
\end{figure*}

\begin{figure*}
  \includegraphics[width=\linewidth]{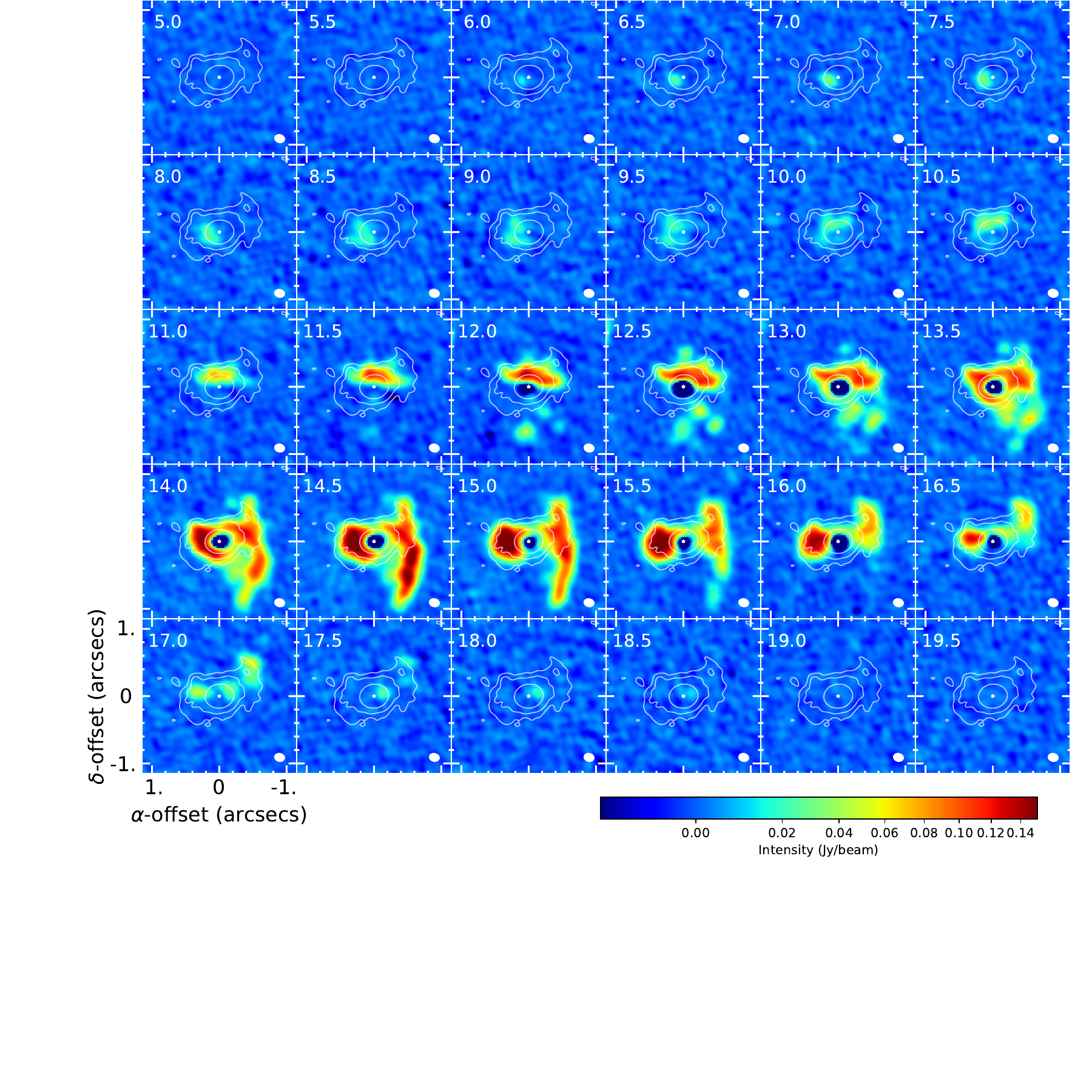}
\caption{Zoom up velocity cube image of the CH$_3$OH~(3$_{1,2}$-2$_{0,2}$) emission toward GGD27-MM1. Velocity is labeled in the top left corner and the synthesized beam appears in the bottom right corner of each channel. Band 6 continuum emission contours are overlapped at levels 5, 15 and 100 times the rms.}
\label{f:ch3ohcube} 
\end{figure*}

\begin{figure*}
  \includegraphics[width=\linewidth]{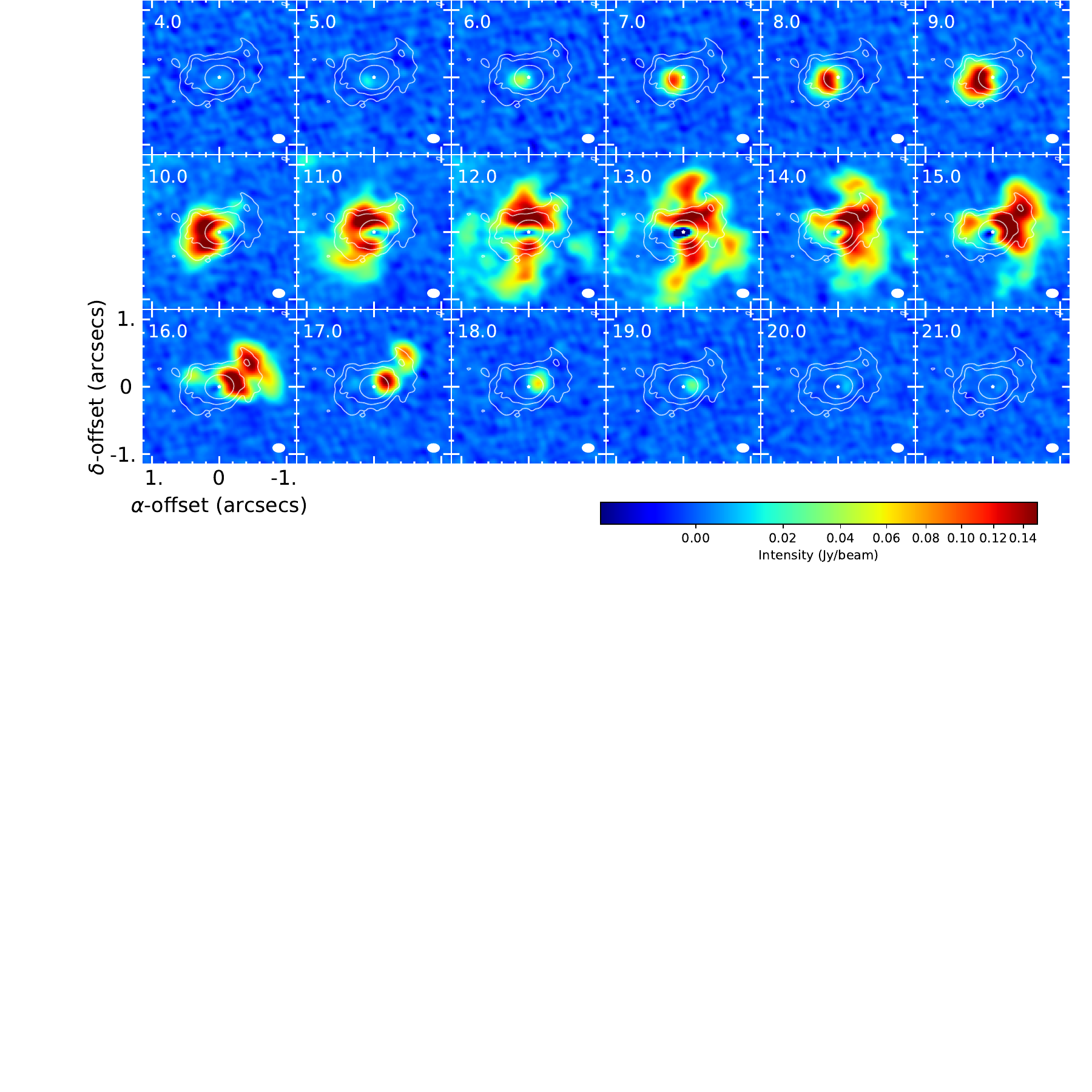}
\caption{Zoom up velocity cube image of the SO$_2$\,(9$_{2,8}$-8$_{1,7}$) emission toward GGD27-MM1. Velocity is labeled in the top left corner and the synthesized beam appears in the bottom right corner of each channel. Band 6 continuum emission contours are overlapped at levels 5, 15 and 100 times the rms.}
\label{f:so2cube} 
\end{figure*}

\begin{figure*}
  \includegraphics[width=\linewidth]{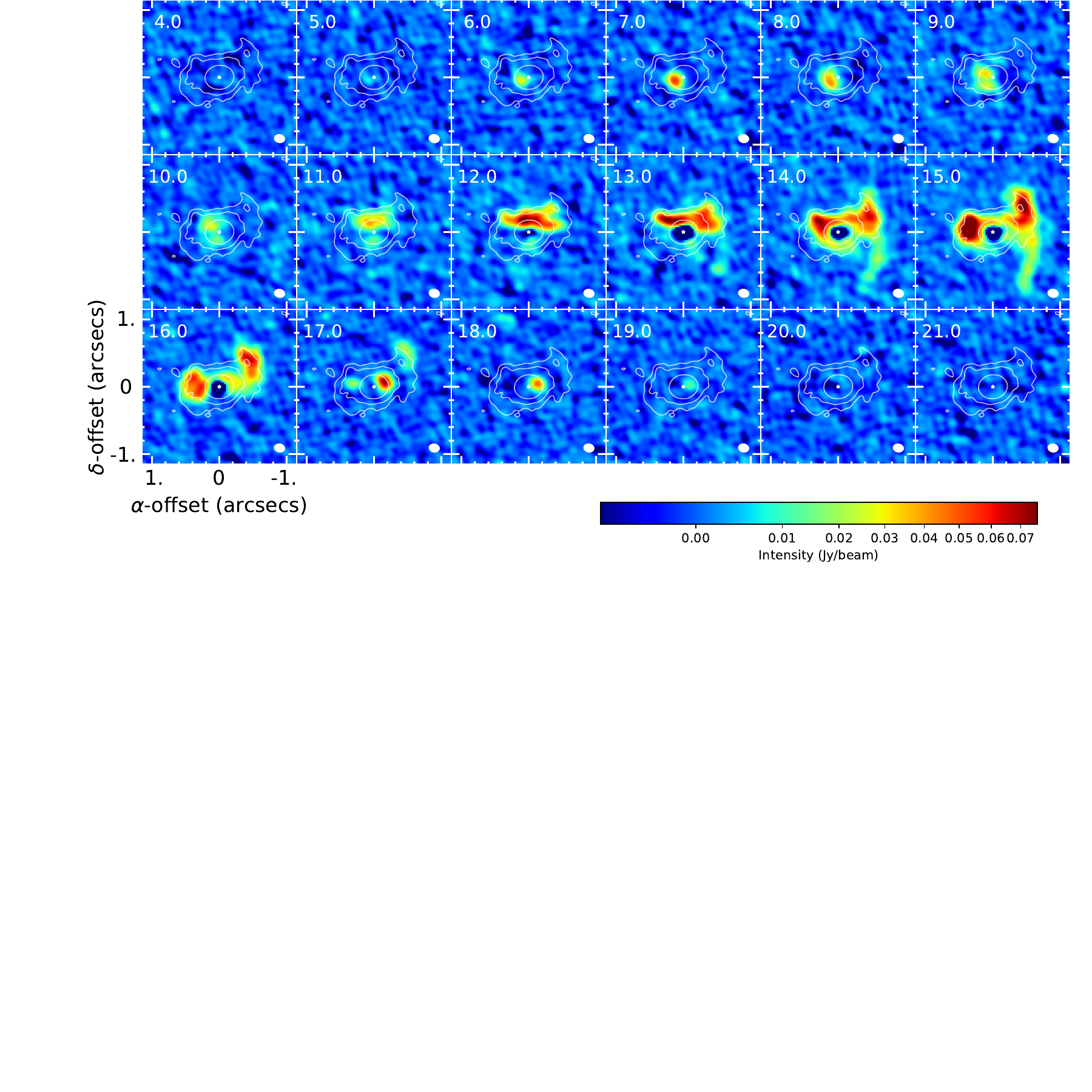}
\caption{Zoom up velocity cube image of the CH$_3$CN~(17$_3$-16$_3$) emission toward GGD27-MM1. Velocity is labeled in the top left corner and the synthesized beam appears in the bottom right corner of each channel. Band 6 continuum emission contours are overlapped at levels 5, 15 and 100 times the rms.}
\label{f:ch3cncube} 
\end{figure*}

\section{$T_k$ and $N_{CH_3CN}$ Fitting Procedure Using CASSIS}\label{sec:appendix_cassis}
\begin{figure*}
\minipage{0.5\linewidth}
\includegraphics[width=\linewidth]{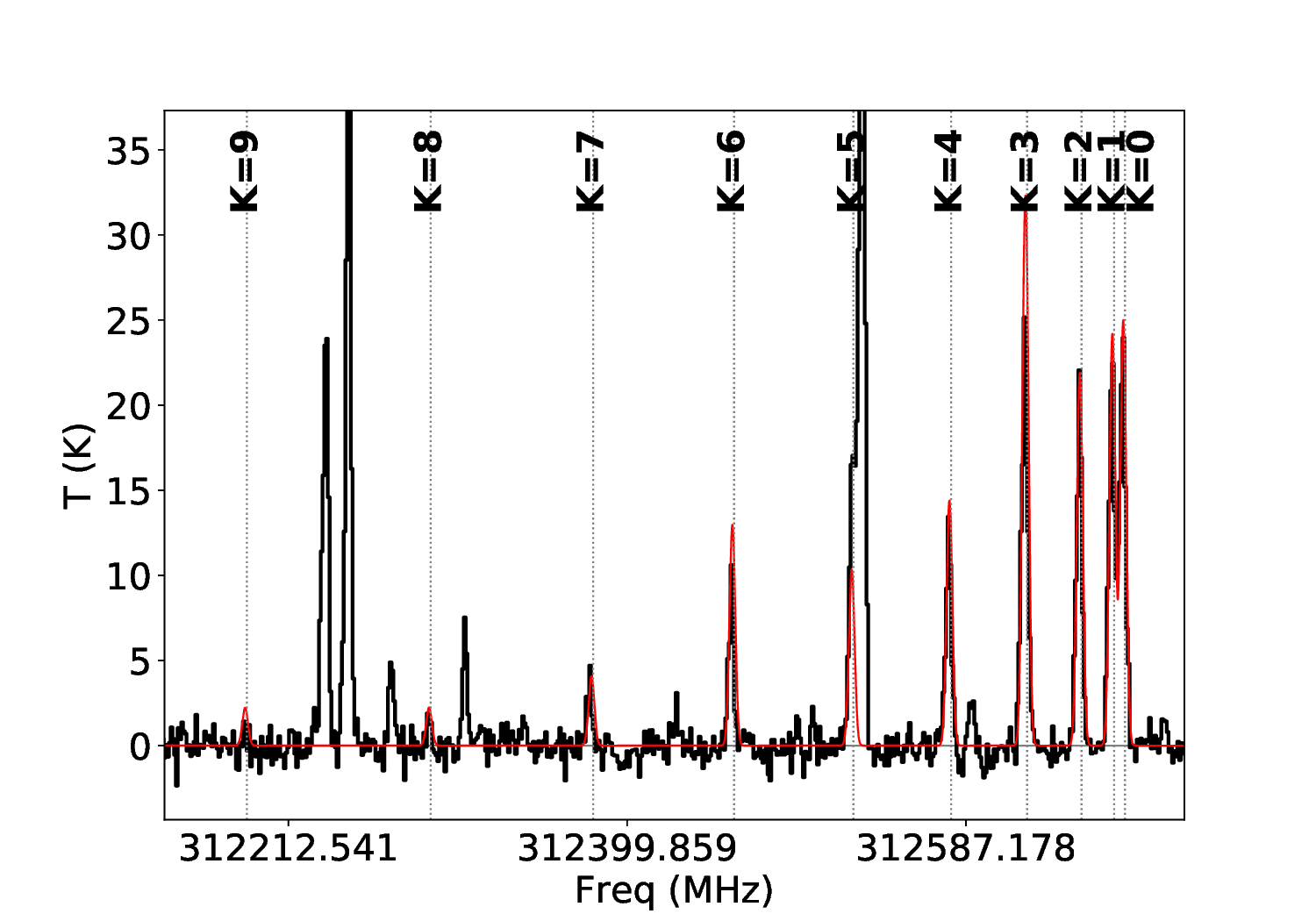}
\endminipage\hfill
\minipage{0.5\linewidth}
\includegraphics[width=\linewidth]{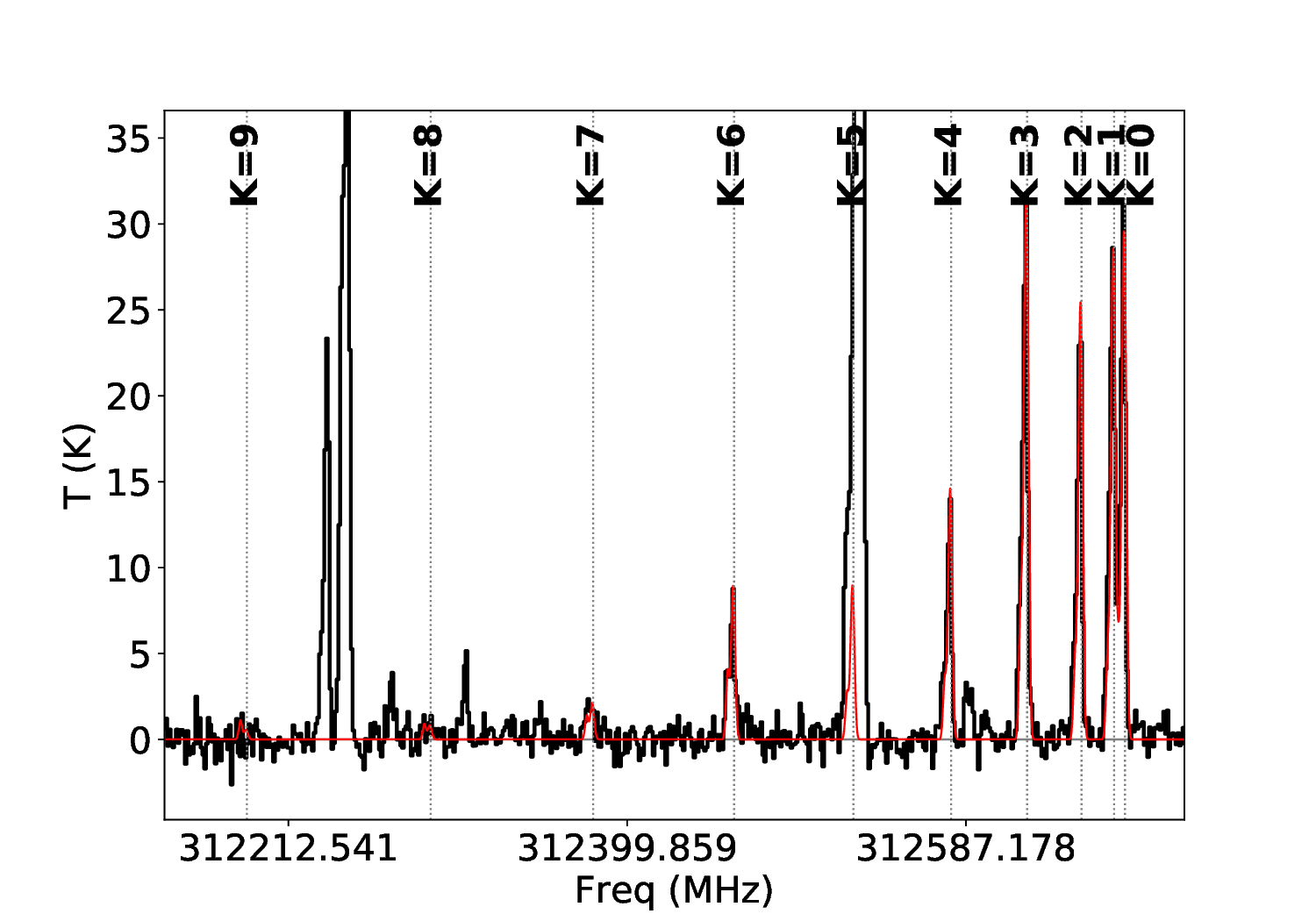}
\endminipage\\\vspace{0.5cm}
\minipage{0.5\linewidth}
\includegraphics[width=\linewidth]{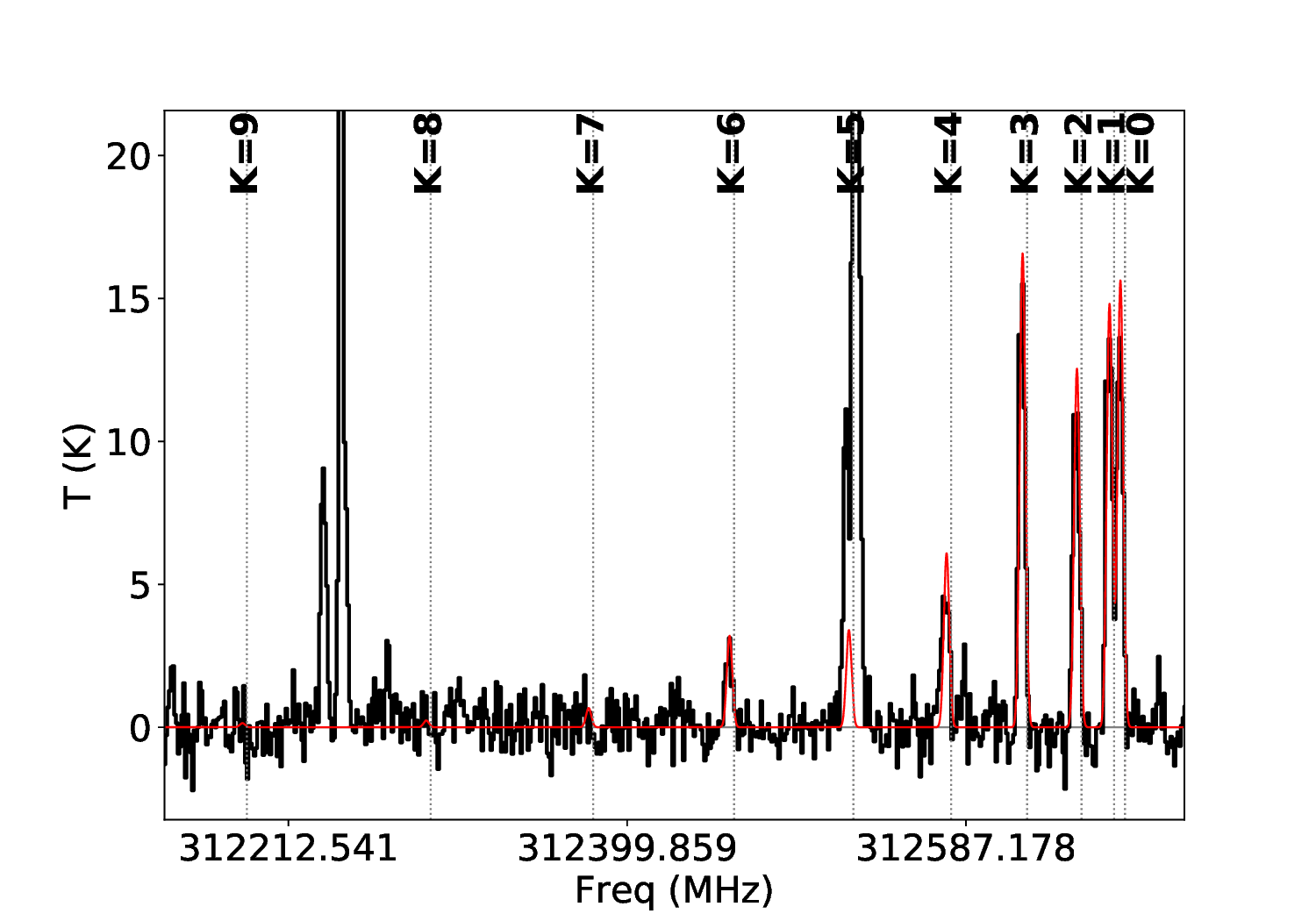}
\endminipage\hfill
\minipage{0.5\linewidth}
\includegraphics[width=\linewidth]{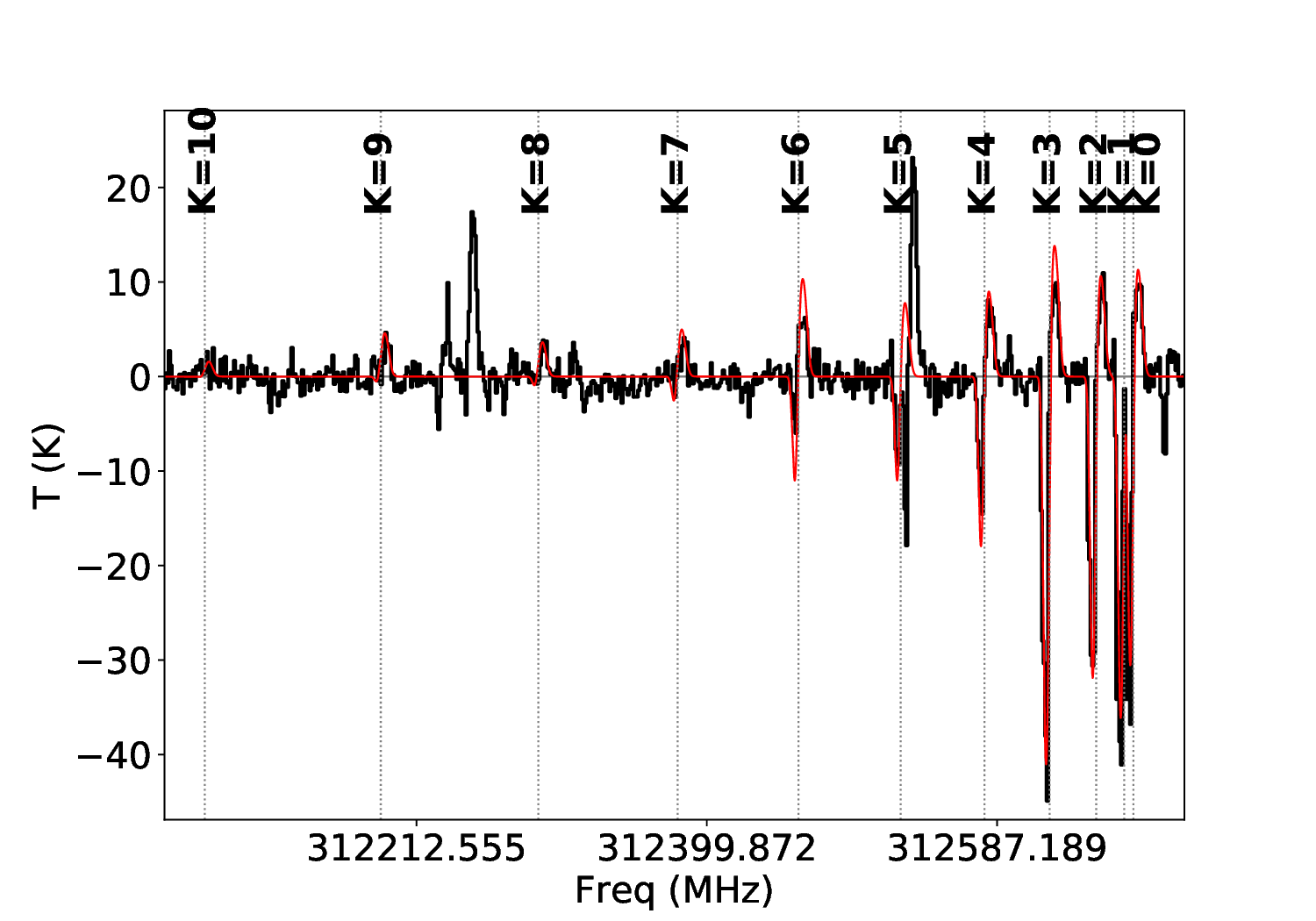}
\endminipage
\caption{Example of spectra fitted manually using CASSIS (pixels 3-9, 4-9, 6-7 and that of the central pixel). The observed spectra is shown by the black histograms, while the CASSIS fitting is displayed by the red curves. The top and bottom left spectra are examples of fits obtained using a single kinematical component; the top right is a fit obtained using two spectral components; the bottom right spectrum corresponds to the central pixel spectrum, which was fit using a constant background of $\sim300$\,K, a blueshifted emitting layer and a foreground absorbing layer slightly redshifted. Note that CH$_3$CN K=5 is contaminated by the SO$_2$ $22_{4,18}-22_{3,19}$ line ($\nu_{rest}=312.543$\,GHz). There are other strong lines (e.g., a CH$_3$OH and a SO$_2$ lines at about 312.24\,GHz) in the spectra, that will be reported somewhere else.}
\label{f:example_fits} 
\end{figure*}

\begin{figure}
\includegraphics[width=\linewidth]{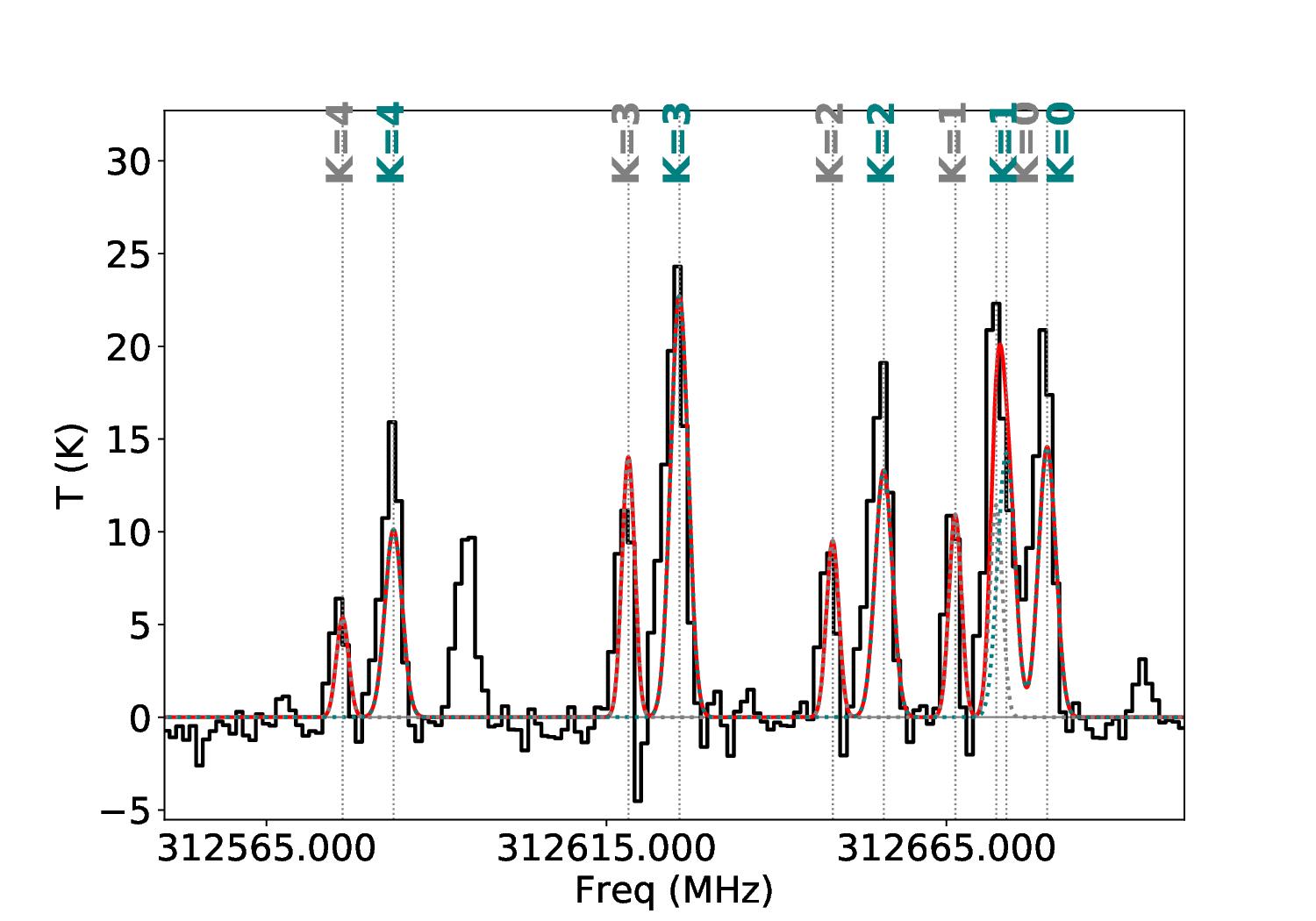}
\caption{Zoom-in example up to K=4 transition of the CH$_3$CN spectrum fitted manually using CASSIS (pixel 3-7). This demonstrates the need for two kinematical components  and the relative goodness of the fit. The observed spectrum is shown by the black histogram, each of the two components has its spectrum plot in grey and blue dotted lines, and the red solid line represents the total fit after adding the two components contributions.}
\label{f:example_fits2} 
\end{figure}

\begin{figure}
\includegraphics[width=\linewidth]{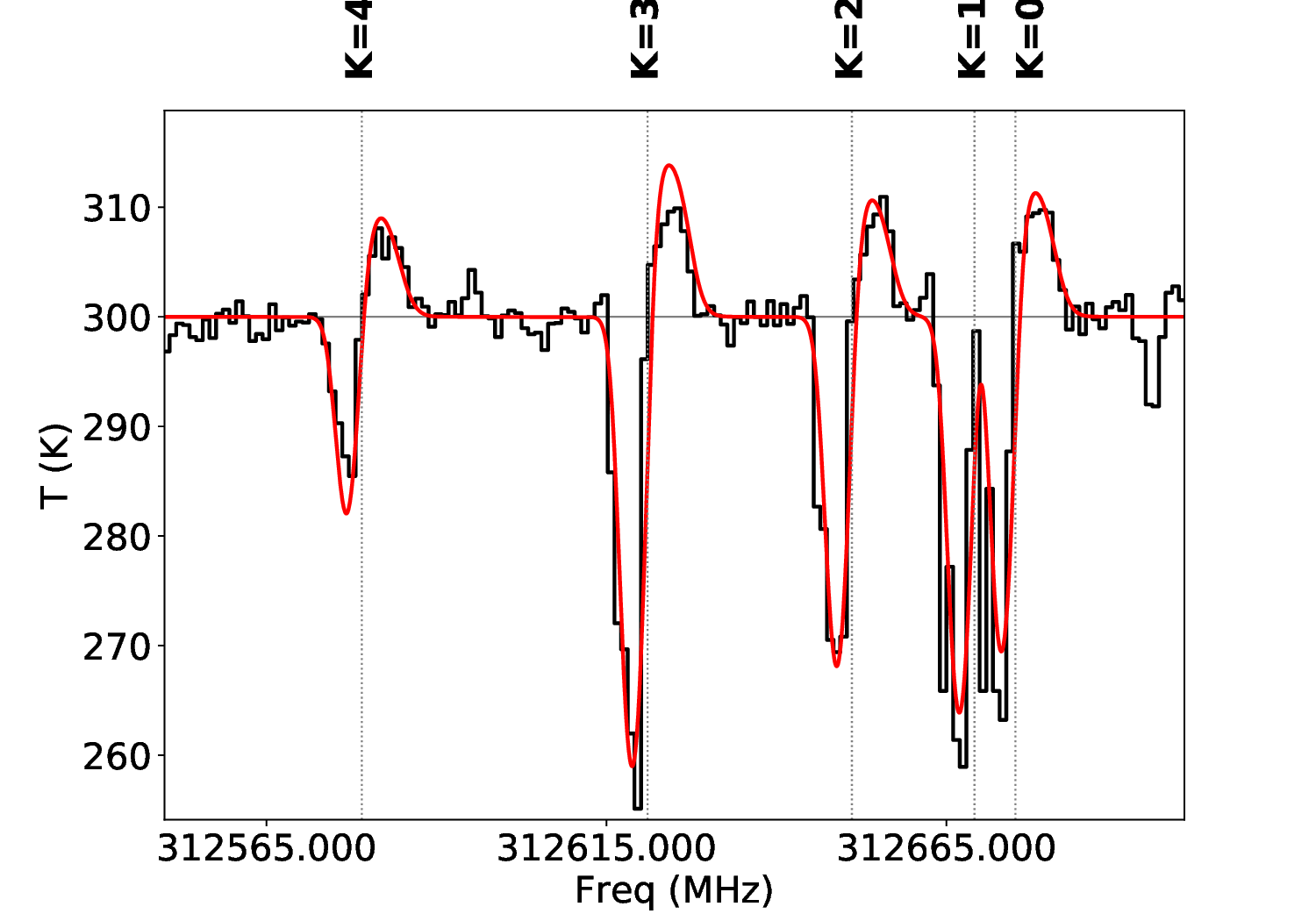}
\caption{Zoom-in example up to K=4 transition of the CH$_3$CN spectrum fitted manually using CASSIS at the central pixel. This demonstrates the relative quality of the fit fixing the background temperature at 300\,K and using two interactive emitting/absorbing layers. The observed spectrum is shown by the black histogram and the red line represents the model.}
\label{f:example_fits3} 
\end{figure}

We analyzed the CH$_3$CN data using CASSIS \citep{2015Vastel}. For each beam-sized pixel we fit one or two spectral components manually, except for the central pixel (see Fig. \ref{f:example_fits}, \ref{f:example_fits2}, and \ref{f:example_fits3}). We take into account lines with peaks over the 4$\sigma$ level ($\sigma=0.72$\,K), avoiding the K=5 transition, whose emission is blended with SO$_2$\,(22,4,18-22,3,19). We fixed the size of all the spectral components as if they would fill the beam ($0\farcs15$) and set the telescope to ALMA with a 1300\,m baseline. Our trials sample a range of excitation temperatures (10--600\,K), column densities ($8\times10^{13}$--$8\times10^{15}$\cmd), line widths (0.1--5\kms), and central velocities around the 12.1\kms system velocity. We estimate that uncertainties may be 20\,K for the excitation temperatures, a 10-20\% in the CH$_3$CN column densities, and $0.1-0.25$\kms for both the line width and the offset velocities. The pixels with multiple spectral components have been fitted using a multi-component approach (\lq\lq interacting\rq\rq\xspace layer in CASSIS). After all the pixels were fitted, we identified two separate kinematical and temperature components (disk and envelope), used to construct maps of temperature, column density and velocity for each component. The K=9 ladder line is detected in several pixels. At the central pixel, even the K=10 line is detected (Fig. \ref{f:example_fits}).

\section{Infall Model Comparison}\label{sec:appendix_model}
The analytic accretion flow model for the material in rotating and collapsing to a central gravitational object was presented by \citet{2009Mendoza}. This model, is an expansion of the model presented by \citet{1976Ulrich}, the differences are that the radius of the rigid body rotating cloud is finite and the particles on the border of the cloud, they can have a radial velocity component, that is, the velocity aim to the center of the coordinates system. The model Mendoza's model depends of the two parameters associated with these differences, $\mu$ and $\nu$. In such a way if these parameters are $\mu=0$ and $\nu=0$, the Mendoza's model converges to the Ulrich's model. Under these differences, the Mendoza's produces orbits described by conic sections and the Ulrich's model produces parabolic orbits. If the infalling material does not shock with its counterpart or with the accretion disk, this material could return to the border of the cloud, following the trajectory described by a parabolic orbit or by the conic orbit.

\begin{figure*}
\includegraphics[width=\linewidth]{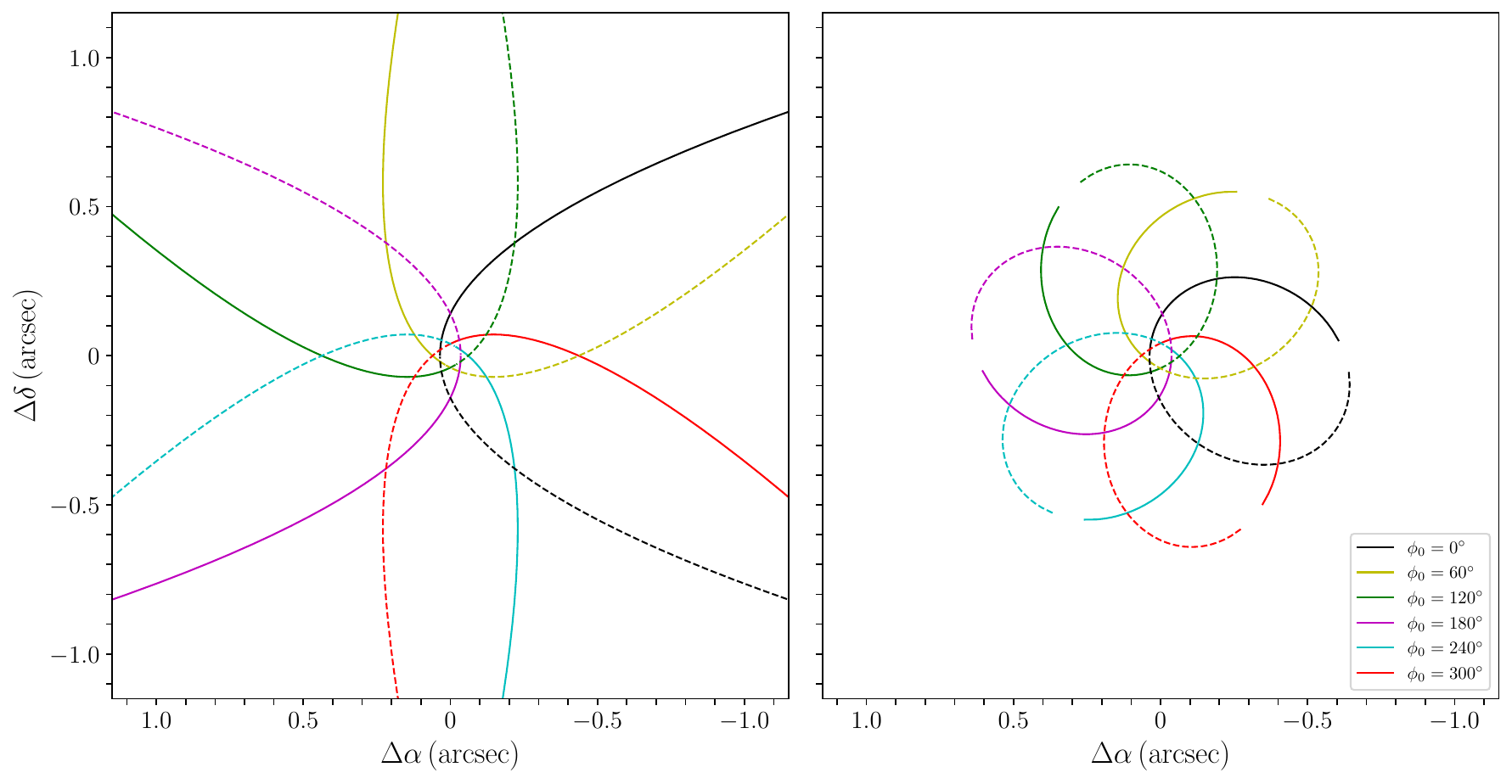}
\caption{Comparison between the orbits projected onto the plane of the sky for an inclination angle $i=0^\circ$, the inner equatorial radius $r_{in}$ as the envelope radius (790 au), the initial polar angle $\theta_0=30^\circ$, and different initial azimuthal angles. Left panel: Ulrich's model. Right panel: Mendoza's model. For the Mendoza's model, we use the parameters $\mu=0.45$ and $\nu=0.15$.  The solid lines represent the infalling material, while the dashed lines denote material coming out}
\label{f:streamt0} 
\end{figure*}

Figure \ref{f:streamt0} presents the orbits projected onto the plane of the sky where we do not consider an inclination angle $i=0^\circ$ and we plot all trajectory of the material, we use an initial polar angle of $\theta_0=30^\circ$ and different initial azimuthal angles, from $\phi_0=0^\circ$ to $\phi_0=300^\circ$ with a difference of $60^\circ$. In order to dimension these orbits, we assume the inner equatorial radius equal to the envelope radius, $r_{in}=790$ au. The left panel corresponds to the parabolic orbits, the Ulrich's model, and the right panel presents the conic orbits, the Mendoza's. For the model Mendoza's model we use the parameters $\mu=0.45$ and $\nu=0.15$, since these parameters are not zero, the conic orbits are smaller and curved that the parabolic orbits because the radius of the cloud is smaller.

\begin{figure*}
\includegraphics[width=\linewidth]{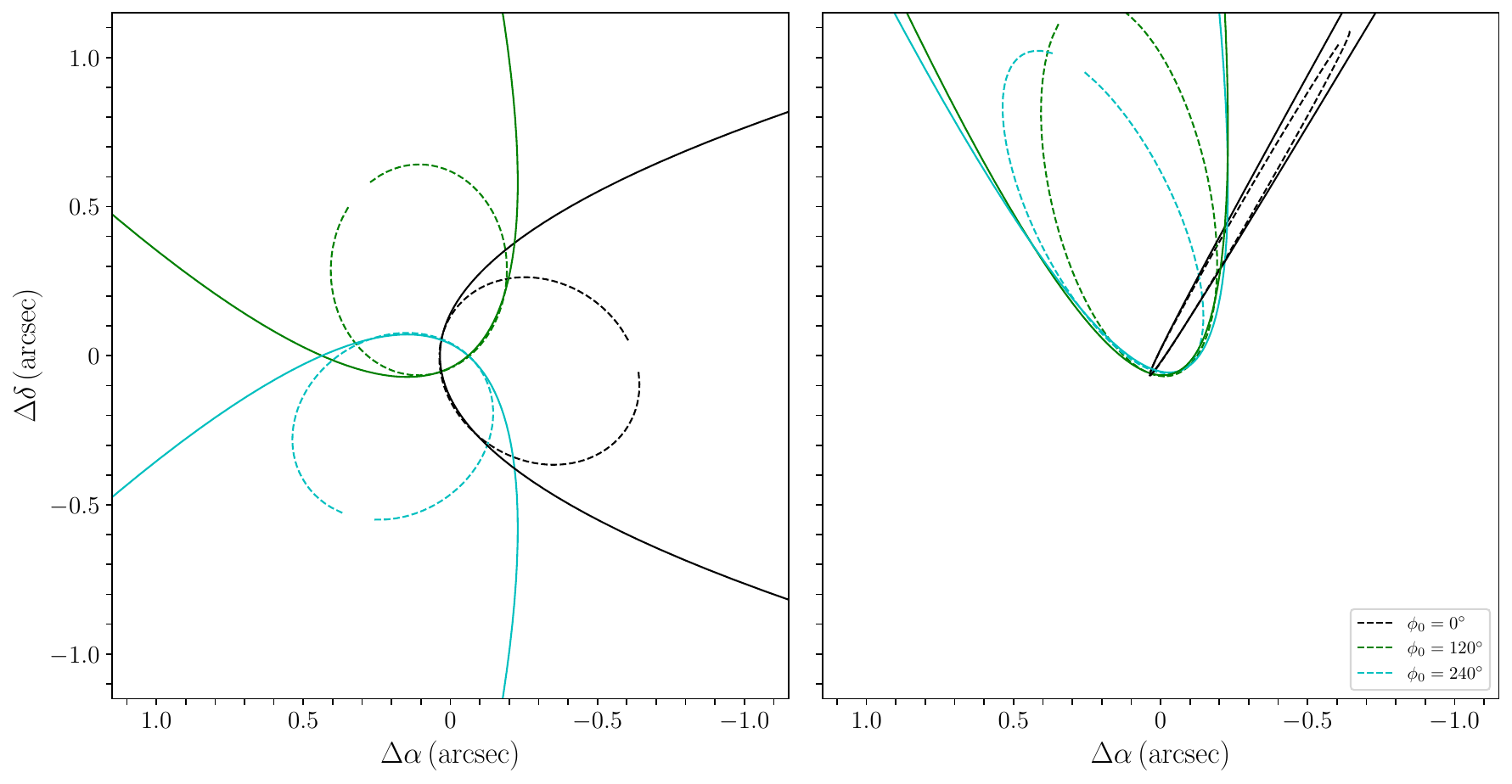}
\caption{Comparison between the orbits projected onto the plane of the sky for an initial polar angle $\theta=30^\circ$ and the envelope radius ($r_{in}=790$ au). Parabolic orbits, following the Ulrich's model are shown in solid lines. Conic orbits, following the Mendoza's model are shown in dashed lines. Left panel: Orbits for an inclination angle respect to the plane of the sky $i=0^\circ$. Right panel: Orbits for an inclination angle respect to the plane of the sky $i=30^\circ$. For the Mendoza's model we consider the parameters $\mu=0.45$ and $\nu=0.15$.}
\label{f:streami0} 
\end{figure*}

For the same parameters $\mu$ and $\nu$, the same angle $\theta_0$, the initial azimuthal angles from $\phi_0=0^\circ$ to $\phi_0=240^\circ$ with a difference of $120^\circ$, and the same inner radius $r_{in}$, the Figure \ref{f:streami0} shows an comparison between the both models. The solid lines represent the orbits of the Ulrich's model and the dashed lines are the orbits produced by the Mendoza's model. The left panel shows an orbits projected onto the plane of the sky for an inclination angle $i=0^\circ$ and the right panel presents orbits with an inclination angle of $i=30^\circ$.

\begin{figure*}
    \includegraphics[width=\linewidth]{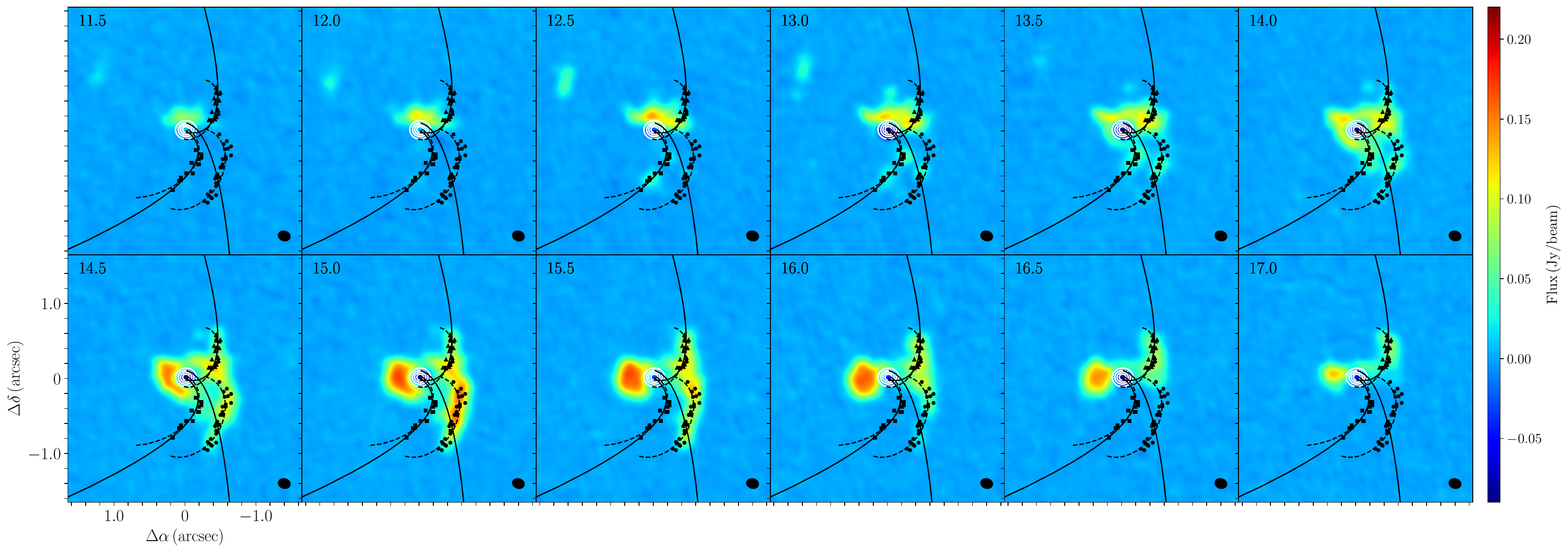}
    \caption{Ulrich (solid lines) and Mendoza (dashed lines) orbits overapped with moment 0 of CH$_3$OH $3_{(1,2)}-2_{(0,2)}$ emission. The parameters $\mu$ and $\nu$ and the angles $\phi_0$ and $\theta_0$ for the streamers S1, S2, and S3 (see Figure \ref{f:sketch}) are shown in Table \ref{t:parametersmodels}. The models using the inner equatorial radius $r_{in}$, as the molecular disk radius (357 au).}
    \label{f:ch3ohfitting1}
\end{figure*}

\begin{figure*}
    \includegraphics[width=\linewidth]{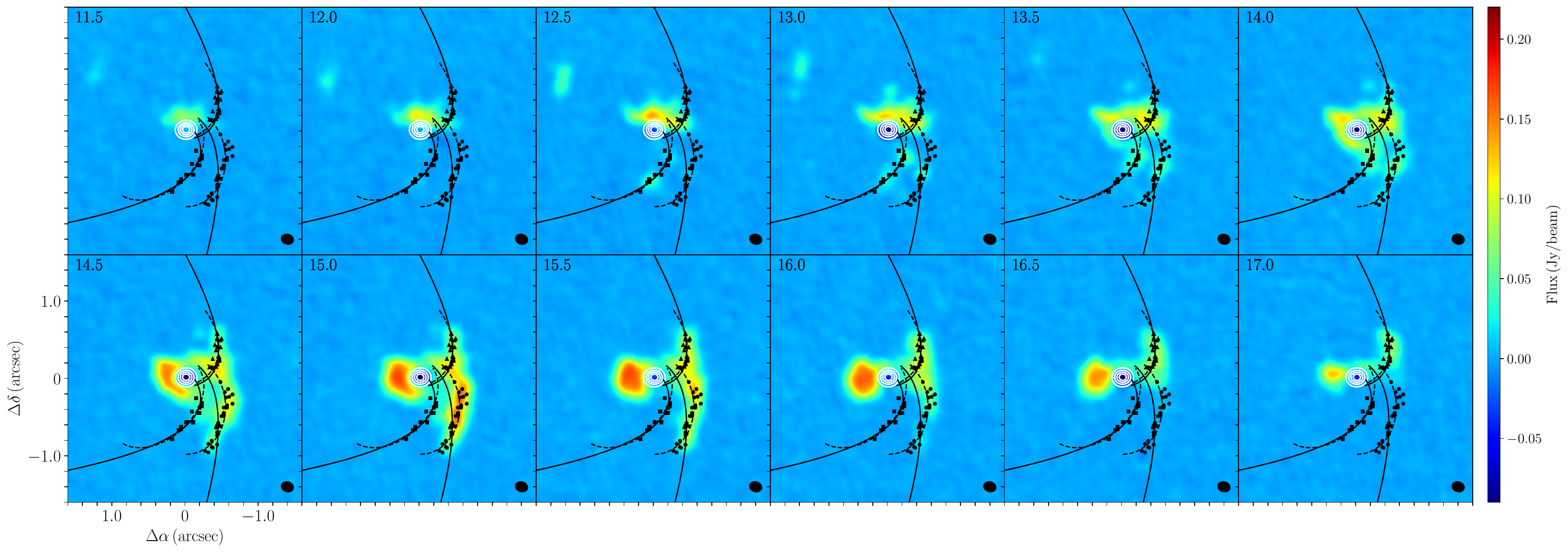}
    \caption{Ulrich (solid lines) and Mendoza (dashed lines) orbits overapped with moment 0 of CH$_3$OH $3_{(1,2)}-2_{(0,2)}$ emission. The same description as Figure \ref{f:ch3ohfitting1} but models using the envelope radius $r_{in}=$ 790 au.}
    \label{f:ch3ohfitting2}
\end{figure*}

The models mentioned above can reproduce the streamers S1, S2, and S3 (see Figure \ref{f:sketch}) as show in Figures \ref{f:ch3ohfitting1} and \ref{f:ch3ohfitting2}. The Ulrich's model (solid lines) and the Mendoza's model (dashed lines) orbits are overlapped with the integrated intensity emission (moment 0) of CH$_3$OH $3_{(1,2)}-2_{(0,2)}$ emission for different velocity channels. We use the parameters $\mu$ and $\nu$ and the angles $\phi_0$ and $\theta_0$ of the Table \ref{t:parametersmodels}. The Figure \ref{f:ch3ohfitting1} is for inner equatorial radius of $r_{in}=357$ au (the molecular disk radius) and the Figure \ref{f:ch3ohfitting2} is for $r_{in}=790$ au (envelope radius). The triangles, circles, and squares in Figure \ref{f:ch3ohfitting1} and Figure \ref{f:ch3ohfitting2} mark the position of the condensations identified in the CH$_3$OH velocity cube.

\section{Flared Disk Model Details }\label{sec:appendix_flared_disk}

In this Appendix we detail how the H$_2$S disk model is constructed. We are assuming a protoplanetary disk around a Young Stellar Object (YSO), observed in some molecular transition, and we want to model the integrated intensity (zeroth moment) of the line. The disk has three different components:
%
i) a central dust disk, with a radius $\R{dust}$ smaller than the inner radius of the
molecular disk. It has no line emission, and is seen in absorption after
subtracting the continuum emission;
ii) a molecular disk, which is flared and illuminated by the central YSO. Only
the two faces of the disk are hot and have line emission. The flared disk
extends from an inner radius $\R{inn}$ to an outer radius $\R{out}$;
and iii) cold gas that fills the volume of the flared disk, with a non-zero optical
depth, responsible for the absorption of the continuum emission of the dust
disk.

\subsection{Model parameters}
The model depends on a number of six parameters describing the geometry of the system, and five describing the integrated intensities and optical depth. The geometrical parameters are:
\begin{description}
\item[$\theta$] Position angle of the projection of the disk axis on the sky, i.e., of the minor axis of the ellipse projected on the sky. With this definition, the position angle of the major axis of the ellipse is $90^\circ+\theta$.

\item[$i$] Inclination angle of the disk, i.e., angle between the disk axis and the line-of-sight. For a face-on disk, $i=0$.

\item [$\R{dust}$] Radius of the dust disk. The dust disk is assumed to be flat, with a half-width $\H{dust}=\R{dust}\tan\phi$.

\item[$\R{inn}$, $\R{out}$] Inner and outer radii of the molecular disk.

\item[$\phi$] Flare angle of the molecular disk. The width of the disk increases linearly with radius. For a radius $r$, the half-height of the disk is given by $h=r\tan\phi$.
\end{description}
The parameters that describe the physical conditions of the disk are:
\begin{description}
\item[$\I{dust}$] Integrated intensity of the dust emission. The dust emission is assumed to be optically thick.

\item[$\I{face}$] Integrated intensity of the face of the molecular disk.


\item [$\I{cold}$] Integrated intensity of the cold gas, assumed to be uniform. The cold gas fills a cylindrical volume of radius $\R{out}$ and half-width 
$\H{cold}=\R{cold}\tan\phi$.

\item [$\tau_0$] Optical depth of the cold gas for a geometrical length
$\R{out}$. For a given path-length $l$ inside the cold gas, the optical depth is
given by $\tau_0(l/\R{out})$.
\end{description}

\subsection{Calculation of the intensity}
The model calculates, for every position of the model map, the
intersections of the different disk components with the line-of-sight. 
The surfaces limiting the different components can be cylindrical (outer and
inner sides of a disk), flat (top and bottom lids of a cylindrical disk), or
conical (faces of a flared disk). 
For each one of these surfaces the intersection with the line-of-sight is found.

For every position of the model map, once the intersections of the line-of-sight
have been found, they are sorted from farthest to closest to the observer. 
In this way, the line-of-sight is partitioned in segments of path inside 
the different disk components (assumed to be homogeneous) or the vacuum. 
For each part of the path, the exiting integrated intensity is calculated
from the radiation transfer equation,
\beq
I= \I{bg} e^{-\tau} + S (1-e^{-\tau}),
\eeq
where
$\I{bg}$ is the integrated intensity calculated from the previous path segment
($\I{bg}$ is taken as 0 for the first intersection);
$S$ is the source function of the component, $\I{dust}$ or $\I{cold}$, or
0 for the vacuum;
and 
$\tau$ is the optical depth, $\infty$ for the dust disk, $0$ for the
vacuum, and $\tau_0(l/\R{out})$ for a geometrical path $l$ in the cold disk.
The case of an intersection with the face of the flared disk is special, and the integrated intensity of the face, $\I{face}$, is added to the exiting intensity.
Finally, when the total intensity is obtained, the dust continuum emission
$\I{dust}$ is subtracted if the line-of-sight intersects the dusk disk.

Once the integrated intensity map is obtained, the model map to be compared with the observation is calculated as the convolution of the intensity map with a Gaussian beam equal to the synthesized beam of the observation.

\subsection{Fitting procedure}
The fitting engine is the same as that used in \citet{2017Estalella}.
The procedure samples the multi-dimensional parameter space using a Sobol pseudo-random sequence.
For each sample, the model is computed, obtaining the integrated intensity map, the residual map, and the residual rms. 
The total number of samples of a fitting run is evenly distributed between a number of loops, a number of seeds, and a number of descendants for each seed.
For the first loop, initial values and search ranges are assigned to the fit parameters, and the seeds of the parameter space within the ranges of each parameter are selected randomly around the initial values. 
For the next loops, the descendants of each seed are selected randomly in the parameter space around the seed. 
The descendants of all seeds for which the chi-square fit residual is lower are taken as seeds for the next loop, for which the search ranges are decreased a constant factor. 
The procedure stops after the number of loops is completed.

The uncertainty in the values derived for the parameters is estimated as the increment of the parameters for which the chi-square residual does not exceed the minimum value by an amount $\Delta(m,\alpha)$, where $m$ is the number of parameters fitted and $\alpha$ is the significance level (0.68 for a 1-$\sigma$ uncertainty). The value of $\Delta(m,\alpha)$ is given by a chi-square probability distribution with $m$ degrees of freedom.
For a more detailed description see \citet{2017Estalella}.

\end{document}